\documentclass[a4paper,12pt,twoside]{article}
\pdfoutput=1
\usepackage{abstract}

\usepackage[utf8]{inputenc}
\usepackage[T1]{fontenc}

\usepackage[english]{babel}

\usepackage[a4paper,top=2cm,bottom=2cm,left=3cm,right=3cm,marginparwidth=1.75cm]{geometry}
\usepackage{amsmath}
\usepackage{graphicx}
\usepackage{hyperref}
\hypersetup{colorlinks=true, allcolors=blue,pdfproducer={}}
\usepackage{fancyhdr}
\usepackage{wrapfig}
\usepackage{subcaption}
\usepackage[braket]{qcircuit}
\usepackage{url}
\usepackage{siunitx}
\usepackage{multirow}

\title{Proceedings of the IFJ PAN\\Particle Physics Summer Student\\Alumni Conference 2022\\(Kraków, 9 -- 10 July 2022)}
\author{Dominik Derendarz, Rafał Staszewski, Maciej Trzebiński}
\date{October 2022}

\input{papers.cls}

\newcommand{\ResCnt}{\setcounter{figure}{0} \setcounter{table}{0} \setcounter{section}{0} \setcounter{equation}{0}}

\makeatletter
\def\@maketitle{%
    \newpage
    \null
    \vskip 2em%
    \begin{center}%
    \includegraphics[width=1.0\textwidth]{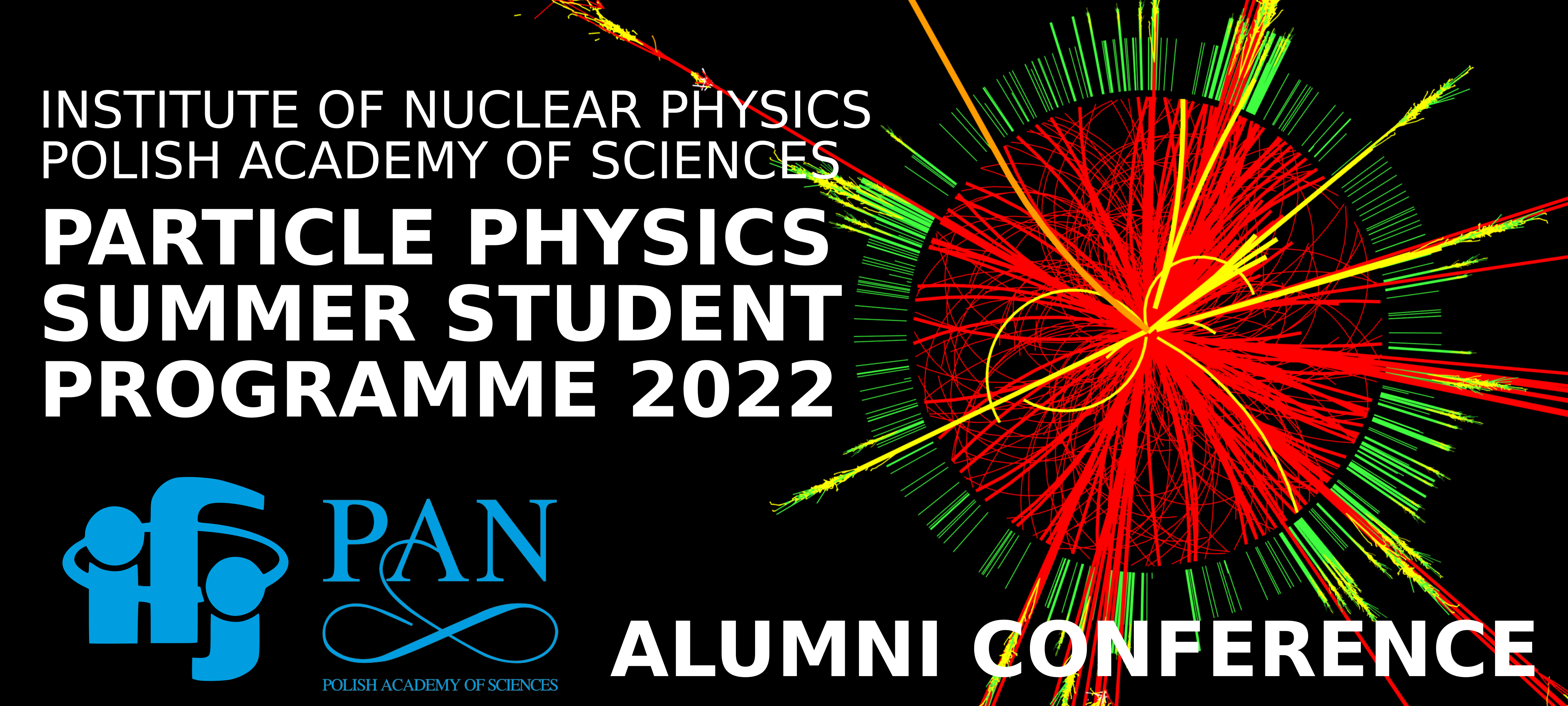}
    \let \footnote \thanks
    \vskip 5em%
    {\linespread{1.3}\Huge\bfseries \@title \par}%
    \vskip 8em%
    {\Large Edited by:\\[2pt]
      \begin{tabular}[t]{c}%
        \@author
      \end{tabular}\par}%
    \vskip 8em%
    {\Large \@date}
  \end{center}
  \par
  \vskip 1.5em}
\makeatother

\begin{document}

\begin{titlepage}
  \maketitle
  \thispagestyle{empty}
  \setcounter{page}{0}
\end{titlepage}

\newpage
\thispagestyle{empty}
\vspace*{20cm}
\noindent Published by The Henryk Niewodnicza\'nski Institute of Nuclear Physics Polish Academy of Sciences\\[0.4cm]
Krak\'ow 2022\\[0.4cm]
Reviewers: dr Domink Derendarz, dr Rafa{\l} Staszewski, dr Maciej Trzebi\'nski\\[0.4cm]
ISBN: 978-83-63542-30-6\\
https://doi.org/10.48733/978-83-63542-30-6
\newpage

\setcounter{tocdepth}{0}
\thispagestyle{empty}
\tableofcontents 
\clearpage

\newpage
\thispagestyle{empty}
\mbox{\ }
\setcounter{page}{0}
\newpage

\begin{papers}

\maketitle

\begin{abstract}
The idea of particle physics summer student programme was born around 2012. We realised that at the Institute of Nuclear Physics Polish Academy of Sciences a very nice offer for students, including possibilities to continue as a PhD, is not really advertised. Simply, there is no BSc. nor MSc. programme at IFJ PAN and the Institute is often unknown to students. Although there always were the possibilities to do internships or BSc./MSc. theses under the supervision of employees of the Institute, not many students were responding to calls. Our solution was to propose an annual, well recognised and highly rated summer student programme!
\end{abstract}

\section{First Edition, 2013}

The first edition, limited to a few students from Krakow universities was organised in 2013. It had a very local character, with advertisements done via posters pinned by us in the physics departments of Krakow universities and 1 or 2 seminars given to physics students. There are almost no documents in our archives and details of this first edition remain mainly in the memories of Organisers.

At that time we had 7 students from Jagiellonian University, AGH University of Science and Technology and Krakow University of Technology. Already at the first edition the general structure of the programme was established: introductory lectures and exercises in the first week followed by the work with tutor on a project.

It should be noted that a tradition of barbecues in the apiary was born already in 2013. BBQ turned out to be an excellent occasion to integrate students with the ``throwing ball'' game to break the first ice.

In our opinion the first edition was quite successful and encouraged us to continue and make everything bigger na better!

\section{Second Edition, 2014}

With a decision to make the programme an annual event, we created a dedicated webpage and e-mail. Target participants were students from Poland. The advertisement was done via poster (see Fig. \ref{PPSS2014} left) and seminars. Programme started in the second week of July and was very ambitious (see Fig. \ref{PPSS2014} right). Especially ROOT exercises, during which students were supposed to write own Monte Carlo generator and make analysis (event selection). The duration of the programme was very flexible: students could declare between 3 and 6 weeks and, except the first common week, take part through all of the Summer. Twenty nine (29!) topics were proposed by 8 scientists from IFJ PAN ATLAS Experiment Department (NZ14). Seventeen students from 6 universities\footnote{These were: AGH University of Science and Technology, Gdansk University of Technology, Krakow University of Technology, Wroclaw University of Science and Technology, Jagiellonian University, Pedagogical University of Krakow.} participated.

\begin{figure}[!htbp]
    \centering
    \includegraphics[width=0.32\textwidth]{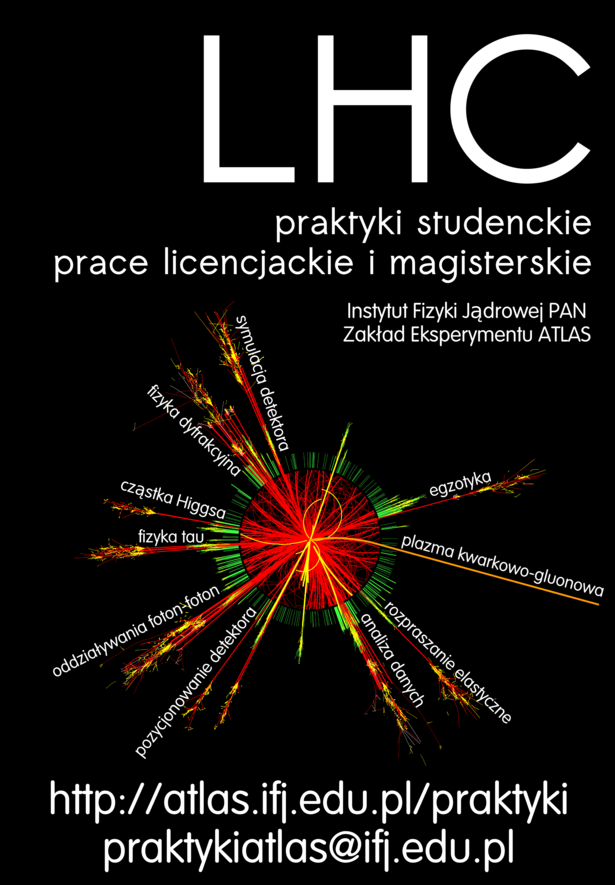}
    \includegraphics[width=0.67\textwidth]{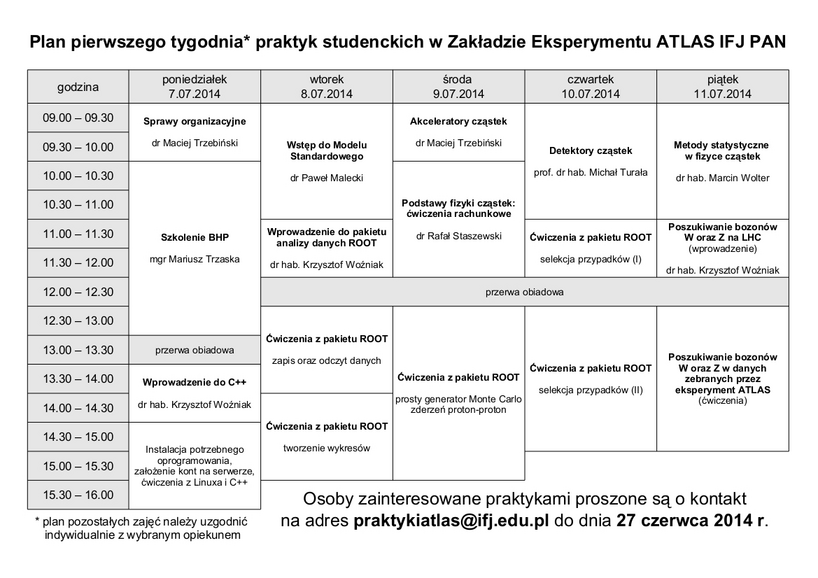}
    \caption{Poster advertising 2014 edition (\textbf{left}) and plan of the first week (\textbf{right}).}
    \label{PPSS2014}
\end{figure}

At the end of 2014 edition we decided to make an anonymous survey. The output (see Fig. \ref{PPSS2014_survey}) was encouraging. Overall organisation and introductory lectures got high marks. ROOT exercises earned lower scores indicating need of change -- extended tutorial and online cheat sheets were planned for the future editions.

\begin{figure}[!htbp]
    \centering
    \includegraphics[width=0.32\textwidth]{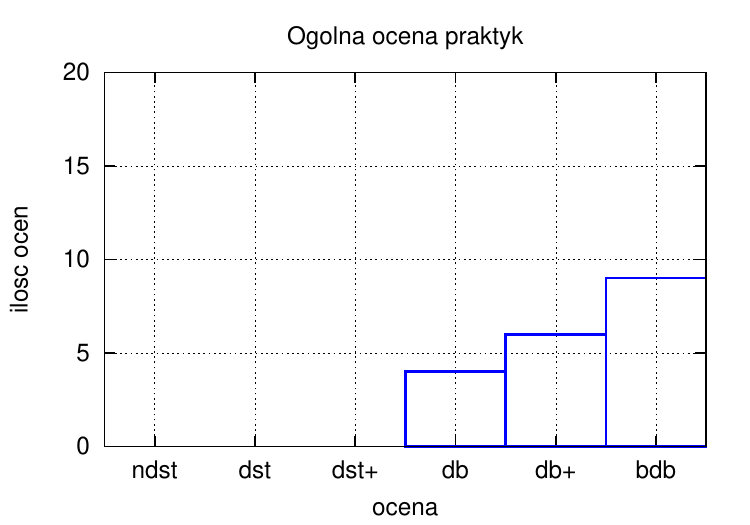}
    \includegraphics[width=0.32\textwidth]{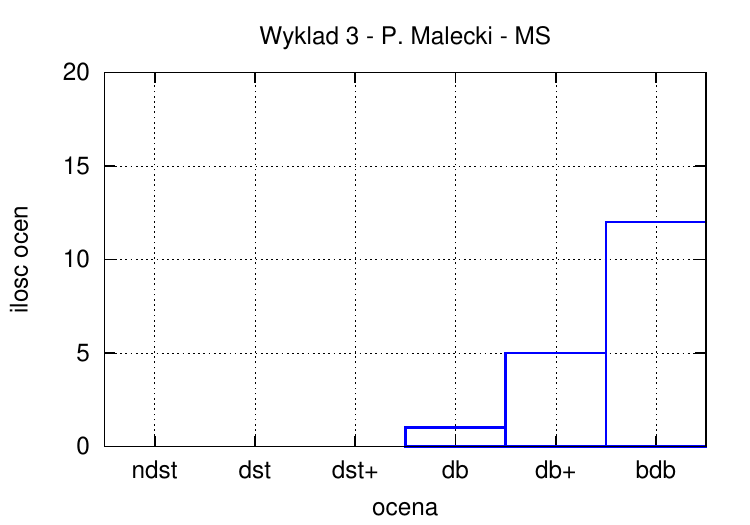}
    \includegraphics[width=0.32\textwidth]{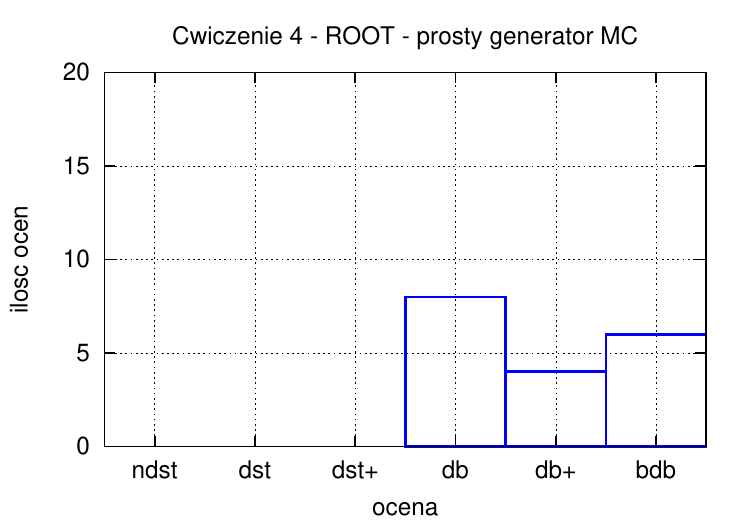}
    \caption{PPSS 2014 anonymous survey: overall opinion about programme (\textbf{left}), Standard Model lecture rating (\textbf{centre}), opinions about one of ROOT exercises (\textbf{right}).}
    \label{PPSS2014_survey}
\end{figure}

\section{Third Edition, 2015}

Growing interest and presence of students from outside Krakow triggered us to search for the financial support. In 2015, it was first edition of PPSS with a small external funding coming from Polish Academy of Sciences (PAS)\footnote{It should be mentioned that Director General of IFJ PAN always strongly supported PPSS programme. Not only administratively but also financially.}.

The third edition of PPSS started on July, 6$^{th}$ with a usual week of lectures -- see Fig. \ref{PPSS2015}. This time we had 40 participants from nine Universities and one high school -- V LO Bielsko-Bia{\l}a. The programme started to be a bit more organised, \textit{e.g.} we defined the duration to be either 3 or 4 weeks in July. As 2015 was a transition period, some individual arrangements with tutors were still possible and a few students participated in August or September.

\begin{figure}[!htbp]
    \centering
    \includegraphics[width=0.32\textwidth]{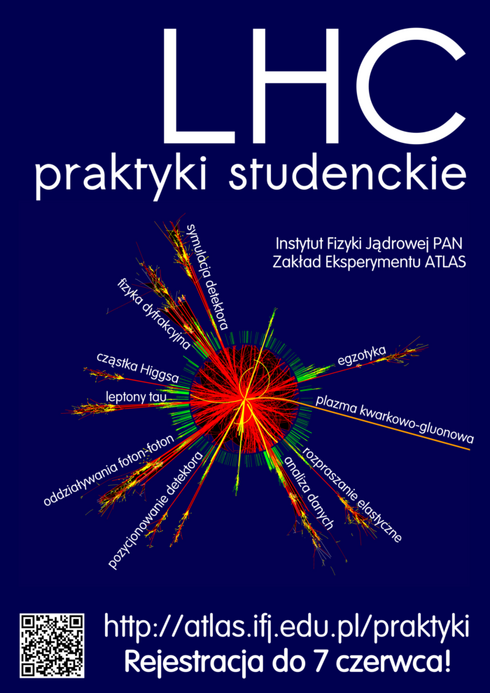}
    \includegraphics[width=0.67\textwidth]{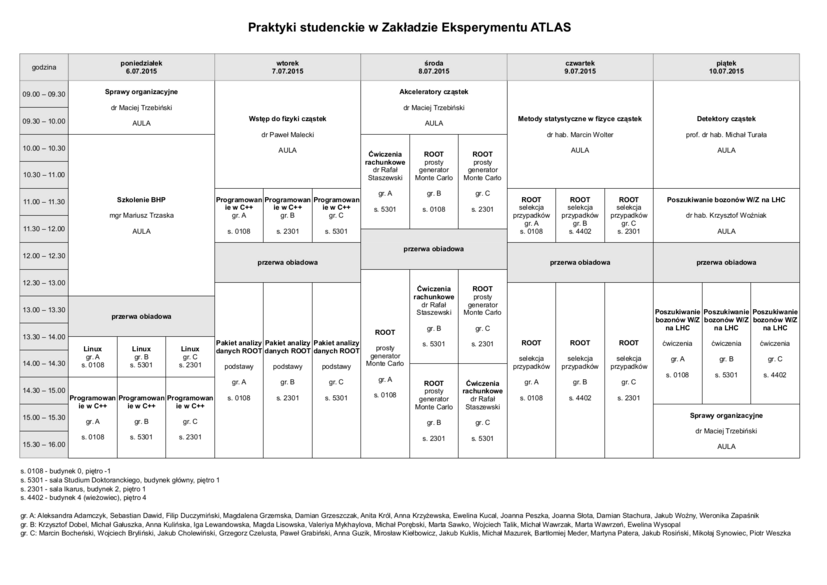}
    \caption{PPSS 2015: poster (\textbf{left}) and plan of the first week (\textbf{right}).}
    \label{PPSS2015}
\end{figure}

Students participating in the programme in July could count on financial support to cover stay in the AGH dorms\footnote{Also, we were able to make some small gadgets: pens and mugs with logo. Making them become a tradition.}. This edition was also the first with mini-conference. In fact there were two conferences: after 3$^{rd}$ and 4$^{th}$ week of the programme. We also decided that there should be a small competition -- the best presenters were awarded.

In the first conference, on July 24$^{th}$, jury\footnote{Jury was composed of Organisers and tutors attending the conference.} awarded Ms. Joanna Peszka (project \textit{Badanie w{\l}a\'sciwo\'sci przypadk\'ow eksluzywnej produkcji jet\'ow} under supervision of dr M. Trzebi\'nski). Joanna also got a prize from the audience. In addition, jury decided to give two distinctions: to Mr. Pawe{\l} Grabi\'nski, Mr. Bart{\l}omiej Meder and Ms. Valeryia Mykhaylova (project \textit{Uczenie maszynowe w fizyce wysokich energii}; supervisor: dr hab. M. Wolter) and to Mr Damian Stachura (\textit{Monitorowanie ewolucji cz\k{e}stotliwo\'sci w czasie sygna{\l}\'ow trygerowych detektora ALFA}; dr hab. K. Korcyl). In the 2$^{nd}$ conference, on July 31$^{st}$, jury gave two prizes, ex aequo to: Mr. Jakub Cholewi\'nski (\textit{Analiza zrekonstruowanych \'slad\'ow w danych pp 13 TeV}; dr hab. K. Wo\,zniak) and Mr. Filip Duczymi\'nski and Mr. Micha{\l} Por\k{e}bski (\textit{Synchronizacja urz\k{a}dze\'n Arduino z serwerem czasu}; dr R. Staszewski and dr M. Trzebi\'nski). Votes from the audience also resulted in ex aequo with awards going to: Mr. Jakub Rosi\'nski, (\textit{Symulacja halo dla wi\k{a}zki proton\'ow w akceleratorze LHC}; dr M. Trzebi\'nski) and Martyna Patera \textit{Zastosowanie statystycznego testu Manna-Whitneya-Wilcoxona do okre\'slenia pozycji detektor\'ow ALFA wzgl\k{e}dem wi\k{a}zki}; dr R. Staszewski).

During this edition a pilot programme for the high-school students was lunched. Although all 6 of them did a very nice projects (Filip and Micha{\l} even got a jury prize for their presentation), we judged that we do not have enough tutors to regularly extend the programme to accept high school students. Therefore, this idea was not continued.

Overall organisation as well as lectures were highly scored by participants in the anonymous survey. Students appreciated that the exercises were held in a small groups which allowed direct support of tutors. However, despite this and also new instructions, participants still indicated that there is not enough time to go through all ROOT exercises -- clearly there were places for improvements!

\section{Fourth Edition, 2016}

Fourth edition required registration and selection of participants -- for the first time we had more candidates (69) than places we could offer (40). Thanks to IFJ PAN and Polish Academy of Sciences, 22 students from outside Krakow again profited from the financial support for stay in AGH dorms. It is worth adding that students were coming from 14 universities: AGH University of Science and Technology, Gda\'nsk University of Technology, Krakow University of Technology, Lodz University of Technology, Poznan University of Technology, Warsaw University of Technology, Poznan University of Science and Technology, Jagiellonian University, Nicolaus Copernicus University in Toru\'n, University of Rzesz\'ow, University of Silesia in Katowice, Bydgoszcz University of Science and Technology, University of Warsaw and University of Wroc{\l}aw.

The duration of programme was again either 3 or 4 weeks, but we tried different starting time: August, 22$^{nd}$. As it turned out, this was not a good idea with students home universities and IFJ PAN administration departments being on holidays and students in need to finish the programme before middle of September in order to account it as mandatory course. Anyway, PPSS started smoothly with a week of lectures and exercises (see Fig. \ref{PPSS2016}).

\begin{figure}[!htbp]
    \centering
    \includegraphics[width=0.32\textwidth]{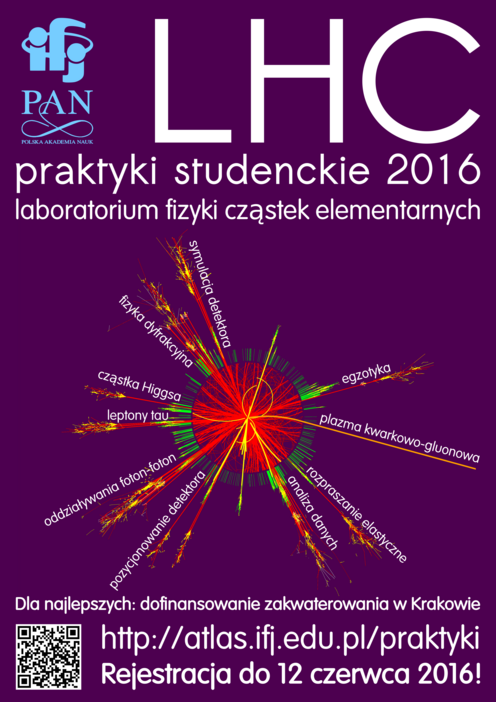}
    \includegraphics[width=0.67\textwidth]{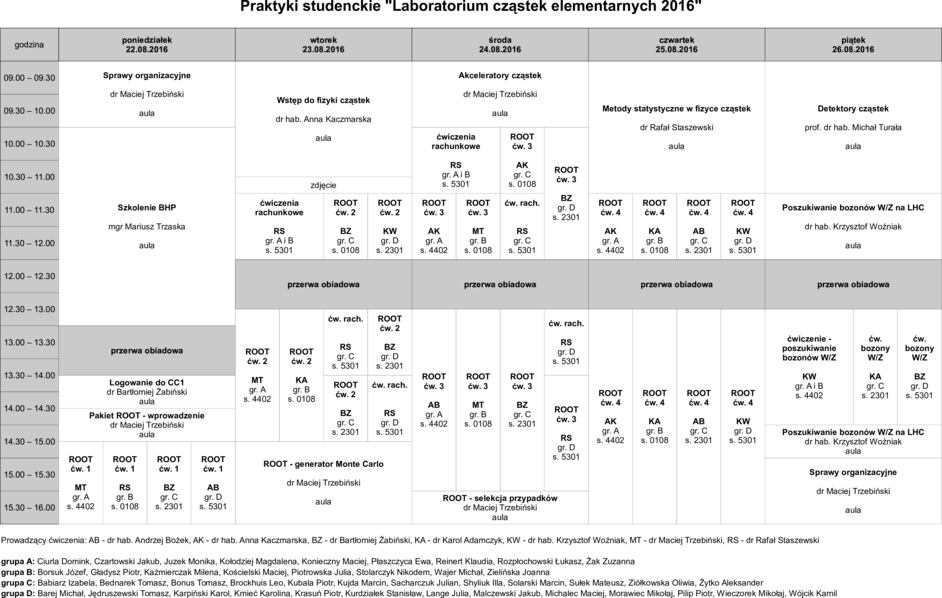}
    \caption{PPSS 2016: poster (\textbf{left}) and plan of the first week (\textbf{right}).}
    \label{PPSS2016}
\end{figure}

This edition was the first one during which projects were done in pairs. Forty participants meant that we had to ,,internally'' extend out programme -- from this edition tutors were not only from the ATLAS Department, but also other IFJ PAN divisions.

The first tour of mini-conference was held on September, 9$^{th}$. First prize from jury and audience was given to Ms. Joanna Zieli\'nska and Mr. Tomasz Bonus for presentation \textit{Analiza pierwszych danych fizycznych zebranych przez detektory AFP} (tutor: dr Rafa{\l} Staszewski). The second prize from jury was given to Mr. Leo Brockhuis and Mr. Maciej Michalec (\textit{Automatyzacja i monitorowanie proces\'ow w wielkoskalowych systemach kontroli eksperymentu ATLAS}; dr Ewa Stanecka) and the second prize from audience went to Ms. Klaudia Reinert and Mr. Nikodem Stolarczyk (\textit{Optymalizacja parametr\'ow analizy wielu zmiennych dla separacji t{\l}a w rozpadach bozonu Higgsa na dwa kwarki pi\k{e}kne}; dr hab. Marcin Kucharczyk). The second tour was held on Friday, a week after. Interestingly, the audience and jury agreed on which two presentations should be awarded, but disagreed who should be on the 1$^{st}$ place: Mr. Maciej Ko\'scielski and Mr. Jakub Malczewski\footnote{\textit{Optymalizacja kryteri\'ow selekcji dla rozpadu $\Lambda c \to p\mu\mu$ za pomoc\k{a} wielowymiarowej analizy danych}; prof. dr hab. Mariusz Witek.} got the 1$^{st}$ prize from audience and the 2$^{nd}$ prize from jury whereas Mr. Karol Karpi\'nski, Mr. Łukasz Rozp{\l}ochowski and Mr. Julian Sacharczuk\footnote{\textit{Oddzia{\l}ywania elektromagnetyczne -- nowe \,zr\'od{\l}o informacji o zderzeniach j\k{a}der atomowych. Studium przygotowawcze dla eksperymentu SHINE w CERN}; dr hab. Andrzej Rybicki.} the opposite.

To address the technical issues with the software we decided that students will work on a common system using the IFJ PAN Cloud Computing 1 cluster. This worked well, except a very good Internet connection had to be provided. In fact, similarly to previous years, the overall programme and lectures was highly scored whereas exercises were indicated to be improved -- see Fig. \ref{PPSS2016_survey}.

\begin{figure}[!htbp]
    \centering
    \includegraphics[width=0.32\textwidth]{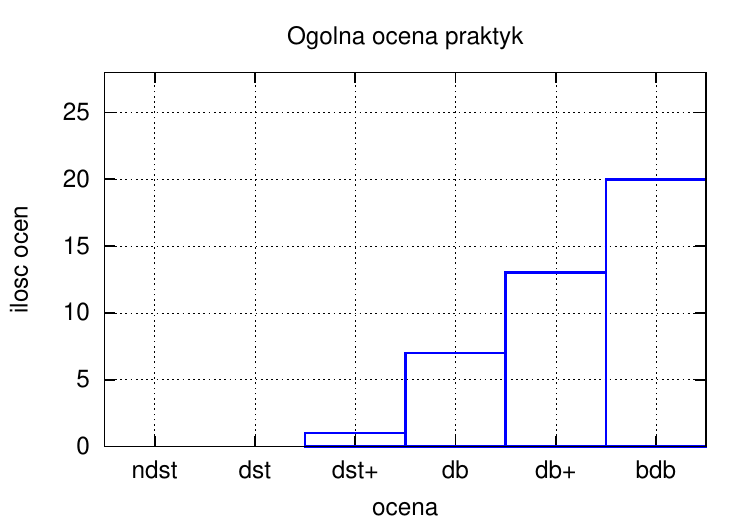}
    \includegraphics[width=0.32\textwidth]{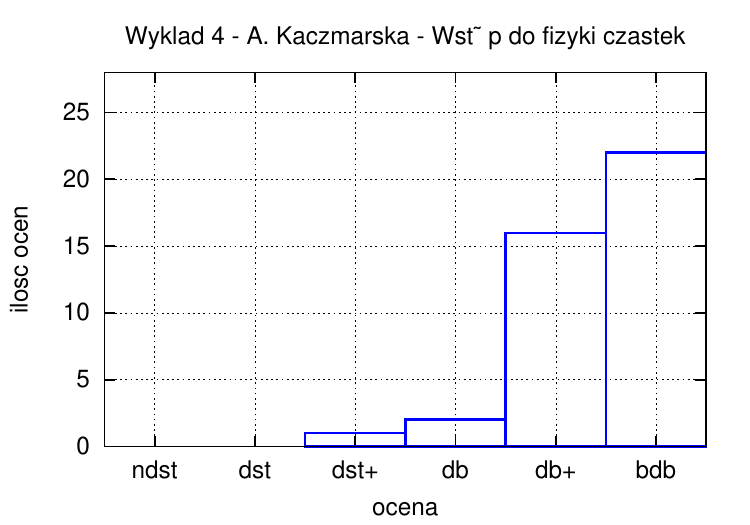}
    \includegraphics[width=0.32\textwidth]{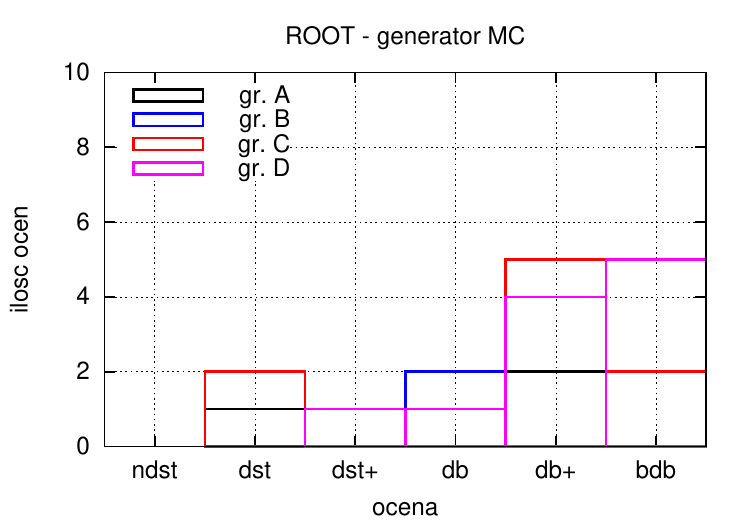}
    \caption{PPSS 2016 anonymous survey: overall opinion about programme (\textbf{left}), Standard Model lecture rating (\textbf{centre}), opinions about one of ROOT exercises (\textbf{right}).}
    \label{PPSS2016_survey}
\end{figure}

\section{Fifth Edition, 2017}

With 2016 experience, we decided that the programme will be organised in July. In 2017 programme started on Monday, July 3$^{rd}$ with a usual set of lectures in the first week -- see Fig. \ref{PPSS2017}. This time we had 33 students selected from 56 candidates. They were coming from 16 universities\footnote{AGH University of Science and Technology, Gda\'nsk University of Technology, Krakow University of Technology, Lodz University of Technology, Poznan University of Technology, Warsaw University of Technology, Poznan University of Science and Technology, Jagiellonian University, Nicolaus Copernicus University in Toru\'n, University of Rzesz\'ow, University of Silesia in Katowice, Bydgoszcz University of Science and Technology, University of Warsaw, University of Wroc{\l}aw, University of Lodz and University of Bialystok.} -- we were a bit surprised that we have that many in Poland!

\begin{figure}[!htbp]
    \centering
    \includegraphics[width=0.32\textwidth]{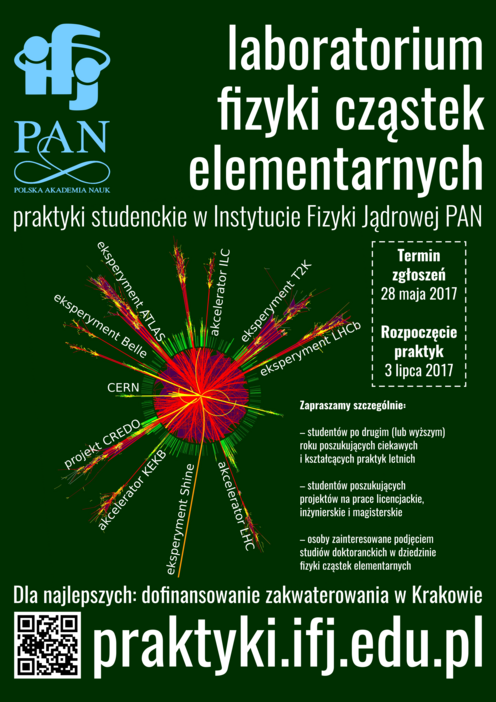}
    \includegraphics[width=0.67\textwidth]{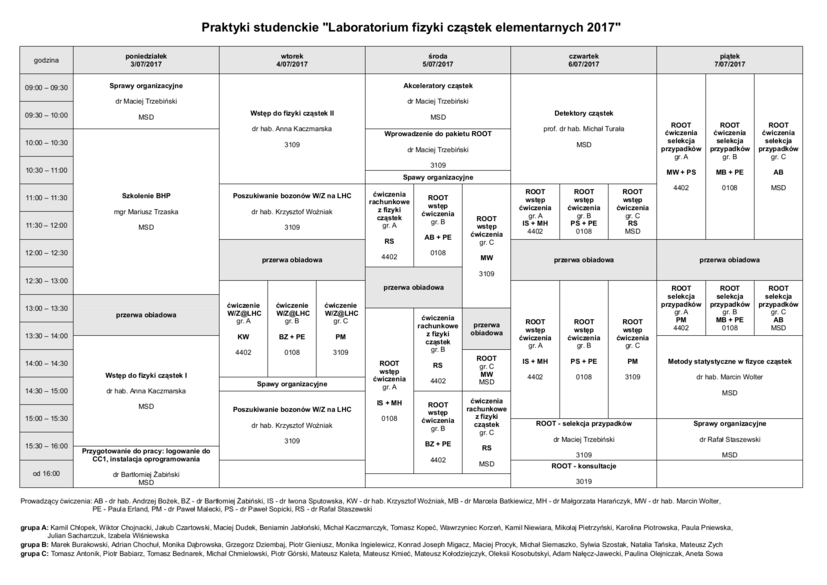}
    \caption{PPSS 2017: poster (\textbf{left}) and plan of the first week (\textbf{right}).}
    \label{PPSS2017}
\end{figure}

Thanks to the financial support from IFJ PAN and PAS a partial coverage of stay at dorms was possible for 22 participants. And again, pens and mugs with logo were prepared.

Having experience from the previous editions we decided to reorganise the ROOT exercises -- more time was given to solve each problem. Also, the tasks were simplified and made more friendly for people not familiar with the high energy physics jargon. This paid off -- in the anonymous survey the marks for exercises oscillated around 4.6 (in 2--5 scale).

In the first mini-conference jury awarded a presentation of Mr. Micha{\l} Chmielowski and Ms. Paula Pniewska (\textit{Pomiar przekroju czynnego dla procesu $Z \to \tau\tau$ w eksperymencie ATLAS}; supervisor: dr Pawe{\l} Malecki). The audience choice was presentation of Mr. Jakub Czartowski and Mr. Micha{\l} Siemaszko (\textit{Badanie sprz\k{e}\.ze\'n bozonu Higgsa i bozon\'ow W z u\.zyciem rozk{\l}ad\'ow k\k{a}towych na{\l}adowanych lepton\'ow}; dr Magdalena S{\l}awi\'nska). In the second mini-conference votes of jury and audience were summed and joint awards were given: 1$^{st}$ place -- Mr. Piotr Babiarz (\textit{Poszukiwanie niewidzialnych cz\k{a}stek ciemnej materii w procesach ekskluzywnych na LHC}; dr Bart{\l}omiej Żabi\'nski and dr Rafa{\l} Staszewski), 2$^{nd}$ place -- Mr. Marek Burakowski and Mr. Beniamin Jab{\l}o\'nski (\textit{Selekcja oddzia{\l}ywa\'n neutrin z wymian\k{a} pr\k{a}d\'ow na{\l}adowanych/neutralnych w eksperymencie T2K}; dr Marcela Batkiewicz and dr Tomasz W\k{a}cha{\l}a) and the 3$^{rd}$ place -- Ms. Sylwia Szostak and Ms. Natalia Ta\'nska (\textit{Optymalizacja inkluzywnej rekonstrukcji strony znakuj\k{a}cej w \'srodowisku eksperymentu Belle II}; dr Karol Adamczyk).

\section{Sixth Edition, 2018}

After successful 2017 edition, with many students from almost all polish universities, the next step was to go international. The sixth edition was decided to be held in English and announced abroad. For that reason we created a Facebook profile: \href{https://www.facebook.com/ifjpanppss}{https://www.facebook.com/ifjpanppss}. Again, a small success was achieved: among 44 candidates, 5 were from abroad. From them, 32 students\footnote{From: AGH University of Science and Technology, Krakow University of Technology, Palacky University Olomouc, Wroc{\l}aw University of Science and Technology, University of Cantabria, Gdansk University of Technology, Jagellonian University, Poznan University of Technology, University of Warsaw and Vilnius University.} were qualified including 3 students from Lithuania, Czech Republic and Spain. We were able to fully support the stay in dorms of 8 participants and another 8 got partial support.

Programme started on July, 2$^{nd}$ with all lectures and exercises held in English. The duration was unified to four weeks for all participants. There were 3 prizes from jury and 3 from audience. They went to: Ms. Joanna Szulc (1st place (jury) and 2nd place (audience); \textit{Studies of photon-jet correlation in the ALICE experiment at the LHC}; dr Adam Matyja), Mr. Pawe{\l} Drabczyk and Ms. Patrycja Pot\k{e}pa (1st place (audience) and 2nd place (jury); \textit{Detection of Cosmic-Ray Ensembles}; dr hab. Krzysztof Wo\,zniak), Ms. Zuzanna Szester and Mr. Micha{\l} Zmy\'slony (3rd place (jury); \textit{Search for New Physics in the ATLAS detector at the LHC}; dr Pawe{\l} Malecki) and Mr. Sebastian Pinocy and Mr. Petr Baron (3rd place (audience); \textit{Quark/gluon jets}; dr Andrzej Si\'odmok).

\begin{figure}[!htbp]
    \centering
    \includegraphics[width=0.32\textwidth]{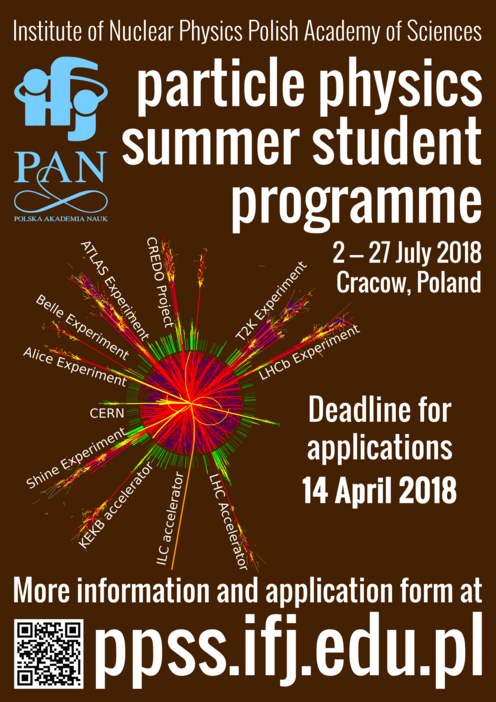}
    \includegraphics[width=0.67\textwidth]{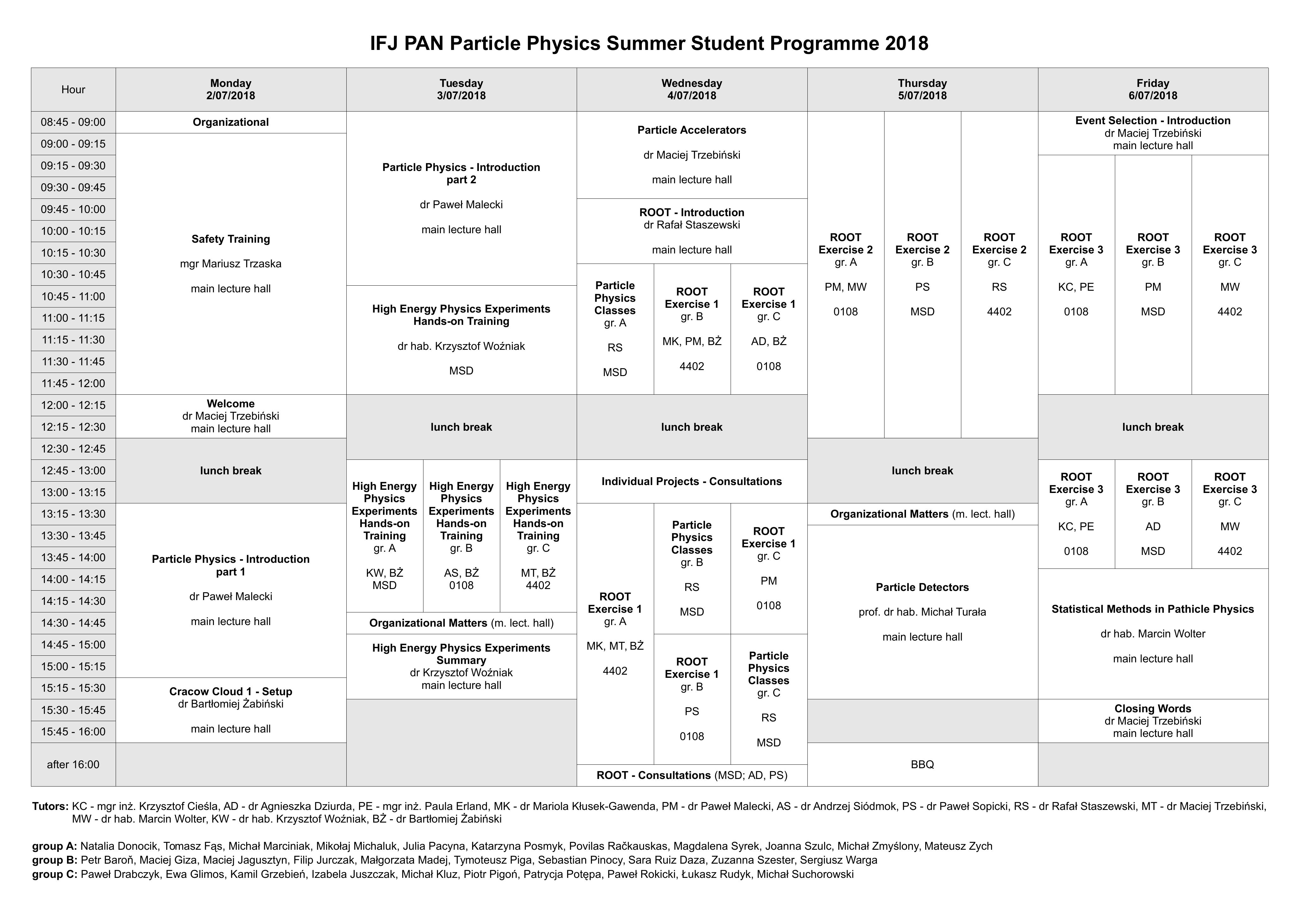}
    \caption{PPSS 2018: poster (\textbf{left}) and plan of the first week (\textbf{right}).}
    \label{PPSS2018}
\end{figure}

In the 2$^{nd}$ week of the programme we tried a new thing: seminars given by tutors on various topics. The goal was to discuss the material which does not fit into the ``usual'' introductory lectures. Seminars were appreciated by students, therefore we decided to add them to the regular programme.

It should be mentioned that around that time Rafal Staszewski got tired of all of the paper questionnaires and a ``manual'' topic assignment and developed a few useful scripts which are still in use. From 2018, students could select topics online, the topic assignment was done by the genetic algorithm and final survey was available on the webpage.

During the 2018 edition we realised that an international advertisement campaign was needed. Fortunately, Maciej Trzebi\'nski was successful in a grant application. Project \textit{Promocja oferty dydaktycznej IFJ PAN w\'sr\'od student\'ow europejskich uczelni} financed by the National Agency for Academic Exchange\footnote{Call: \textit{Nowoczesna Promocja Zagraniczna}, project: PPI/NPZ/2018/1/00106/U/001.} allowed not only to improve the PPSS \href{https:ppss.ifj.edu.pl}{webpage} and Facebook \href{https://www.facebook.com/ifjpanppss}{profile}, but mainly to travel to 13 european universities to advertise the programme. 

\section{Seventh Edition, 2019}

Either the advertisement campaign was successfully or the \textit{word of mouth} worked. The fact is that in 2019 we had 100 candidates including 54 from abroad! The selection was difficult, as we could only offer 21 topics. At the end, we invited 42 students from Poland, Belarus, Czech Republic, Egypt, Hungary, India, Serbia, Slovenia, Spain, Ukraine, United Kingdom and Turkey. Thanks to the financial support we were able to support a stay of 35 people in dorms.

The programme started on July, 8$^{th}$ (see Fig. \ref{PPSS2019}). After a nice BBQ during the first week of lectures and exercises, 3 weeks of projects started. Following the decision from 2018, in the 2$^{nd}$ and 3$^{rd}$ week seminars were organised. The anonymous evaluation survey indicated that programme is tailored quite well, with minor improvements needed.

As usual, on the last day a mini-conference happened. This time the votes were entirely in hands of the audience. The first place was granted ex aequo to Mr. Javier Suarez and Daniel Firak\footnote{\textit{Cluster analysis of data from the silicon pixel detector}; dr in\.z. Piotr Kapusta.} and Ms. Maryna Oleksiienko and Mr. Marcin Jastrz\k{e}bski\footnote{\textit{C++ interfaces for KaTie}; dr hab. Andreas van Hammeren.}. The second place was given to Ms. Milena Simi\'c and Mr. Karol Łukanowski\footnote{\textit{Two for the prize of one!}; dr Bart{\l}omiej Żabi\'nski and dr Magdalena S{\l}awi\'nska.} and the third went to Ms. Nastasija Petkovic and Ms. Maria Liubarska\footnote{\textit{Muon tracks reconstruction for the Baikal Neutrino Telescope}; Konrad Kopa\'nski, dr Jaros{\l}aw Stasielak and dr hab. Robert Kami\'nski.}.

\begin{figure}[!htbp]
    \centering
    \includegraphics[width=0.32\textwidth]{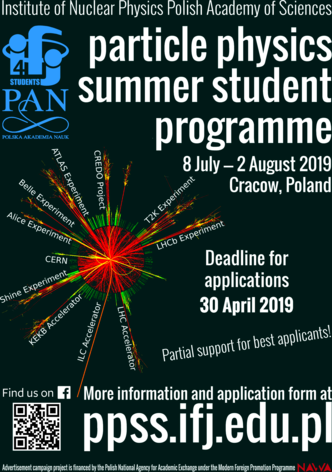}
    \includegraphics[width=0.67\textwidth]{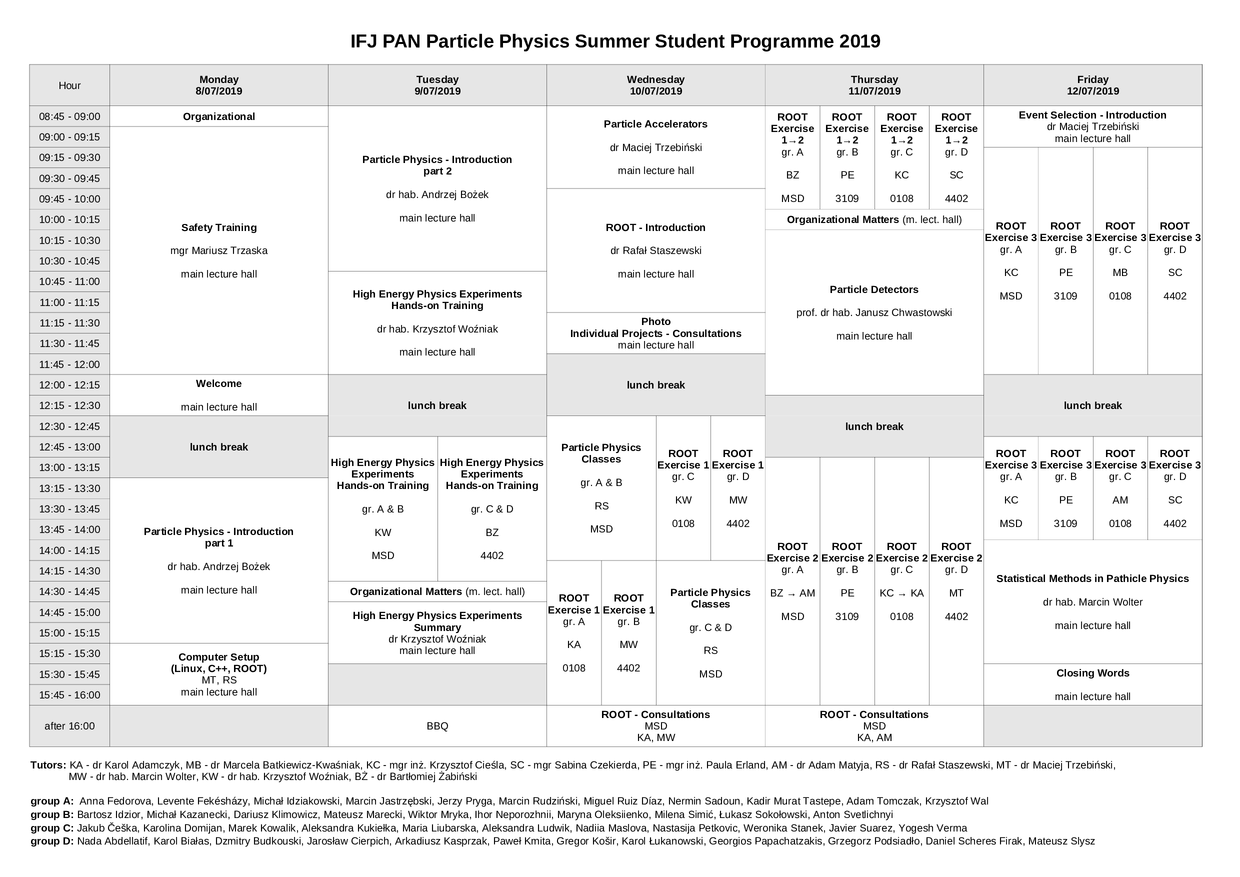}
    \caption{PPSS 2019: poster (\textbf{left}) and plan of the first week (\textbf{right}).}
    \label{PPSS2019}
\end{figure}

In Fall 2019 we decided to send a survey to our alumni -- participants of 2013-2017 editions. We got 30 responses. Quarter of alumni were still students at that time, third of theme where doing PhD and third were employed outside the university. In general, alumni claimed that the knowledge and skill gained during the programme were useful during the studies (36\%), at work (10\%) or both (40\%). All of them would recommend the programme (30\% ``yes'', 70\% ``certainly yes'').

\section{Eighth Edition, 2020}

The preparations for the eight edition were at the full steam, with a financial support already secured, when in March 2020 everything was locked due to COVID-19 pandemic. After quite some discussions, we decided to move the programme to purely virtual mode. This decision had positive (\textit{e.g.} possibility of participation for students from more distant countries) and negative (\textit{e.g.} no direct interactions nor social life in dorms) consequences. The total of 38 students from 14 counties\footnote{India, Poland, Mexico, Ukraine, United Kingdom, Turkey, Republic of Serbia, Cyrpus, Israel, Iran, Georgia, Hungary, Russia and Egypt.} were selected from 61 candidates. It should be mentioned that in 2020 a new Organiser joined: dr Olga Werbycka.

Programme started on July, 6$^{th}$ with the first week adapted to the new reality (see Fig. \ref{PPSS2020}). Everything was moved to the virtual world: lectures and exercises were done via Zoom and recorded, contact with students was via e-mail and various communicators, certificates of participation were in a digital form, \textit{etc}. 

\begin{figure}[!htbp]
    \centering
    \includegraphics[width=0.32\textwidth]{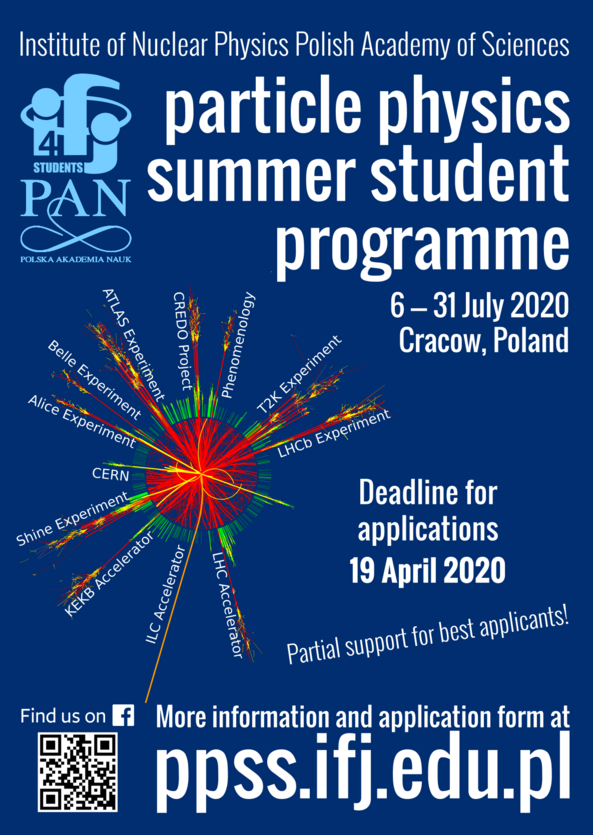}
    \includegraphics[width=0.67\textwidth]{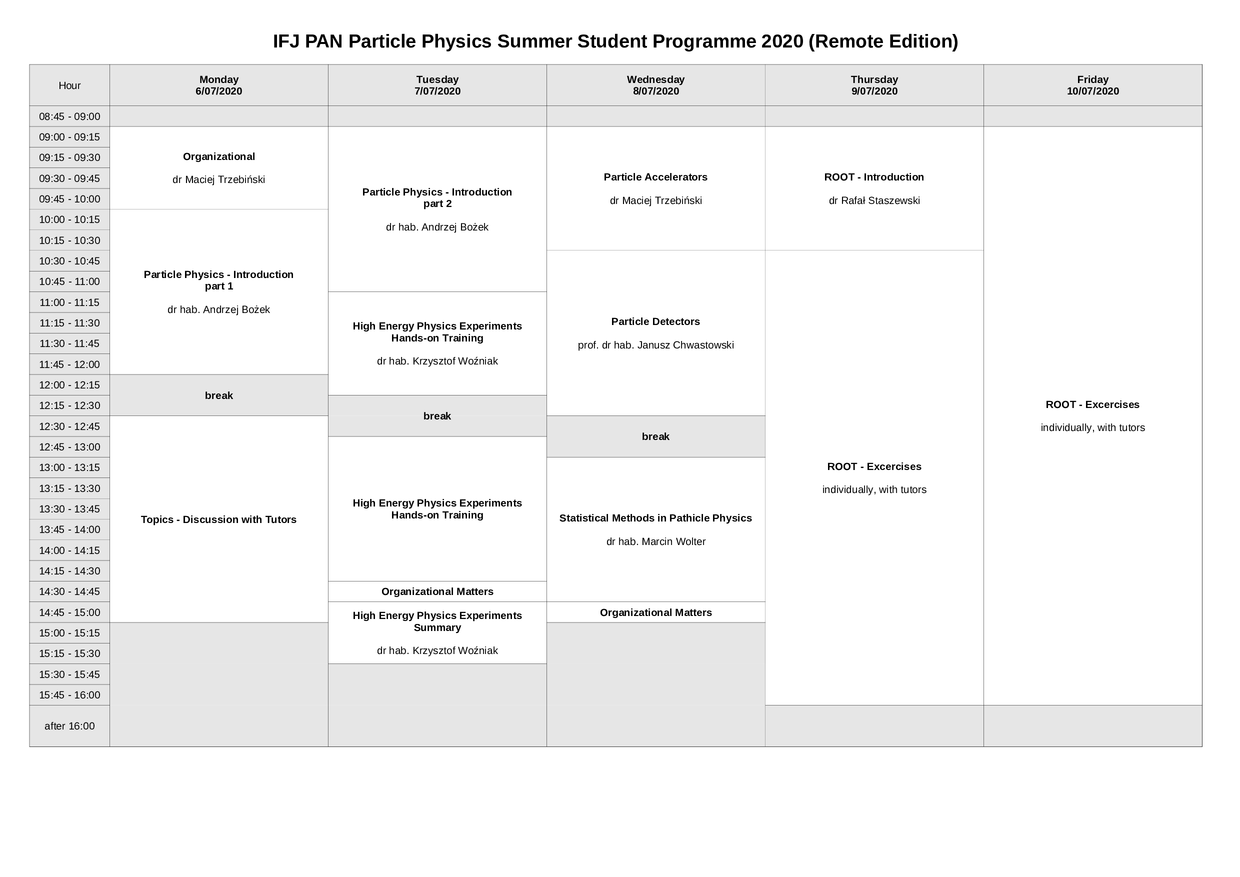}
    \caption{PPSS 2020: poster (\textbf{left}) and plan of the first week (\textbf{right}).}
    \label{PPSS2020}
\end{figure}

At the last day of PPSS the virtual mini-conference was held. Audience decided that the 1$^{st}$ place should go to Ms. Aleksandra Baszak and Mr. Joseph Cave (\textit{Detector Control System for AFP detector in ATLAS experiment at CERN: design of graphical user interfaces}; dr in\.z. El\.zbieta Bana\'s), 2$^{nd}$ to Mr. Filip R\k{e}kawek and Ms. Alexia Mavrantoni (\textit{Detection of Cosmic-Ray Ensembles}; dr hab. Krzysztof Wo\,zniak) and 3$^{rd}$ ex aequo to Mr. Nazar Semkiv and Ms. Arantza Pineda Gonzalez (\textit{Study of semi-inclusive characteristics for $B \to D/D_shX$ decays at Belle II}; dr Olga Werbycka) and Mr. Swapnil Dutta and Mr. Eslam Zenhom (\textit{Fast computations for global analysis of parton distributions}; dr Aleksander Kusina).

\section{Ninth Edition, 2021}

In 2021 till the very end it was not clear if we would be able to welcome students at the Institute. Unfortunately, we again had to make everything purely remotely. From 81 candidates we were able to accept 38 participants from 14 countries\footnote{Poland, Philippines, Norway, India, Iran, Taiwan, Greece, Hungary, Venezuela, United Kingdom, Brazil, Spain, Turkey and Russia.}. Programme was very similar as in 2021 -- see Fig. \ref{PPSS2021}.

\begin{figure}[!htbp]
    \centering
    \includegraphics[width=0.32\textwidth]{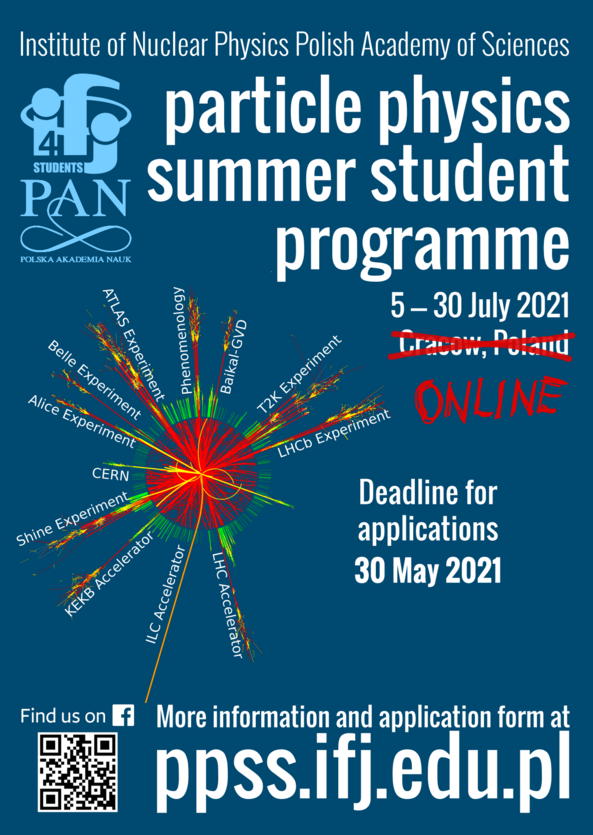}
    \includegraphics[width=0.67\textwidth]{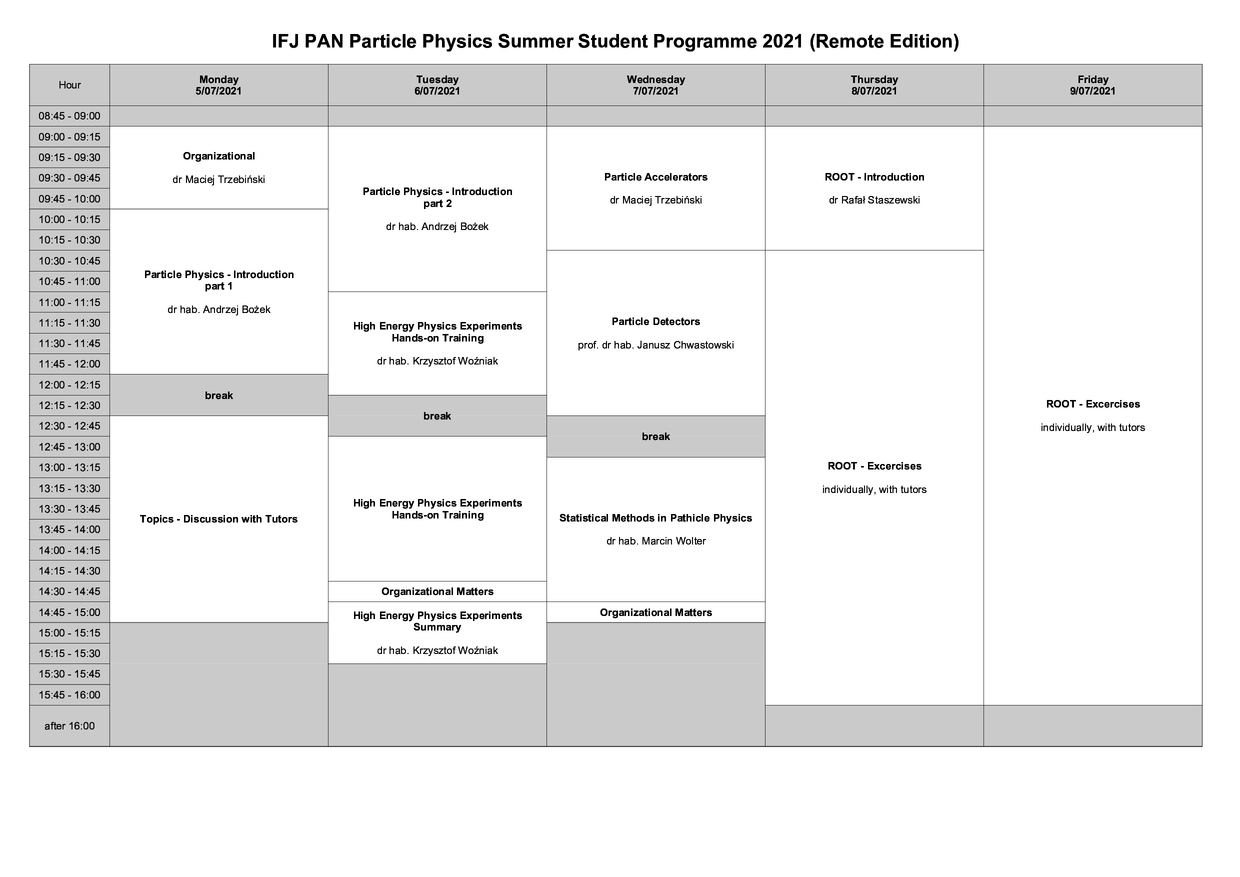}
    \caption{PPSS 2021: poster (\textbf{left}) and plan of the first week (\textbf{right}).}
    \label{PPSS2021}
\end{figure}

The votes during the mini-conference were again in the hands of the audience. The first place went to Mr. Giorgi Asatiani and Ms. Maria Ramos for the presentation \textit{Detection of Cosmic-Ray Ensembles} (supervisor: dr hab. Krzysztof Wo\,zniak). The second place was taken by Mr. Silje Oino and Mr. Bart{\l}omiej Pierzcha{\l}a (\textit{Search for CP violation in decays $B_0 \to D^*D^\pm$ in the Belle II experiment}; dr Olga Werbycka). The third place was ex aequo for Mr. Bhupesh Dixit and Mr. Chin Zhe Tee (\textit{Reconstruction of exclusive jet events}; dr Rafa{\l} Staszewski) and Mr. Yash Arya and Ms. Roxana Busuioc (\textit{Geometry of heavy ion collisions during lepton pair production}; dr Mariola K{\l}usek-Gawenda).

In Fall 2021 another global survey was lunched. This time we asked people participating in years 2018-2021. 71 former participants responded. Majority of them (69\%) were still BSc./MSc. students, 14\% were doing PhD and 13\% were employed outside the university. 48\% judged that the knowledge and skills gained during PPSS programme were useful during studies, 41\% pointed that they were useful also at work and 7\% said that these were useful only at work. From all the respondents only 2 would rather not recommend participation in PPSS, meaning that 97\% were very happy with the programme (70\% ``definitely yes!'', 27\% ``yes''). It should be added that students who applied for a professional training/school/practices after PPSS judged that ``PPSS programme was an important point in my CV'' (72\%), ``knowledge and skills gained during PPSS allowed me to prepare better application'' (64\%) and ``thanks to participation in PPSS I've got a recommendation letter'' (21\%).

After 2021 edition Olga got a few-year stipend abroad. Fortunately, dr. Dominik Derendarz joined!

\section{Tenth Edition, 2022}

With understanding of COVID-19 disease and development of vaccinations, we had high hopes that we will be able to greet students at the Institute. The war in Ukraine complicated things -- access to IFJ PAN was restricted due to the CHARLIE-CRP (potential threat in cyberspace) and BRAVO (potential terrorist threat) alerts in Poland. However, after two years of remote editions we were determined to at least have a hybrid mode.

We received a record number of 141 candidatures. Selection was challenging since we could offer only 30 places locally and 10 remotely. Finally, we had participants from 16 countries\footnote{India, Greece, Poland, Turkey, Serbia, Spain, Czech Republic, Egypt, Sweden, Cyprus, Peru, South Africa, Hungary, United Kingdom, Morocco and China.}. Thanks to the support from IFJ PAN and Polish Academy of Sciences we were able to fully cover the cost of stay at dorms for 25 participants!

Programme started July, 4$^{th}$ (see Fig. \ref{PPSS2022}). A BBQ for the local participants was organised in the first week. The novelty was PPSS Alumni conference held during the weekend. Seminars were organised in the third week and mini-conference, as usual, at the last day -- July, 29$^{th}$.

\begin{figure}[!htbp]
    \centering
    \includegraphics[width=0.32\textwidth]{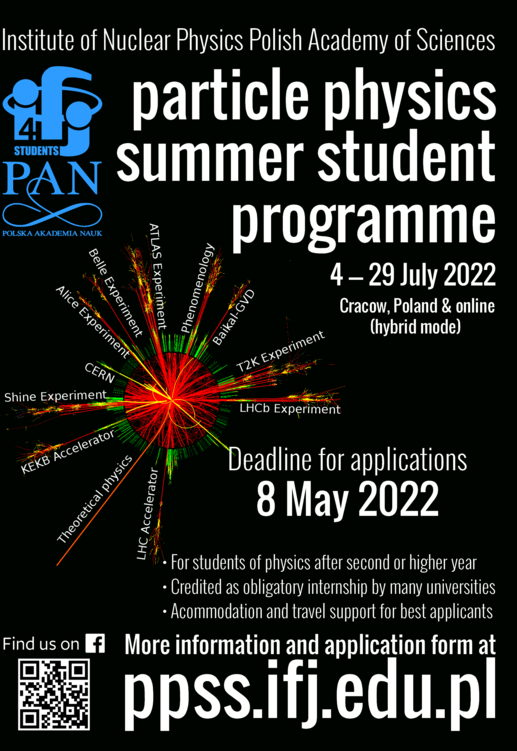}
    \includegraphics[width=0.67\textwidth]{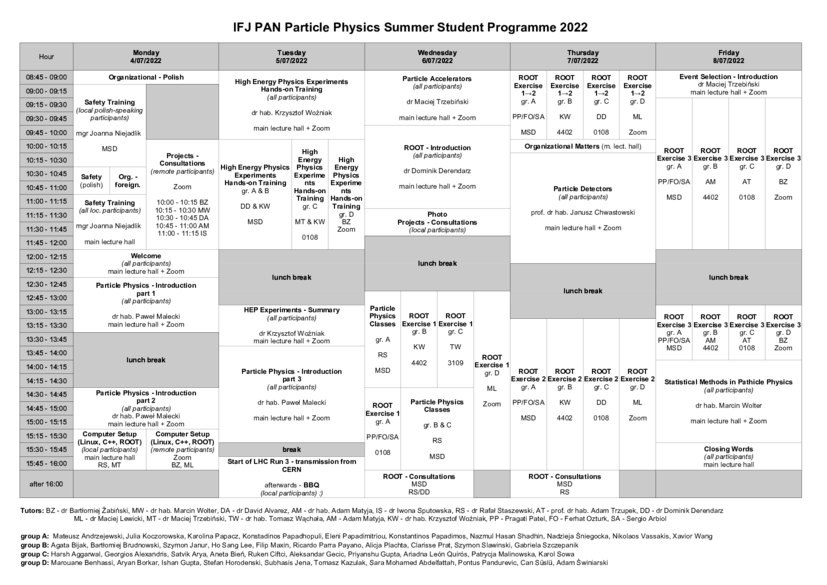}
    \caption{PPSS 2022: poster (\textbf{left}) and plan of the first week (\textbf{right}).}
    \label{PPSS2022}
\end{figure}

Audience decided that the best presentation was given by Mr. Filip Maxin and Mr. Szymon Slawinski\footnote{\textit{Three approaches to B-jet identification in ATLAS experiment from easy to hard}; dr Dominik Derendarz.}. The second place was won by Mr. Priyanshu Gupta and Ms. Clarisse Prat\footnote{\textit{Reconstruction of particle tracks using Deep Neural Networks}; dr hab. Marcin Wolter.} and the third by Ms. Ariadna Le\'on Quir\'os and Ms. Nadzieja Śniegocka\footnote{\textit{Exploring Jet quenching in relativistic heavy ion collisions}; dr Souvik Priyam Adhya.}.

\section{First PPSS Alumni Conference -- 2022}

The idea of the conference for PPSS alumni was born probably around 2019. The concept was to meet again with alumni in person at the Institute. On the one hand this would be an opportunity to remind students about our nice research at the IFJ PAN. On the other, to create an opportunity for the current participants to discuss with alumni. Unfortunately, COVID-19 delayed everything. 

The first PPSS Alumni Conference was organised on 9-10 July 2022. With the funds from IFJ PAN and PAS we were able to partially support stay and travel of speakers coming from outside Krakow. Twenty alumni wished to give talks (7 of them remotely). There was also a ``motivational talk'' given by dr Maciej Trzebi\'nski and advertisement of Krakow Interdisciplinary Doctoral School by Ms. Aleksandra Pacanowska. A very important point in the programme was a discussion session devoted to Msc. and PhD. studies and scientific career held on Saturday. The discussion continued during the conference dinner.

We decided that proceedings, the ones you are reading now, should be prepared. For many participants this would be a nice opportunity to publish their work.

The conference was appreciated by alumni and participants of PPSS 2022. Maybe it will become an annual event?

\section*{PPSS -- Group Photos}

\includegraphics[width=1.0\textwidth]{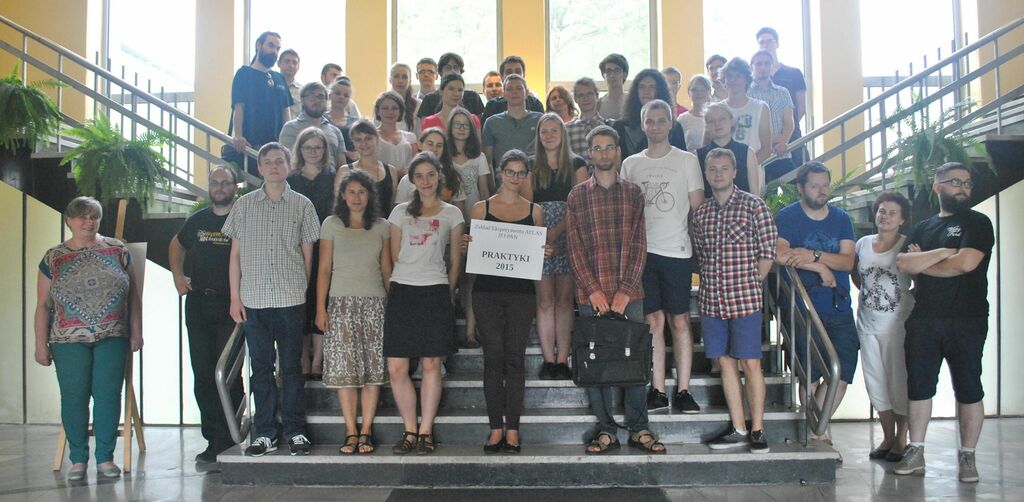}
\includegraphics[width=1.0\textwidth]{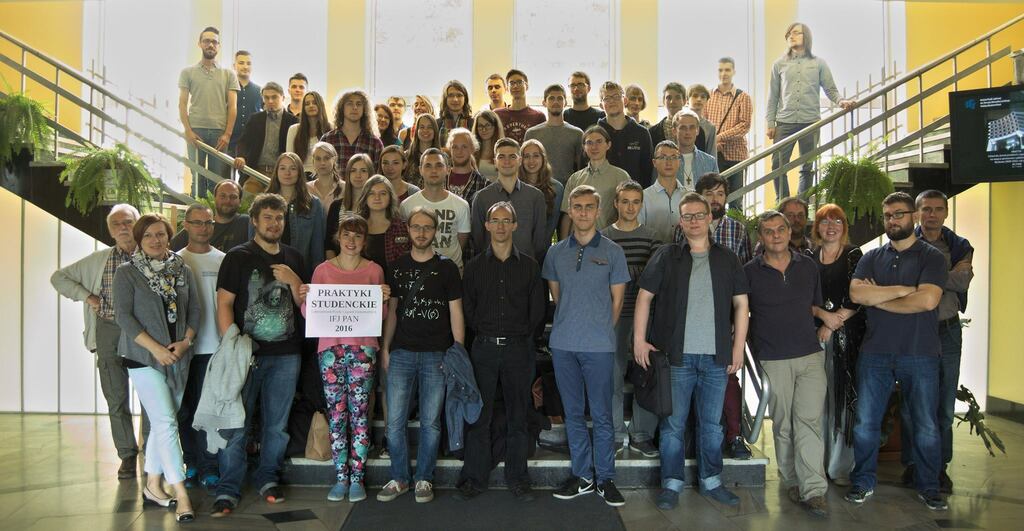}
\includegraphics[width=1.0\textwidth]{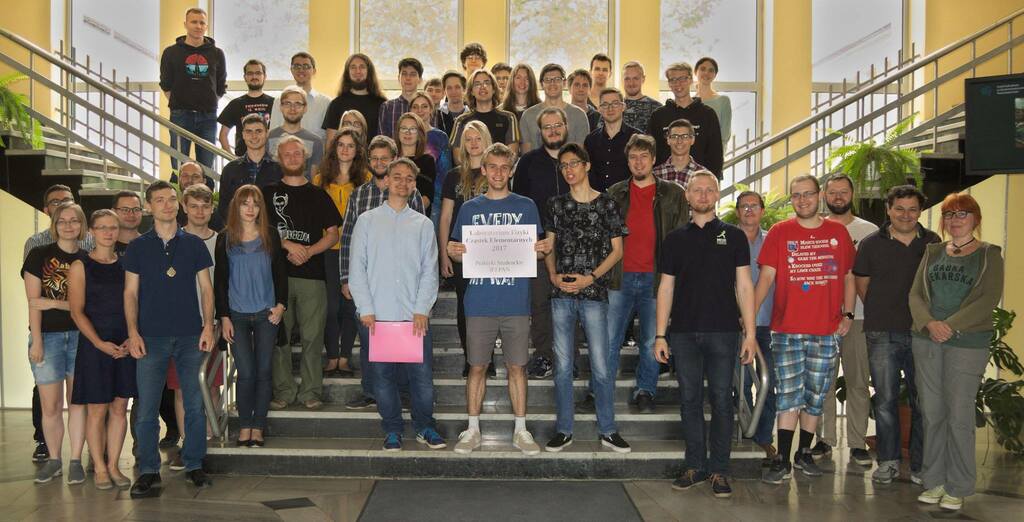}
\includegraphics[width=1.0\textwidth]{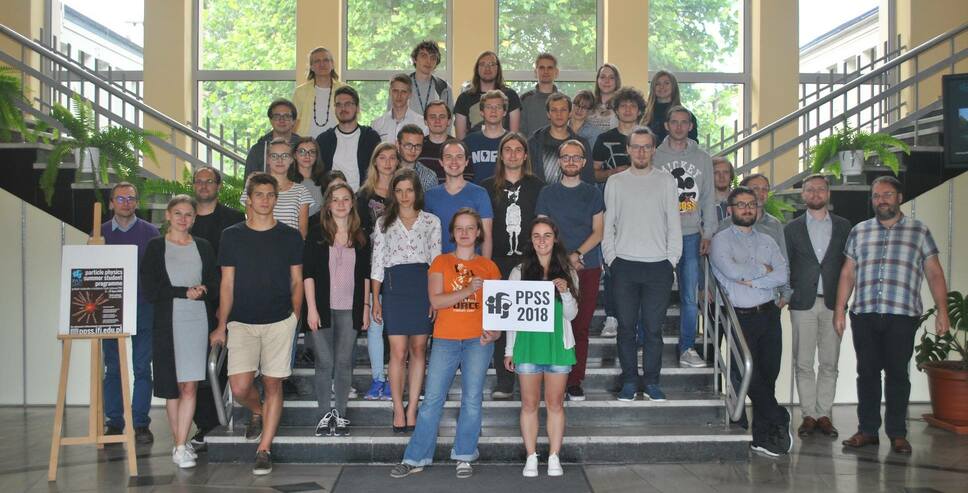}
\includegraphics[width=1.0\textwidth]{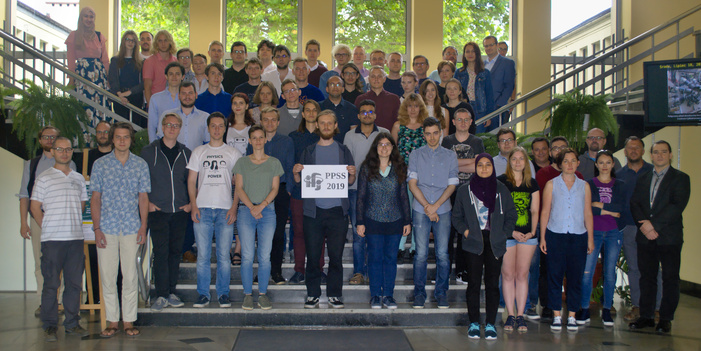}
\includegraphics[width=1.0\textwidth]{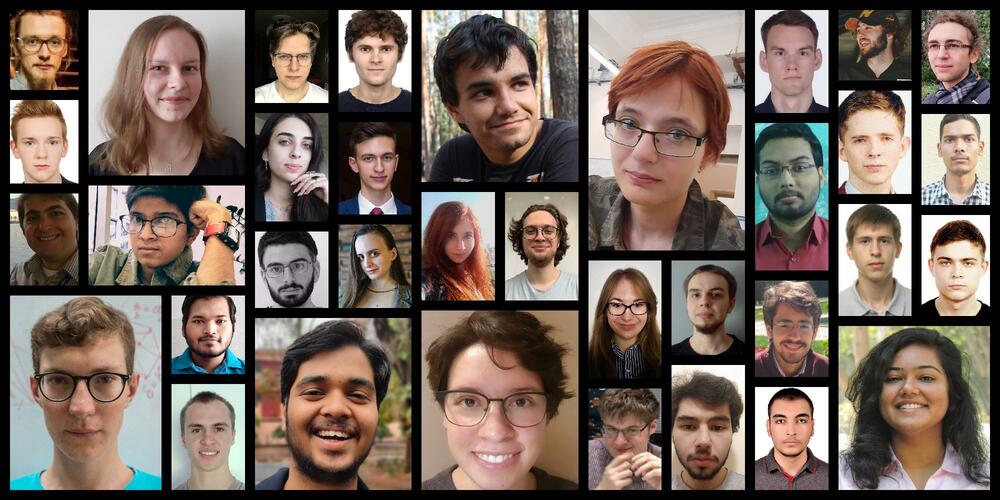}
\includegraphics[width=1.0\textwidth]{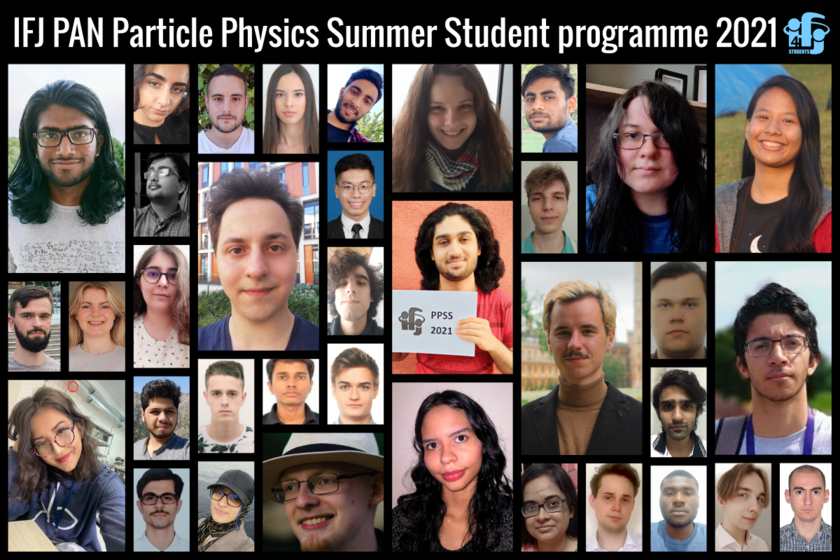}
\includegraphics[width=1.0\textwidth]{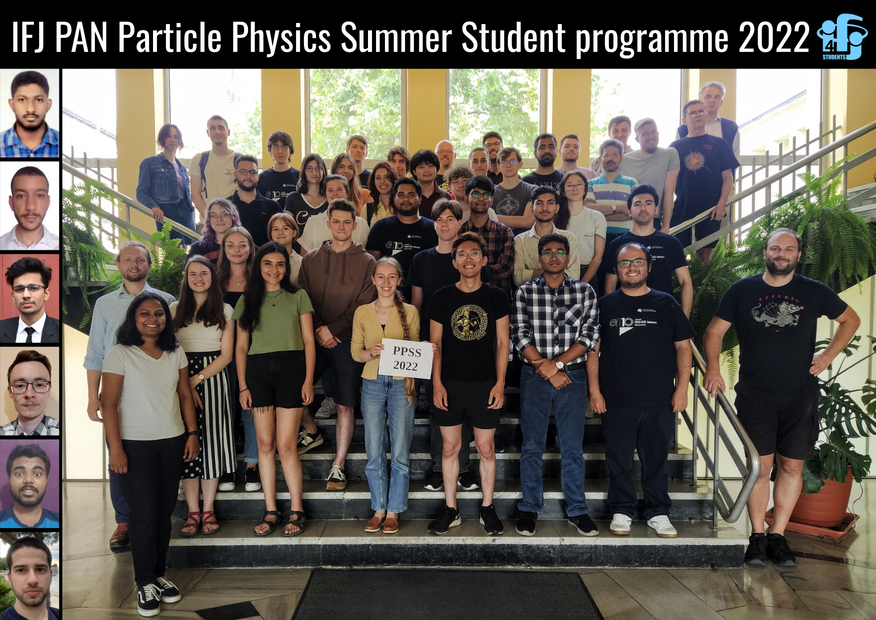}

\section*{PPSS Alumni Conference -- Group Photo}
\includegraphics[width=1.0\textwidth]{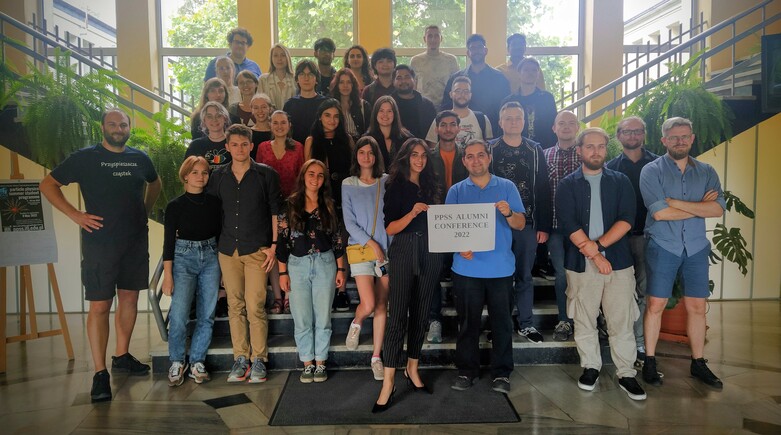}

\section*{PPSS Alumni Conference 2022 -- List of Talks}
\begin{itemize}
    \item Alexia Mavrantoni, \textit{Detection of Cosmic-Ray Ensembles},
    \item Anna Kuli\'nska, \textit{ATLAS Local Trigger Interface (ALTI) firmware development and verification},
    \item S{\l}awomir Tadeja, \textit{Potential of Coupling VR/AR with Digital Twinning: A Short Introduction with Case Studies},
    \item Izabela Babiarz, \textit{Light-cone approach to mesons gamma* gamma* form factors},
    \item Micha{\l} Barej, \textit{QCD phase diagram and factorial cumulants},
    \item Swapnil Dutta, \textit{Sphalerons in Weinberg Salam theory},
    \item Monika Juzek, \textit{My work as a PhD student at the Institute of Nuclear Physics PAN or Background estimation in the search of charged Higgs boson decaying to heavy tau lepton in the ATLAS experiment},
    \item Patrycja Pot\k{e}pa, \textit{Searching for electrons in heavy-ion collisions in the ATLAS experiment at the LHC},
    \item Paula Erland, \textit{Analysis with AFP and ATLAS Detectors},
    \item Sara Ali Mahmoud Ibrahim, \textit{Analysis of Cluster Shapes in the ATLAS Forward Proton Detector},
    \item Bhupesh Dixit, \textit{Estimation of Combinatorial Background in Diffractive Events with Forward Proton Tagging},
    \item Maciej Giza, \textit{Study of beauty to charm hadron decays and proton-proton collision reconstruction at LHCb experiment},
    \item Jakub Malczewski, \textit{Semileptonic $\Lambda_c$ decays at LHCb},
    \item Joanna Peszka, \textit{Fundamental physics tests with antihydrogen in ALPHA Experiment},
    \item Roxana Busuioc, \textit{An insight into semi-central heavy-ion collisions},
    \item Chaitanya Paranjape, \textit{The Higgs plus three-gluon amplitude at one loop with pySecDec},
    \item Valeriya Mykhaylova, \textit{Quasiparticle perspective on transport properties of hot QCD matter},
    \item Sara Ruiz Daza, \textit{Monte Carlo simulations of a beam telescope setup based on a 65 nm CMOS Imaging Technology},
    \item Grzegorz Czelusta, \textit{Quantum simulations of loop quantum gravity},
    \item Ophir Ruimi, \textit{The Plasma Window as a Vacuum-Atmosphere Interface for Measurements of Stellar Neutron-induced Reaction Cross Sections},
    \item Alexia Mavrantoni, \textit{Operating rover as an analogue astronaut}.
\end{itemize}

\ResCnt

\maketitle
\vspace*{-0.5cm}
\begin{abstract}
The combinatorial background is one of the major sources of background in the study of diffractive events. This study aims to validate a method for estimating the combinatorial background using the statistical models. The Monte Carlo samples of diffractive events in pp collisions at the centre of mass energy $\sqrt{s} = 13$ TeV, with the ATLAS detector were used for the study. The method was validated by comparing the results obtained from the statistical models with the background obtained from the Monte Carlo sample. The study was performed for two different types of diffractive events: soft diffractive events and diffractive W-boson production as an example of a hard diffractive process. In both cases, the proposed models provided a good estimation of the combinatorial background. 
\end{abstract}

\vspace*{-0.5cm}
\section{Introduction}
At the LHC, the proton beam is in the form of bunches. Each bunch consists of about $10^{11}$ protons. In a single bunch crossing, interaction of two protons travelling in opposite directions is called a single interaction or an event. When two bunches cross each other, there is always a possibility that more than one collision occurs. In such a case, more than one interaction that occurred is observed as a single event.

\begin{wrapfigure}{r}{0.55\linewidth}
\centering
\includegraphics[width=1.1\linewidth]{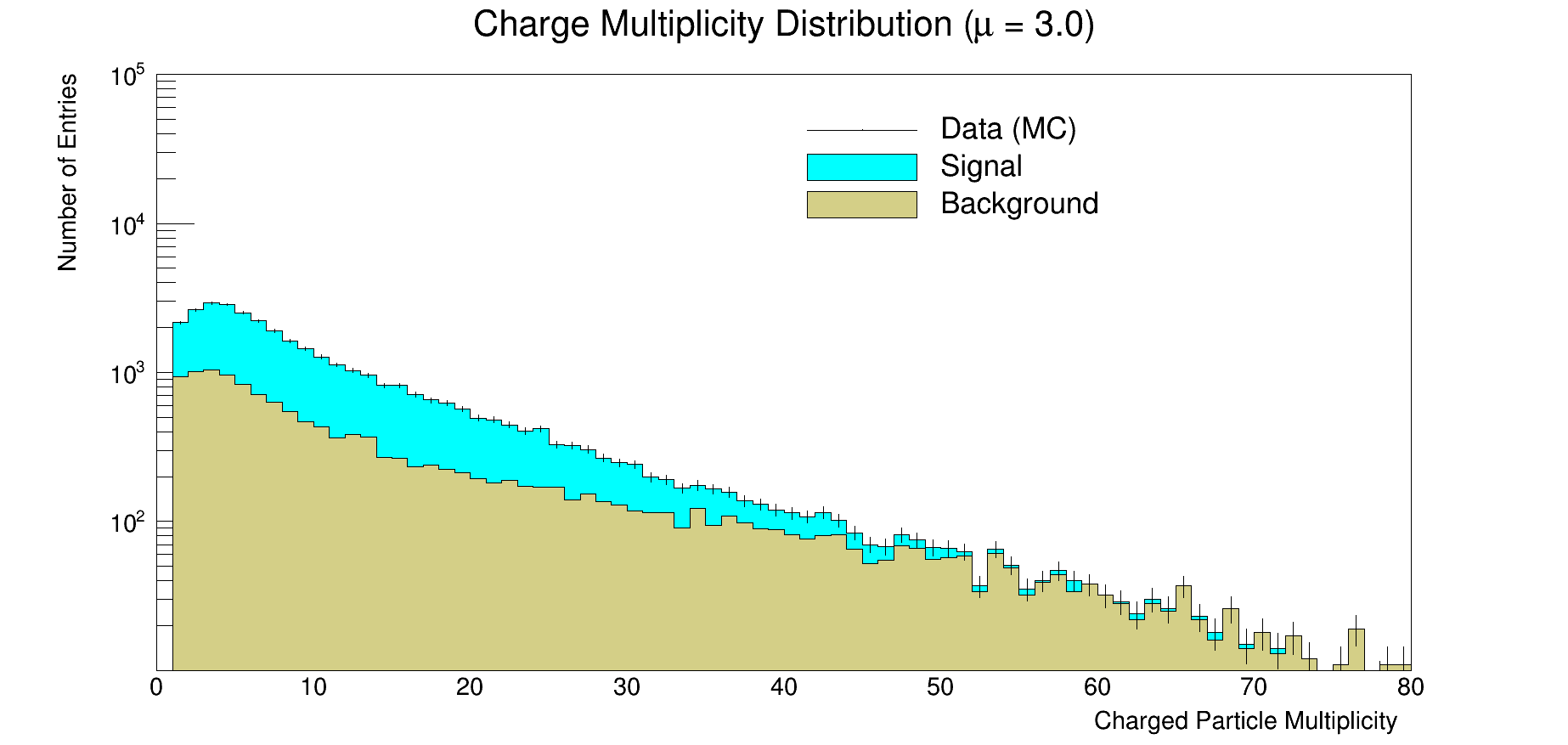}
\caption{Charged particle multiplicity generated by Pythia 8.3 Monte Carlo assuming pile-up of 3. Signal and combinatorial background are shown in blue and brown, respectively.}
\label{exp_soft_2}
\end{wrapfigure}

This is called the event pile-up, $\mu$. If the signal consists of more than one observable (\textit{e.g.} a forward proton, a vertex and a certain number of tracks, \textit{etc.}) then the pile-up of events (each having only a part of the signal-like signature) can produce the same signature as a signal event. This is called the \emph{combinatorial background}. 

For example, let the inner detector of ATLAS \cite{p5} be used to detect the charged particles and the ATLAS Forward Proton (AFP) detectors \cite{p6} to measure forward protons. If one is looking for the signal events that produce a vertex and a forward proton then this signature is also produced from the pile-up of two events: one of which contributes to a vertex and the other to a forward proton. Figure \ref{exp_soft_2} shows this background and the signal for the number of charged particles (charged particle multiplicity) as observable.

\section{Soft Diffractive Case}
\label{3_combinatorial_soft}
This  study is primarily based on the work done by Dr. Sabina Czekierda for her Ph.D. thesis \cite{p3}, where she described a method to estimate the combinatorial background. The present work aims to validate this method using a Monte Carlo simulation. The soft diffractive events were generated in PYTHIA 8.3 \cite{p1} for $pp$ collisions at $\sqrt{s} = 13$ TeV. The signal is defined as a single diffractive event containing a vertex and a forward proton. The cutoffs used for the charged particles are: transverse momentum, $p_{T} > 0.5$ GeV and pseudorapidity, $|\eta| < 2.5$. For the forward protons the fraction of energy lost in the interaction by the proton was required to be $0.01 < \xi < 0.1$.

It was assumed that the pile-up of events follows the Poisson distribution. Thus, the probability that $n_{tot}$ number of pile-up events is given by:
\begin{equation}
    P(n_{\textrm{tot}}, \mu) = e^{- \mu} \frac{\mu^{n_{\textrm{tot}}}}{n_{\textrm{tot}}!},
\end{equation}
where $\mu$ is the average value of pile-up multiplicity.

Let $p_{\textrm{pv}}$ be the probability that single $pp$ interaction produces a forward proton and a vertex, $p_{\textrm{v}}$ -- the probability that it produces a vertex only and $p_{\textrm{p}}$ that it produces only a forward proton. Then, the probability of observing a signal-like signature produced from pile-up is given by:
\begin{equation}
    P_{\textrm{bd}} = P(2, \mu) \cdot p_{\textrm{p}} \cdot p_{\textrm{v}} = \frac{1}{2} e^{-\mu} \mu^{2} \cdot p_{\textrm{p}} \cdot p_{\textrm{v}}.
    \label{2_background_a}
\end{equation}

\noindent The probability of observing a vertex ($n_{vtx}$) without the proton in AFP ($n_{afp}$) is:
\begin{equation}
    P(n_{\textrm{vtx}} = 1, n_{\textrm{afp}} = 0  ) = P(1, \mu) \cdot p_{\textrm{v}} = e^{-\mu} \cdot \mu \cdot p_{\textrm{v}}.
    \label{2_background_b}
\end{equation}

\noindent From equations \ref{2_background_a} and \ref{2_background_b} one gets:
\begin{equation}
    P_{\textrm{bd}} =  P(n_{\textrm{vtx}} = 1, n_{\textrm{afp}} = 0  )\cdot \mu \cdot p_{\textrm{p}}.
    \label{prob_bg_soft}
\end{equation}

The probability of observing an event that contains a vertex and no forward proton, $P(n_{\textrm{vtx}} = 0, n_{\textrm{afp}} = 1)$ can be obtained from the data (MC), but the parameters $\mu$ and $p_{\textrm{p}}$ remain unknown because the detector observes only piled-up events. To estimate these parameters a statistical model is used.

Let the probability of observing $n_{\textrm{vtx}}$ vertices and $n_{\textrm{afp}}$ forward protons in a piled-up event be given by $ P(n_{\textrm{vtx}}, n_{\textrm{afp}} )$. Then:
\begin{flalign}
    P(n_{\textrm{vtx}} = 0, n_{\textrm{afp}} = 1  ) = e^{-\mu} \mu \cdot p_{\textrm{p}},
    \label{2_soft_model_a}\\
        P(n_{\textrm{vtx}} = 1, n_{\textrm{afp}} = 1  ) = e^{-\mu}\mu p_{\textrm{pv}} + e^{-\mu}{\mu}^{2}p_{\textrm{v}}p_{\textrm{p}},
        \label{2_soft_model_b}\\
    P(n_{\textrm{vtx}} = 2, n_{\textrm{afp}} = 1  ) = e^{-\mu}{\mu}^{2}p_{\textrm{v}}p_{\textrm{pv}} + \frac{1}{2}e^{-\mu}{\mu}^{3}{p^{2}_{\textrm{v}}}p_{\textrm{p}}.
    \label{2_soft_model_c}
\end{flalign}

The probabilities on the left hand side of the above equations can be calculated directly from the data. Thus, using the method of least squares the values of the parameters $p_{\textrm{v}}$, $p_{\textrm{p}}$ $p_{\textrm{pv}}$ and $\mu$ on the right hand side of the above equations can be determined. It should be noted that $p_{\textrm{pv}}$, $p_{\textrm{v}}$ and $p_{\textrm{p}}$ add up to the full phase-space (ignoring the empty events):
\begin{equation}
    p_{\textrm{pv}} + p_{\textrm{v}} + p_{\textrm{p}} = 1.
    \label{least_sqaures_explain}
\end{equation}
Usage of Eq. \ref{least_sqaures_explain} leads to the following $\chi^{2}$ definition:
\begin{equation}
    \chi^{2}(\mu, p_{\textrm{pv}}, p_{\textrm{p}}) = \sum^{3}_{i} \frac{(P^{\textrm{measured}}_{i} - P^{\textrm{model}}_{i}(\mu, p_{\textrm{pv}}, p_{\textrm{p}}))^{2}}{\sigma^{2}_{i}}, \mathrm{where}
\end{equation}

\begin{equation}
    \sigma_{i} = \frac{\sqrt{P^{\textrm{measured}}_{i}}}{\sqrt{N_{\textrm{tot}}}}
\end{equation}
and the index $i = 1, 2 \ \mathrm{and} \ 3$ corresponds to the equations \ref{2_soft_model_a}, \ref{2_soft_model_b} and \ref{2_soft_model_c}, respectively.

\begin{wrapfigure}{r}{0.55\linewidth}
\centering
\includegraphics[width = 1.1\linewidth]{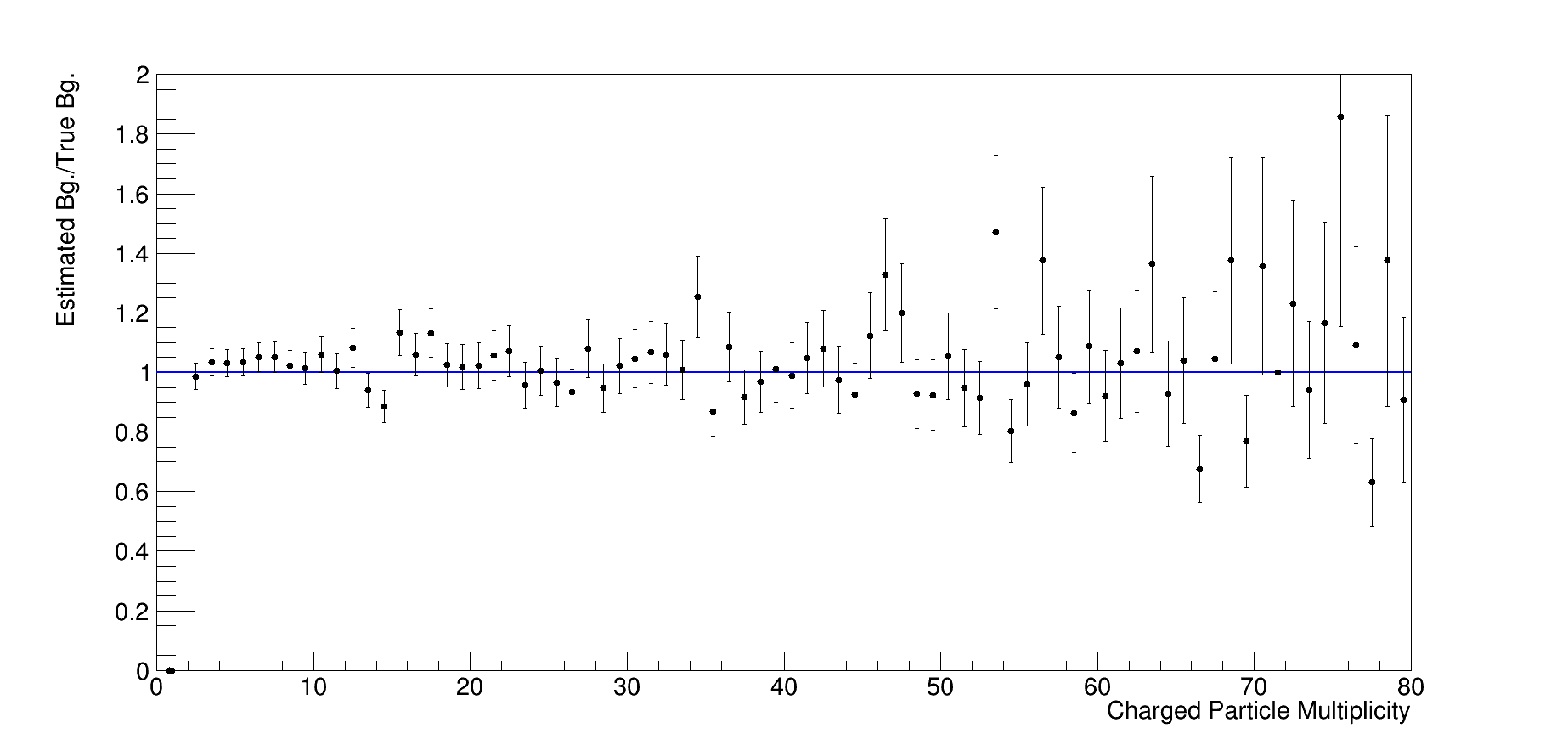}
\caption{Ratio of estimated to the true background for diffractive events generated by Pythia 8.3 MC for $\mu = 3$.}
\label{r_soft_cb_3}
\end{wrapfigure}

The Minuit package \cite{p4} available in CERN ROOT program was used to find the minimum value of $\chi^{2}$ and that of the parameters. Once the parameters are known, the equation \ref{prob_bg_soft} can be used to estimate the background. For this study the charged particle multiplicity of an event is used as an observable. The expected number of events in the combinatorial background for a given range of charged particle multiplicity can be obtained by calculating the expectation value of background for that range. Figure \ref{r_soft_cb_3} shows the ratio of estimated background (calculated from the above procedure) and the true background (known from MC). The ratio being close to 1 within reasonable error limits validates the method used to estimate the background in the study.

\section{Hard Diffractive Case}
In the hard diffractive production of $W$ boson, taken as an example, an event that contains a $W$ boson and a forward proton coming from the same vertex was considered as a signal. The pile-up of an event containing a $W$ boson and a vertex with an event containing a forward proton will produce the same signature. This is the combinatorial background for the hard diffractive event.

The MC sample of hard diffractive $W$ boson production in $pp$ collisions at $\sqrt{s} = 13$ TeV were generated using PYTHIA 8.3 \cite{p1}. The cutoffs used for the charged particles were $p_{T} > 0.5$ GeV and $|\eta| < 2.5$ and $0.02 < \xi < 0.1$ for the forward protons. Similarly to the previous study of soft diffractive events, it was assumed that the pile-up of events follows Poisson distribution.\\

Let $\mu$ be the pile-up multiplicity, $p_{\textrm{p}}$ be the probability of observing an event having only a forward proton, $p_{\textrm{v}}$ be the probability of observing an event having only a vertex, $p_{\textrm{wv}}$ be the probability of observing an event having a W boson and a vertex and $p_{\textrm{wpv}}$ be the probability of observing an event having a W boson, a forward proton, and a vertex. The probability of observing an event from the combinatorial background is:
\begin{equation}
    P_{\textrm{bg}} = P(2, \mu) \cdot p_{\textrm{p}} \cdot p_{\textrm{wv}} = e^{-\mu} \cdot \mu^{2} \cdot p_{\textrm{p}} \cdot p_{\textrm{wv}}
    \label{3_hard_a}
\end{equation}
\noindent and the probability of observing an event with a $W$ boson and a vertex is given by:
\begin{equation}
    P(n_{\textrm{w}} = 1, n_{\textrm{p}} = 0, n_{\textrm{v}} = 1) = P(1, \mu) \cdot p_{\textrm{wv}} = e^{-\mu} \cdot \mu \cdot p_{\textrm{wv}}.
    \label{3_hard_b}
\end{equation}

\noindent From equations \ref{3_hard_a} and \ref{3_hard_b} one will get:
\begin{equation}
    P_{\textrm{bg}} = P(n_{\textrm{w}} = 1, n_{\textrm{p}} = 0, n_{\textrm{v}} = 1) \cdot \mu \cdot p_{\textrm{p}}.
    \label{prob_bg_hard}
\end{equation}
To estimate the parameters $\mu$ and $p_{\textrm{p}}$ the following model was used:
\begin{align}
    P(n_{\textrm{w}} = 1, n_{\textrm{p}} = 0, n_{\textrm{v}} = 1) &= e^{-\mu} \cdot \mu \cdot p_{\textrm{wv}},\\
    P(n_{\textrm{w}} = 1, n_{\textrm{p}} = 1, n_{\textrm{v}} = 1) &= e^{-\mu} \cdot \mu \cdot p_{\textrm{wpv}} + e^{-\mu} \cdot \mu^{2} \cdot p_{\textrm{p}} \cdot p_{\textrm{wv}},\\
    P(n_{\textrm{w}} = 1, n_{\textrm{p}} = 1, n_{\textrm{v}} = 2) &=  e^{- \mu} \cdot \mu^{2} \cdot (p_{\textrm{wpv}} \cdot p_{\textrm{v}} + p_{\textrm{wv}} \cdot p_{\textrm{pv}} + \mu \cdot p_{\textrm{wv}} \cdot p_{\textrm{v}} \cdot p_{\textrm{p}}),\\
    P(n_{\textrm{w}} = 1, n_{\textrm{p}} = 0, n_{\textrm{v}} = 2) &=  e^{-\mu} \cdot \mu^{2} \cdot p_{\textrm{wv}} \cdot p_{\textrm{v}},\\
    P(n_{\textrm{w}} = 1, n_{\textrm{p}} = 2, n_{\textrm{v}} = 1) &=  e^{-\mu} \cdot \mu^{2} \cdot p_{\textrm{wpv}} \cdot p_{\textrm{p}} + \frac{1}{2} \cdot e^{-\mu} \cdot \mu^{3} \cdot p_{\textrm{wv}} \cdot p^{2}_{\textrm{p}}.
\end{align}

\begin{wrapfigure}{r}{0.55\linewidth}
\includegraphics[width = 1.1\linewidth]{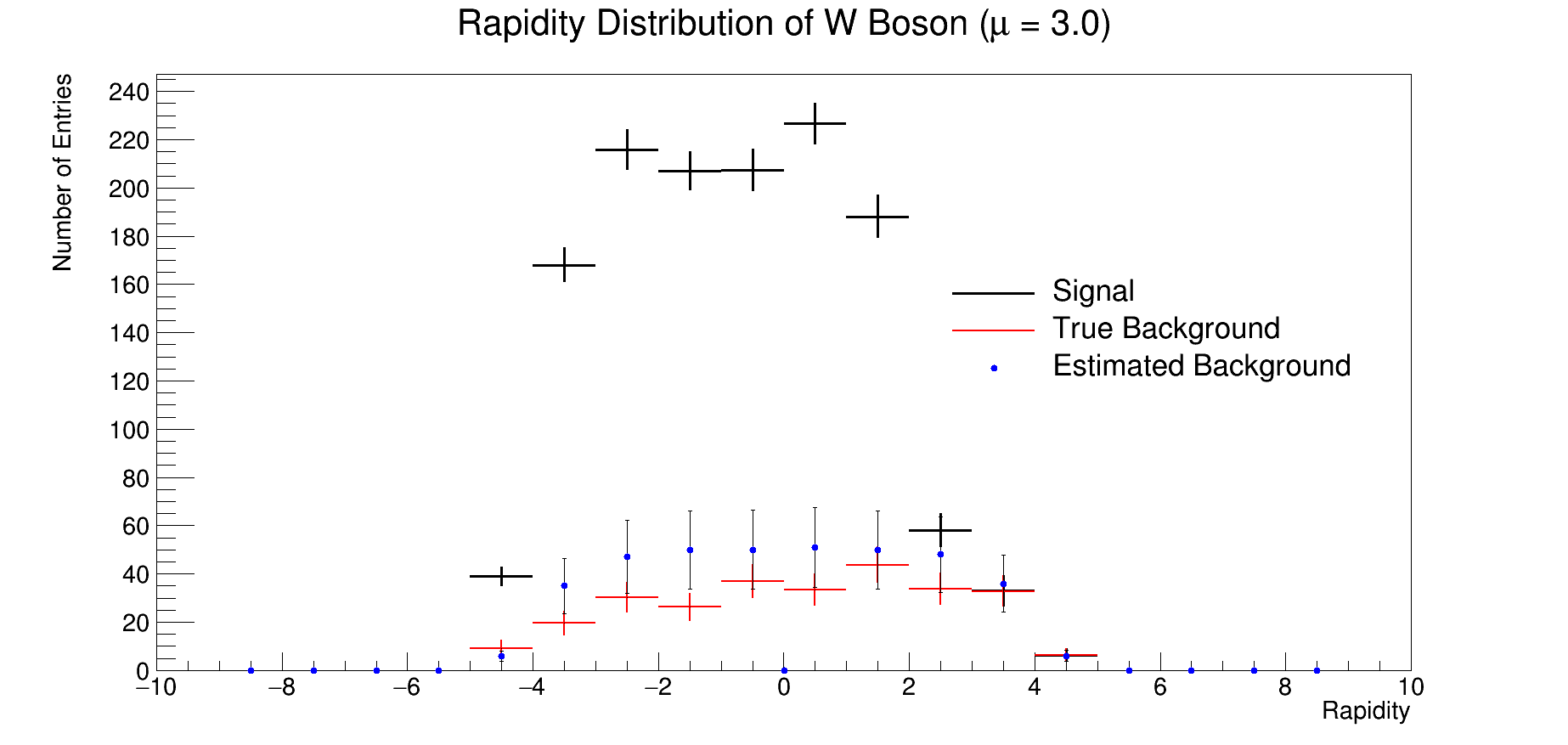}
\caption{Rapidity of $W$ boson for estimated (blue dots) and true (magenta points) background for hard diffractive events generated by Pythia 8.3 MC for $\mu = 3$.}
\label{results_hard}
\end{wrapfigure}

\noindent Note that $n_{\textrm{w}}$, $n_{\textrm{p}}$ and $n_{\textrm{v}}$ represent the number of $W$ bosons, forward protons and vertices, respectively.

Using the least-square method, the parameters $\mu$ and $p_{\textrm{p}}$ can be obtained (\textit{cf.} Fig. \ref{3_combinatorial_soft}). Once the parameters are known, the combinatorial background can be estimated by using Eq. \ref{prob_bg_hard}. Figure \ref{results_hard} shows the signal, estimated background the true background against the rapidity of $W$ boson. At low event multiplicities significant background from non-diffractive $W$ bosons was observed which was not accounted for in this study.

\section{Conclusion}

The proposed statistical models provide a good description of the pile-up process and can be used for the estimation of combinatorial background for soft and hard diffractive events.

\ResCnt

\maketitle
\let\thefootnote\relax\footnote{Work done during PPSS 2020, together with Mr. Filip R\k{e}kawek, under the supervision of dr hab. Krzysztof Wo\'zniak.}
\addtocounter{footnote}{-1}\let\thefootnote\svthefootnote

\begin{abstract}
Using COsmic Ray SImulations for KAscade (CORSIKA), five different altitudes were picked at which particles were detected. At each altitude the percentage of particles in different radius of the detectors were shown. Finally, the average number of muons, photons and electrons at each altitude was calculated.
\end{abstract}

\section{Introduction}
In some theoretical models \cite{theoredical2002} it is predicted that the interaction of ultra high energy cosmic ray particles such as photons may happen far from the Earth. If it is so, such high energy cosmic ray particles should create a lot of (tens/hundreds/thousands) particles forming cascades in the atmosphere. This phenomena are called super pre-showers. The study of such events has an impact on the fundamental particle physics, ultra-high energy astrophysics and cosmology \cite{example2016}.

\begin{wrapfigure}{r}{0.45\linewidth}
\centering
\includegraphics[width=0.9\linewidth]{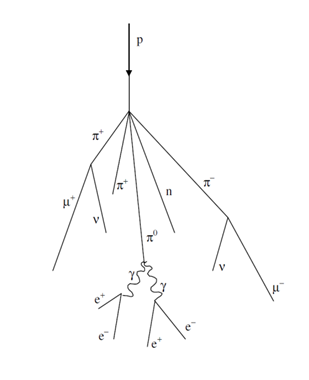}
\caption{An illustration of the shower of particles produced by cosmic ray collision.}
\label{fig_diagram}
\end{wrapfigure}

A typical cosmic ray diagram is shown on Figure \ref{fig_diagram}: high energy proton ($p$) interacts with atmosphere creating particles (here pions, $\pi^\pm$ and neutrons, $n$), these can further interact or decay creating a particle shower -- cascade.

Studies of such showers are done by dedicated experiments. One of them is KASCADE \cite{kascade}. It consists of 252 scintillator detector stations, distributed in a regular grid with 13 m spacing covering an area of 200x200 $m^2$.

It is worth mentioning that there are projects like the Cosmic-Ray Extremely Distributed Observatory (CREDO) \cite{credo} which aims to combine data from all available detectors. For this purpose even a very simple detectors can be used, including individual smartphones, to detect muons arriving to the surface of the earth.

\section{CORSIKA Program}
Data analysed in this project was extracted from CORSIKA \cite{Heck1998CORSIKAAM} -- COsmic Ray SImulations for KAscade. Its purpose is to simulate in detail the development of air showers in the atmosphere. All secondary particles produced in the air are tracked until they interact of decay. The following parameters can be set up:

\begin{itemize}
\item primary particle type (recognizes 50 elementary particles),
\item energy range (up to $10^{20}$ eV),
\item primary angle of incidence,
\item the atmospheric parameter -- the density variation of the atmosphere,
\item observation levels (up to altitude of 30 km),
\item simulation of electromagnetic cascades (on/off).
\end{itemize}

\section{Simulation of Cascades}

Five different altitudes were picked to detect the particles: 110 m, 2000 m, 5000 m, 10000 m and 15000 m. It is expected that muons are created below 15 km. CORSIKA simulations were run at the above observation levels for three value of energies: 10 TeV, 100 TeV and 1000 TeV. For each energy 100 cosmic-ray showers were generated. Both low and high energy particles were studied. The generated files were analysed using CERN ROOT framework\cite{root}.

As an example, the multiplicity of muons present in cascade caused by 10 TeV particle and registered at altitude of 110 m is shown in Fig. \ref{fig_number_of_muons} (left). Beside their multiplicity, it is interesting to see how muons are located wrt. position of particle which initiated cascade. The amount of muons present at 5000 m is shown in Fig. \ref{fig_number_of_muons} (right). The inner red circle, with radius of about 1.2 km, contains 90\% of muons. 95\% of muons are expected to be within a radius of 1.7 km. It should be noted that the amoun tof muons decreases rapidly with increase of distance for cascade center $(0, 0)$.

\begin{figure}[!htbp]
\centering
\includegraphics[width=0.5\textwidth]{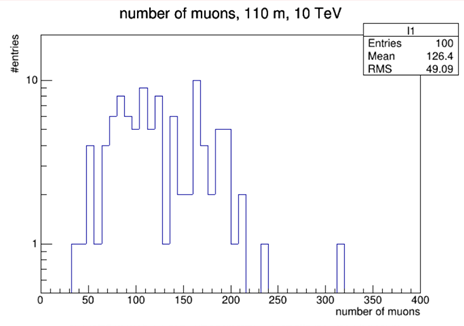}
\includegraphics[width=0.38\textwidth]{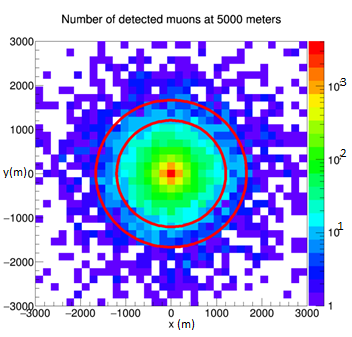}
\caption{CASCADE simulation. \textbf{Left:} multiplicity of muons present in cascade caused by 10 TeV particle and registered at altitude of 110 m. \textbf{Right:} location of muons present in particle shower wrt. its center $(0, 0)$ registered at the altitude of 5000 m.}
\label{fig_number_of_muons}
\end{figure}

\subsection{Studies of Muon Multiplicities}
Results discussed above illustrate situation for a given altitude ad energy of incoming particle. It is interesting to study how situation changes for different initial conditions and observation points.

As shown on figure \ref{fig_height_radious} (left), the average number of muons increases with the energy of cosmic-ray particle initiating a shower. It is not a surprise as more energy means more secondary particles produced, many of which are muons. This number first grows while going upward through the atmosphere, but then between 2 and 4 km decreases so much that at 15 km it gets lower than on the ground. Possible explanation is that muons are created at altitude not much higher then around 15 km because of the atmosphere density

\begin{figure}[!htbp]
\centering
\includegraphics[width=0.48\textwidth]{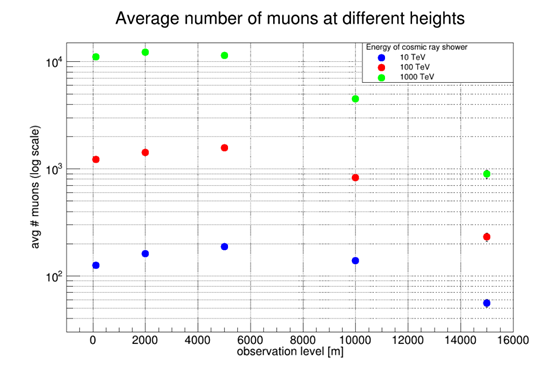}
\includegraphics[width=0.48\textwidth]{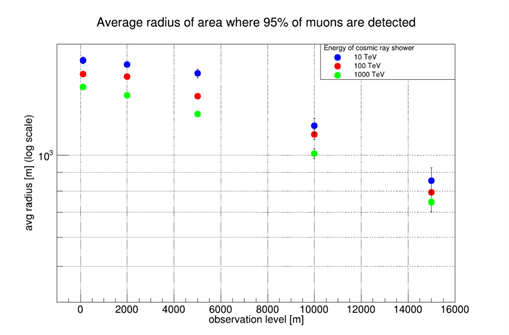}
\caption{CORSIKA simulation for cascades initiated by particle of 10 (blue), 100 (red) and 1000 TeV (green points): average number of muons (\textbf{left}) and average radius of area containing 95\% of muons at various altitudes.}
\label{fig_height_radious}
\end{figure}

As shown of Fig. \ref{fig_height_radious}, radius of the area where muons were detected decreases with an increase of both the observation level and the energy of cosmic-ray shower. The higher the observation is conducted, the shorter time it takes to develop a cascade in the atmosphere. As a result the secondary particles are less spread. It is worth stressing that this is expected accordingly to the special relativity theory: as the energy of the primary particle increases the secondary particles are generally emitted at smaller angles thus their distance from the center of the cascade is on average smaller.

\subsection{Comparison with Photons and Electrons}
Finally, it is interesting to compare multiplicity of muons to the one of electrons and photons. As one can immediately judge from Fig. \ref{fig_electrons_photons}, there are much more photons and electrons than muons at each observation level. The fact that initially (at 15 km) there were more electrons than photons but this changes at lower altitudes can be explained by annihilation of positrons. Positrons are produced by the decay of muons happening in about 15 km. That in result decreases their number and simultaneously increases number of photons in the cascade.

\begin{figure}[!htbp]
\centering
\includegraphics[width=0.65\textwidth]{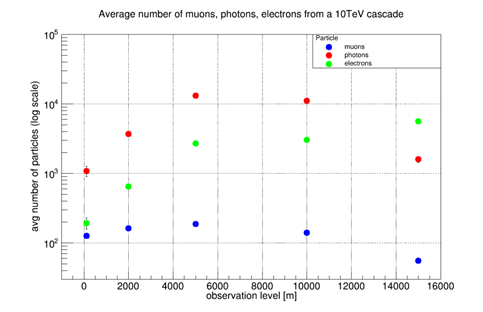}
\caption{Average number of muons, photons and electrons at different altitudes and energies of particles initiating cascades.}
\label{fig_electrons_photons}
\end{figure}

\section{Summary}
The interaction of ultra high energy cosmic ray particles with atmosphere creates a lot of particles which form cascades. The study of such events was done using CORSIKA program. Five different altitudes and three energies of incoming particles were considered. At each altitude the percentage of particles in different radius of the detectors were shown. Finally, the average number of muons, photons and electrons at each altitude was calculated.

\ResCnt

\maketitle

\begin{abstract}
Searches for the New Physics and potential signs of phenomena beyond the Standard Model are a significant part of today's particle physics experiments. The heavy $\tau$ lepton is used as an important signature for many analyses at the Large Hadron Collider (LHC) at CERN. In analyses using $\tau$ leptons, estimation of backgrounds arising from jets misidentified as hadronically decaying $\tau$ leptons becomes a crucial issue. This article presents an usage of the data-driven fake factor method for background modelling, employed by the ATLAS experiment at the LHC, which is based on the ongoing $H^{\pm}\to\tau^{\pm}\nu_{\tau}$ analysis. 
\end{abstract}

\section{Introduction}

The discovery of the Higgs boson in 2012 by ATLAS \cite{ATLAS} and CMS \cite{CMS} Experiments at CERN complemented the Standard Model (SM) as the most successful particle physics theory of all times. Despite its undeniable success at describing fundamental particles and their interactions, the SM does not explain significant phenomena observed today, like matter-antimatter asymmetry or neutrino masses. These problems induce the scientists to search for extensions to the SM, called Beyond Standard Model (BSM) or New Physics theories, that would fill the gaps in our understanding of the Universe. Examples of such theories like two-Higgs-doublet models (2HDM) and the Minimal Supersymmetric Standard Model (MSSM) predict existence of more Higgs bosons, including a charged Higgs boson $H^{\pm}$ \cite{MSSM}.
In this article, the focus is put on the case where such a boson would decay into a $\tau$ lepton and a $\nu_{\tau}$ neutrino: $ H^{\pm}\to\tau^{\pm}\nu_{\tau}$. 

\section{Tau lepton}

The $\tau$ lepton, discovered in 1975, provides not only unique tools for studying SM Higgs boson properties, but also is an important element of searches for New Physics phenomena at LHC at CERN. The $\tau$ lepton has a mass of $m_{\tau}$ = 1.777 GeV, what makes it the only lepton heavy enough to decay into both: hadrons (65\% of decays) and light leptons (35\% of decays). 

In the ATLAS Experiment, reconstruction and identification of hadronically decaying tau leptons consist of two steps. Firstly, $\tau$ candidates are recreated from jets reconstructed in the calorimeters and then tracks reconstructed in the tracker are attached to them. The second step
is identification based on the machine learning, which aims at distinguishing of $\tau$ candidates coming from true $\tau$ and those coming
from quark/gluon jets. The most recent method is recurrent neural networks (RNN)\cite{RNN}, which uses variables related to calorimeter deposits and the reconstructed tracks.

\section{Search of the charged Higgs boson}

The observation of charged Higgs boson would be the confirmation of BSM theories and would lead to new perspectives on the completion of Standard Model. In this search, the potential production process of the charged Higgs boson is associated with top/b-quarks through $gg\to...$\footnote{Type of process depends on $H^{\pm}$ mass.}$\to(W^{\pm}\to\Bar{b})(H^{\pm}b)$. The $H^{\pm}\to\tau^{\pm}\nu_{\tau}$ channel is important as one of the major decay modes of $H^{\pm}$. Only hadronically decaying $\tau$ leptons are analysed.

Depending on the decaying mode of the $W$ boson produced simultaneously with $H^{\pm}$, two different channels are possible:  $\tau$-jet, when the $W$ decays to hadrons and $\tau$-lep, in case of $W$ decaying to leptons. Full description of the $H^{\pm}\to\tau^{\pm}\nu_{\tau}$ analysis using 2015-2016 dataset collected by the ATLAS Experiment can be found in Ref. \cite{HplusNote}. 

\subsection{Estimation of background from misidentified $\tau$ leptons}

The dominant background processes can be categorized based on the object that gives rise to reconstructed and identified hadronically decaying $\tau$ candidate. The $\tau$ candidates considered in this analysis are identified through their hadronic decays, which are characterized by the presence of mostly one or three charged tracks (called prongs), accompanied by a neutrino and possibly neutral pions. Although a hadronic $\tau$ decay with only one charged hadron can be mimicked by electrons and muons, the dominant background source for $\tau$
identification are quark- or gluon-initiated jets, as a result of their large production cross-section. These backgrounds are poorly modelled because of statistical limitations in the sample of simulated events (e.g. multi-jet processes). Therefore, an approach based on data, denoted the fake factor (FF) method, is used to estimate backgrounds arising from jets misidentified as $\tau$ candidates.

\subsection{Fake factor method}

The fundamental idea of this data-driven method is simple: select a control sample (enriched with events of the background being estimated) and then use an extrapolation factor to relate these events to the background in the signal region (SR).
For this purpose, an $\mathrm{anti-\tau}$ selection is defined by requiring the $\tau$ candidate to fail the identification criteria of the nominal selection (based on RNN identification). The extrapolation factor, called the fake factor (FF), is defined as the ratio between the number of jets reconstructed as $\tau$ candidates and fulfilling the nominal $\tau$ identification criteria ($\mathrm{N^{CR}_{\tau-id}}$) to the number of corresponding candidates failing the identification criteria ($\mathrm{N^{CR}_{anti-\tau-id}}$). The FF is measured in a dedicated control region (CR) enriched with fake $\tau$s as follows:

\begin{equation}
    \mathrm{FF = \frac{N^{CR}_{\tau-id}}{N^{CR}_{anti-\tau-id}}}
\end{equation}

\noindent Then, total number of background events from jets in signal region $\mathrm{N^{SR}_{\tau-fakes}}$ is computed as:
\begin{equation}
    \mathrm{N^{SR}_{\tau-fakes} = N^{SR}_{anti-\tau-id}(data) \times FF}
\end{equation}

\subsection{Considering quark-gluon jet composition}

Fake factors usually strongly depend on the type of jets. Therefore it is needed to ensure the same composition of quark-gluon jets in the control region as is expected in the signal region. In $H^{\pm}\to\tau^{\pm}\nu_{\tau}$ analysis, these proportions are not known, hence the fake factors are extracted in two control regions - enriched in either gluon-initiated (Multijet CR) or quark-initiated jets (W+jet CR). Then, the FFs for each CR are linearly combined using a free parameter $\mathrm{\alpha_{MJ}}$ to obtain final fake factors:  

\begin{equation}
    \mathrm{FF = \alpha_{MJ}\times FF_{MJ} + (1-\alpha_{MJ})\times FF_{W+jet}}
\end{equation}

The $\mathrm{\alpha_{MJ}}$ represents the gluonic jet fraction and is estimated using template-fit method. It is based on the variable that allows distinguishing jet origin ($\tau$ jet width in this search). Then, for each $\tau$ $p_{T}$ bin $i$ templates of that variable for Multijet and W-jet CRs are produced and their linear combination is fitted to the normalized distribution measured in the signal region, by varying the $\mathrm{\alpha_{MJ}}$ in every bin of $p_{T}$ and minimizing the $\chi^{2}$ distribution for each channel separately. Example of the template-fit procedure for 1-prong $\tau$ candidate in $\tau$-jet SR is shown in Fig.\ref{fig:ff_alpha}.

\begin{figure}[ht]
  \begin{center}
\includegraphics[width=0.95\textwidth]{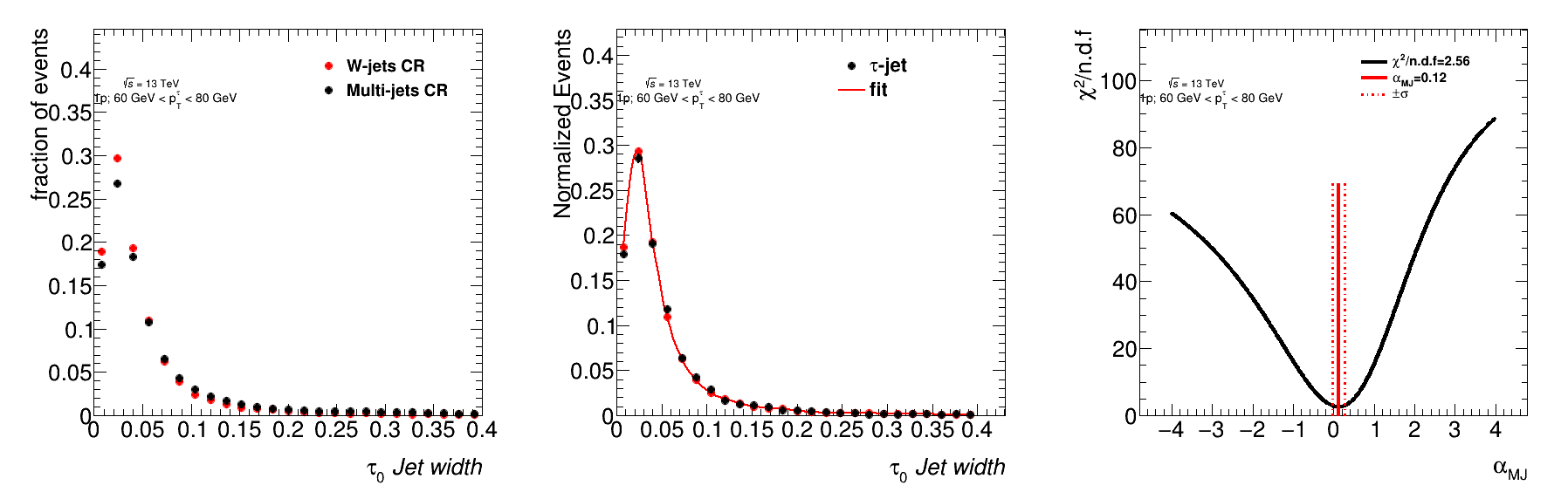}
  \end{center}
  \caption{
Example of the $\mathrm{\alpha_{MJ}}$ estimation process in $\tau$-jet SR in three steps: preparing $\tau$ jet width templates for Multijet and W+jet CRs (left), fitting linear combination of CRs templates to distribution in the SR (center) and $\chi^{2}/ndf$ distribution of the fit as a function of $\mathrm{\alpha_{MJ}}$ (right).
   }
  \label{fig:ff_alpha}
\end{figure}

\subsection{Validation of background estimation}

Validation of the fake factor method, presented in Fig. \ref{fig:ff_validation}, is conducted in the control region with the same particle restrictions as a signal region $\tau$+electron, but with additional requirement of same-sign charges of $\tau$ and electron (same-sign $\tau$+e CR), what ensures high number of fake $\tau$ candidates. A pile-up of simulated background processes and estimated misidentified jets (violet area) is compared with experimental data (black dots).  As can be seen, the FF method provides effective background estimation. 

\begin{figure}[ht]
  \begin{center}
\includegraphics[width=0.6\textwidth]{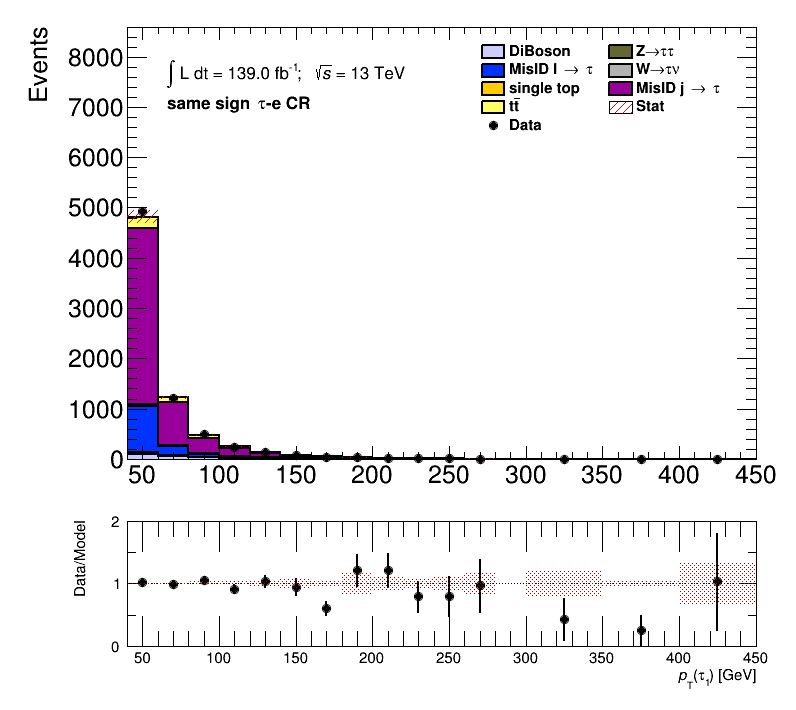}
  \end{center}
  \caption{
Data compared with fake $\tau$s and simulation in same-sign $\tau$+e CR. Only statistic error is included.
   }
  \label{fig:ff_validation}
\end{figure}

\section{Summary}

Tau lepton is an important part of many BSM physics searches at LHC. Due to its special characteristics and large number of decay channels, the tau lepton requires advanced reconstruction and identification techniques. This article shows the common approach for estimation of misidentified hadronic $\tau$ decays in ATLAS analyses, called the fake factor method. It estimates entire background from all sources and can be implemented for various types of searched processes. Although the presented results, based on the ongoing $H^{\pm}\to\tau^{\pm}\nu_{\tau}$ analysis, are preliminary and additional refinement is needed, the usage of the method enables precise background estimation.

\ResCnt

\maketitle

\begin{abstract}
The search for the first-order phase transition between the hadronic matter and quark-gluon plasma is discussed. The factorial cumulants and cumulants in the selected subsystem of the relativistic heavy-ion collision are calculated assuming global baryon number conservation and short-range correlations.
\end{abstract}

\section{Introduction}

It is commonly known that the atomic nuclei consist of protons and neutrons (hadrons) which are built up by the confined quarks and gluons. The early Universe is believed to have been fulfilled with deconfined quarks and gluons, the quark-gluon plasma (QGP). The lattice quantum chromodynamics (QCD) numerical calculations showed that at the baryon chemical potential $\mu_B \approx 0$ there is a rapid but continuous crossover between the hadronic matter and QGP with a pseudo-critical temperature of about 155~MeV \cite{Bazavov:2009zn}. The relativistic heavy-ion collisions experiments at the Relativistic Heavy Ion Collider (RHIC) of Brookhaven National Laboratory (BNL) and the Large Hadron Collider (LHC) of the European Organization for Nuclear Research (CERN) provided the signatures of QGP such as the azimuthal asymmetry of particle production, known as elliptic flow \cite{STAR:2000ekf}, or the jet quenching effect \cite{PHENIX:2001hpc}.

Because of the so-called sign problem \cite{Aarts:2015tyj}, lattice QCD methods are limited to a small $\mu_B$ regime. Consequently, the regions of the QCD phase diagram of strongly interacting matter with greater $\mu_B$ are still not well explored \cite{Bzdak:2019pkr}. However, many effective models predict a first-order phase transition with the corresponding critical endpoint between the hadronic matter and QGP \cite{Braun-Munzinger:2015hba,Stephanov:2004wx}. The search for this phase transition is a considerable theoretical and experimental challenge in high-energy physics these days. 

In order to look for the phase transition and the critical point, the regions of higher $\mu_B$ can be reached experimentally by performing the heavy-ion collisions at various collision energies, for example, the Beam Energy Scan program at RHIC and the experiments of NA61/SHINE
Collaboration. The fluctuations of the baryon number, electric charge, and strangeness are sensitive to the critical phenomena, therefore they are studied at different collision energies. These fluctuations are often described using the cumulants, $\kappa_n$, which naturally appear in statistical mechanics. On the other hand, the factorial cumulants, $\hat{C}_n$, might be easier to interpret since they represent the integrated multiparticle correlation functions, and vanish in absence of correlations \cite{Bzdak:2019pkr}. The factorial cumulants can be converted into cumulants using the formula 
\begin{equation} \label{eq:fact-cum-2-cum}
    \kappa_n = \textstyle \sum_{k=1}^{n} S(n,k) \hat{C}_k\,,
\end{equation}
where $S(n,k)$ is the Stirling number of the second kind \cite{Friman:2022wuc}.

The recent results of the STAR and HADES collaborations \cite{STAR:2021fge,HADES:2020wpc} show that the $\kappa_4/\kappa_2$ ratio in the central Au+Au collisions depends non-monotonically on the collision energy. This might be a signature of the critical phenomena. However, other effects, not related to phase transitions, e.g., impact parameter fluctuations and baryon number conservation can also produce significant fluctuations.  In this proceedings, I show how to calculate analytically the factorial cumulants originating from the baryon number conservation and short-range correlations.

\section{Method}

Consider the relativistic heavy-ion collision system divided into two subsystems, inside and outside the acceptance \cite{Barej:2022jij}. For simplicity, we assume that only baryons are present and there are no antibaryons. This makes the reasoning applicable to low energies. Then, the probability that there are $n_1$ baryons in the first subsystem and $n_2$ baryons in the second one, assuming the global baryon number conservation, is 
\begin{equation}
    P_B(n_1,n_2) = A P_1(n_1) P_2(n_2) \delta_{n_1+n_2,B}\,,
\end{equation}
where $A$ is a normalization constant, $P_1(n_1)$ and $P_2(n_2)$ are the probabilities without baryon number conservation and the Kronecker delta requires the total number of baryons to be equal to the conserved value $B$. $P_1$ and $P_2$ include only short-range correlations \cite{Bzdak:2019pkr},
\begin{equation} \label{eq:short-range-fact-cum}
    \hat{C}_k^{(i)} = \alpha_k \langle n_i \rangle \,, \quad i=1,2\,,
\end{equation}
where $\alpha_k$ is the $k$-particle correlation strength ($\alpha_1=1$). Therefore, the probability that there are $n_1$ baryons inside the acceptance assuming the baryon number conservation and short-range correlations is
\begin{equation}
    P_B(n_1) = \textstyle \sum_{n_2} P_B(n_1, n_2)\,.
\end{equation}

Then, the corresponding factorial cumulant generating function (for the first subsystem with baryon number conservation) is calculated \cite{Barej:2022jij}:
\begin{equation}
\begin{split}
    G_{(1,B)}(z) &= \ln \left[ \textstyle \sum_{n_1} P_B(n_1) z^{n_1}  \right] \\ 
    &= \ln \left[ \frac{A}{B!} \left. \frac{d^B}{dx^B} \exp \left( \sum_{k=1} \frac{(x z - 1)^k \hat{C}_k^{(1)} + (x-1)^k \hat{C}_k^{(2)} }{k!} \right) \right|_{x=0} \right]\,.
\end{split}
\end{equation}
Using Faà di Bruno's formula one can obtain
\begin{equation} \label{eq:g_b-bell}
   \begin{split}
	G_{(1,B)}(z) &= \ln\biggl[\frac{A'}{B!} {\rm{Bell}}_B \biggl(  \sum_{k=0}^{\infty}\frac{(-1)^k}{k!} \left[\hat{C}_{k+1}^{(1)}z + \hat{C}_{k+1}^{(2)}\right], \\
	& \sum_{k=0}^{\infty}\frac{(-1)^k}{k!} \left[\hat{C}_{k+2}^{(1)}z^2 + \hat{C}_{k+2}^{(2)}\right],  
	\ldots, \sum_{k=0}^{\infty}\frac{(-1)^k}{k!} \left[\hat{C}_{k+B}^{(1)} z^B + \hat{C}_{k+B}^{(2)}\right] \biggr) \biggr] \,,
	\end{split}
\end{equation}
where ${\rm{Bell}}_B$ is the $B^{th}$ complete exponential Bell polynomial and $A'$ is the constant not relevant for further calculations. The factorial cumulants in the first subsystem with baryon number conservation and short-range correlations are calculated as
\begin{equation} \label{eq:fact-cum}
    \hat{C}_k^{(1,B)} = \left. \frac{d^k}{dz^k} G_{(1,B)}(z) \right|_{z=1}\,.
\end{equation}

\section{Results}
The factorial cumulants are obtained using Eq. \eqref{eq:fact-cum} from the factorial cumulant generating function \eqref{eq:g_b-bell} with Eq. \eqref{eq:short-range-fact-cum}. The approximated factorial cumulants for small $\alpha_k$ in the limit of large $B$ read\footnote{For details of calculations see \cite{Barej:2022jij}.}
\begin{flalign}\label{eq:c1-many-b-large}
	\hat{C}_1^{(1,B)} &= f B  \,,&&
\end{flalign}
\begin{flalign}\label{eq:c2-many-b-large}
	\hat{C}_2^{(1,B)} &\approx f B\mlb[ -f + \mfb \alpha_2 \mr]  \,,&&
\end{flalign}
\begin{flalign}\label{eq:c3-many-b-large}
	\hat{C}_3^{(1,B)} &\approx f B \mlb[ 2f^2 -6 \mfb f \alpha_2 + \mfb (1-2f) \alpha_3 \mr]   \,,&&
\end{flalign}
\begin{flalign}\label{eq:c4-many-b-large}
	\hat{C}_4^{(1,B)} &\approx f B \; [ -3!f^3 +36\mfb  f^2 \alpha_2 -12\mfb  f(1-2f) \alpha_3 +\mfb (1-3\mfb f) \alpha_4 ] \,,&&
\end{flalign}
where $f$ is a fraction of particles in the first subsystem and $\mfb = 1-f$. Note that for large $B$, the factorial cumulants are proportional to baryon number $B$. Moreover, $\hat{C}_n^{(1,B)}$ is not influenced by $\alpha_k$ with $k > n$. For example, in $\hat{C}_3^{(1,B)}$ only 2- and 3-particle short-range correlations (represented by $\alpha_2$ and $\alpha_3$) are significant and higher-order ones are suppressed.

The cumulants are calculated from the factorial cumulants using Eq. \eqref{eq:fact-cum-2-cum}. The cumulants in the subsystem with baryon number conservation, $\kappa_n^{(1,B)}$, expressed in terms of the global short-range cumulants (in both subsystems combined) without baryon number conservation, $\kappa_m^{(G)}$, are given by:
\begin{flalign} \label{eq:k1-1}
	\kappa_1^{(1,B)} &= f \kappa_1^{(G)}  \,,&&
\end{flalign} 
\begin{flalign} \label{eq:k2-1}
	\kappa_2^{(1,B)} &\approx \mfb f \kappa_2^{(G)}  \,,&&
\end{flalign} 
\begin{flalign} \label{eq:k3-1}
	\kappa_3^{(1,B)} &\approx \mfb f (1-2f)\kappa_3^{(G)}  \,,&&
\end{flalign}
\begin{flalign}\label{eq:k4-1}
	\kappa_4^{(1,B)} &\approx \mfb f \mlb[ \kappa_4^{(G)} - 3\mfb f \mlb(\kappa_4^{(G)} + \mf{(\kappa_3^{(G)})^2}{\kappa_2^{(G)}} \mr)  \mr]  \,.&&
\end{flalign}
These results are in agreement with net-baryon cumulants obtained in a different approach, using statistical mechanics in the thermodynamic limit and with subensemble acceptance in Ref. \cite{Vovchenko:2020tsr}. We note that the formulas from \cite{Vovchenko:2020tsr} do not include dependence on the baryon or antibaryon number. This fact can explain the agreement between baryon and net-baryon number cumulants. 

\section{Summary}
The global baryon number conservation can generate baryon number correlations. The mathematical model to calculate these (long-range) correlations in the subsystem with short-range correlations assumed was presented in this paper. The factorial cumulant generating function \eqref{eq:g_b-bell} enables the calculation of the factorial cumulants [Eqs. (\ref{eq:c1-many-b-large})-(\ref{eq:c4-many-b-large})] and cumulants [Eqs. (\ref{eq:k1-1})-(\ref{eq:k4-1})]. This method should also allow for the calculation of the first correction to cumulants in the limit of large $B$. It would be important to take also antibaryons into account.

\section*{Acknowledgements}
This work was partially supported by the Ministry of Science and Higher Education, and by the National Science Centre, Grant No. 2018/30/Q/ST2/00101.

\ResCnt

\maketitle

\begin{abstract}
This paper describes the selection of Single Diffractive Jets (SD JJ) in data collected with the ATLAS Forward Protons (AFP) detectors. The method to distinguish signal events will be shown. Finally, background subtraction will be applied to reveal the presence of SD JJ in the sample.
\end{abstract}

\section{Introduction}

Located at the Large Hadron Collider (LHC) \cite{LHC}, the ATLAS experiment \cite{ATLAS_paula} has been designed with the goal of measuring the products of proton--proton collisions. Although its central part has a full azimuthal angle coverage and a large acceptance in pseudorapidity ($|\eta|<$4.9), some scattered particles escape detection. This is especially unfortunate for a certain group of physics processes, in particular the diffractive physics.

Diffractive processes can be characterised by the presence of: a rapidity gap (a space in rapidity where no particles are produced) and protons scattered at very small angles. The signature is due to the nature of such interactions in which the exchanged object is a colour singlet: a Pomeron (in QCD: two gluons + h.o. terms) for the strong interactions. 

Because of those characteristic observables one can think of two methods of detecting such processes. The first approach focusses on studies of the rapidity gap. It is a classical recognition method. Unfortunately, the gap may be destroyed by \textit{e.g.} particles coming from from pile-up\footnote{Multiple proton-proton collisions happening during the same bunch crossing. The pile-up multiplicity is indicated as $\mu$.} or may be produced outside the acceptance of the ATLAS central detector. The second method is based on detecting scattered protons. The advantage of this approach is that the protons are measured directly, hence it can be used in the non-zero pile-up environment. However, it requires the additional forward detectors installed far away from the interaction point (IP). Around ATLAS IP those are: the ATLAS Forward Proton \cite{AFP} system and Absolute Luminosity For ATLAS (ALFA) detectors\footnote{ALFA detectors are not topic of this note and will not be discussed further.}.

\section{Data Sample}

Single Diffractive Di-Jet\footnote{Commonly di-jet is defined as two jets comming from the same vertex.} production (SD JJ) is characterised by the presence of two jets and forward proton as well as a rapidity gap between them. In order to prevent the gap being populated by events coming from pile-up, in this work only the periods of data taking (so-called runs) with a very small $\mu\sim$ 1 were considered.


The first step of the event selection starts already during the data taking. Algorithms for very quick but crude selection are called the Level 1 (L1) triggers and are the first part of ATLAS trigger chain. After that selected events are passed to the second part called High Level Triggers (HLT). These algorithms have more time for the event selection thus tend to be more sophisticated.

In this analysis, the interest is on events with a forward proton on either side of the ATLAS detector (called A or C) and presence of at least two jets. In order to select such events the combination of triggers selecting events with jets (with minimum transverse momentum 10 or 20 GeV) and/or protons (signal in AFP detectors) were used.

\section{Signal Selection}\label{sec:sig_selection}

In SD JJ events it is expected to see some dependence between central system and forward proton. In order to see such relationship one can look at energy lost by proton during collision and compare it to the energy loss calculated from objects produced in the central system. Table~\ref{tab:energy_loss} presents various possible measurements of energy losses. One can use two jets with the highest p$_T$, but also use response of ATLAS calorimeters (reconstructed clusters of calorimeter cells) or information from ATLAS inner detector (reconstructed tracks).

\begin{table}[htbp!]\centering
\caption{Various definitions of energy loss. The sign in formulas indicates the side on which forward proton is produced (plus/minus is for A/C side of ATLAS). The transverse momentum of the object is denoted as $p_T$ and its rapidity as $y$.}
\begin{tabular}{ @{} l c p{7cm} @{} }
\hline
\textbf{Object} & \textbf{Formula} & \textbf{Description} \\ \hline
proton & $\xi_p^{\pm}=1-\frac{E_{proton}^{A/C}}{E_{beam}}$ & values limited by the acceptance of AFP detectors (0.03$<\xi_p<$0.1)\\[1ex]
clusters & $\xi_{cl}^{\pm}=\frac{1}{2 \cdot E_{beam}}\sum_{cl}p_T^{cl} \exp(\pm y^{cl})$ & values expected to be similar to $\xi_p$ \\[4ex]
tracks & $\xi_{trk}^{\pm}=\frac{1}{2 \cdot E_{beam}}\sum_{trk}p_T^{trk} \exp(\pm y^{trk})$ & likely to be smaller than $\xi_p$ due to central detector acceptance ($|\eta_{trk}|<$ 2.4)\\[1ex]
dijet & $\xi_{dijet}^{\pm}=\exp(\pm y_{dijet})M_{dijet}\frac{1}{2E_{beam}}$ & anticipated to be smaller than $\xi_p$ since only part of central system is taken\\[1ex]
\hline
\end{tabular}
\label{tab:energy_loss}
\end{table}

Selected samples contain both signal and background events (see the next section). In order to reduce the background-to-signal ratio, a few selection steps were done during this analysis. First, a very effective cut is selection of events having only one primary vertex reconstructed without any additional vertices. This significantly reduces the pile-up events. Due to the nature of the studied process, it is justified to look at the events containing exactly two jets and one reconstructed proton.

At this point in analysis the question appears if the objects used for the event selection are of good quality, \textit{i.e.} if they are true objects and not \textit{e.g.} noise. The presented results are based on jets, protons and clusters. The selection listed in Table~\ref{tab:quality_cuts} was used to satisfy ``good'' quality conditions. The cuts on quality of jets and protons influence the number of events whereas the cut on clusters only affects the distribution of $\xi_{cl}$. After all cuts discussed in this section, the selected events will be referred to as the "signal" sample.

\begin{table}[htbp!]\centering
\caption{Selection applied on jets, clusters and protons.}
\begin{tabular}{ @{} l p{12cm} @{} }
\hline
\textbf{Object} & \textbf{Selection criteria}\\ \hline
Proton & $0.02 < \xi_p < 0.2$\\[1ex]
Jet & being calibrated, come from the primary vertex, have p$^{jet}_T>$20 GeV, $|$timing$|<$12.5 ns and $|\eta^{jet}|<2.4$\\[1ex]
Cluster & p$^{cl}_T>$200 MeV, $|\eta^{cl}|<4.8$, $|$timing$|<$12.5 ns and the most significant sampling (\textit{i.e.} the one with the largest energy) must have $\sigma_{samplings}>3$\\[1ex]
\hline
\end{tabular}
\label{tab:quality_cuts}
\end{table}

\section{Background Subtraction}
As it was mentioned before, some dependence between the central system and the forward proton is expected. Such relationship was seen in the correlation between $\log_{10}(\xi_{cl})$ and proton $x$-position in a very low-$\mu$ 2016 data~\cite{ATL-PHYS-PUB-2017-012}. Therefore, it is expected to see similar dependence in studied 2017 low-$\mu$ data.

On Figure~\ref{fig:bck_sub} (left) one can see the above-mentioned correlation from the "signal" sample. The $\log_{10}(\xi_{cl})$ is on the X axis and the forward proton $x$ position is on the Y axis\footnote{For simplicity only events with proton on side C were shown.}. If the plotted events contain only SD JJ production the clear correlation would be visible. However, the "signal" sample is still dominated by the background.

The main background for SD JJ events comes from the Non-Diffractive jets (ND JJ). The sample containing "background" events was prepared by using random events containing two jets. There were no requirement on forward proton. Of course, it is possible to see SD JJ events with this crude selection but such an addition is negligible due to differences in cross-section.

\begin{figure}[htbp!]\centering

\includegraphics[width=0.45\textwidth]{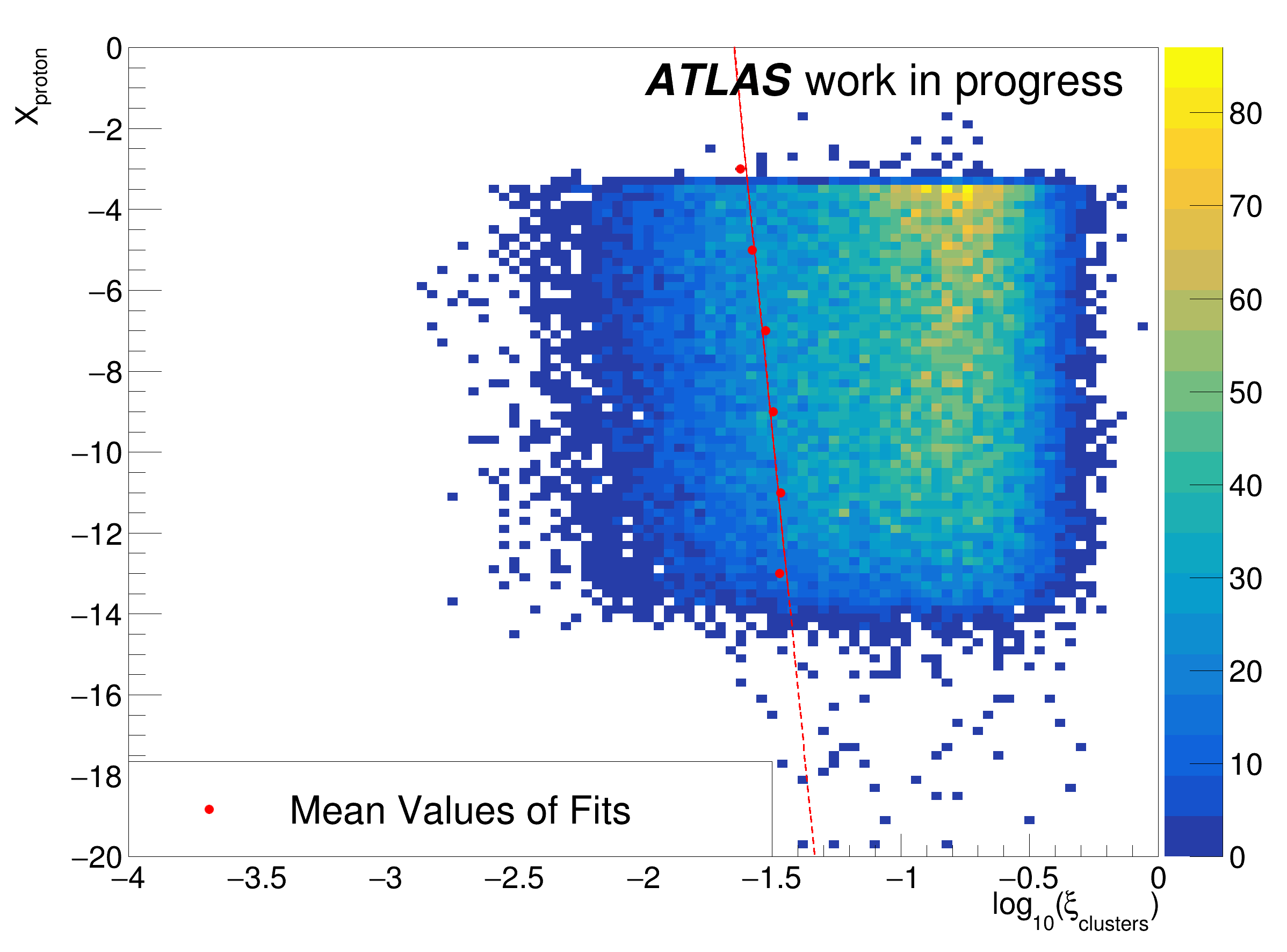}
\includegraphics[width=0.45\textwidth]{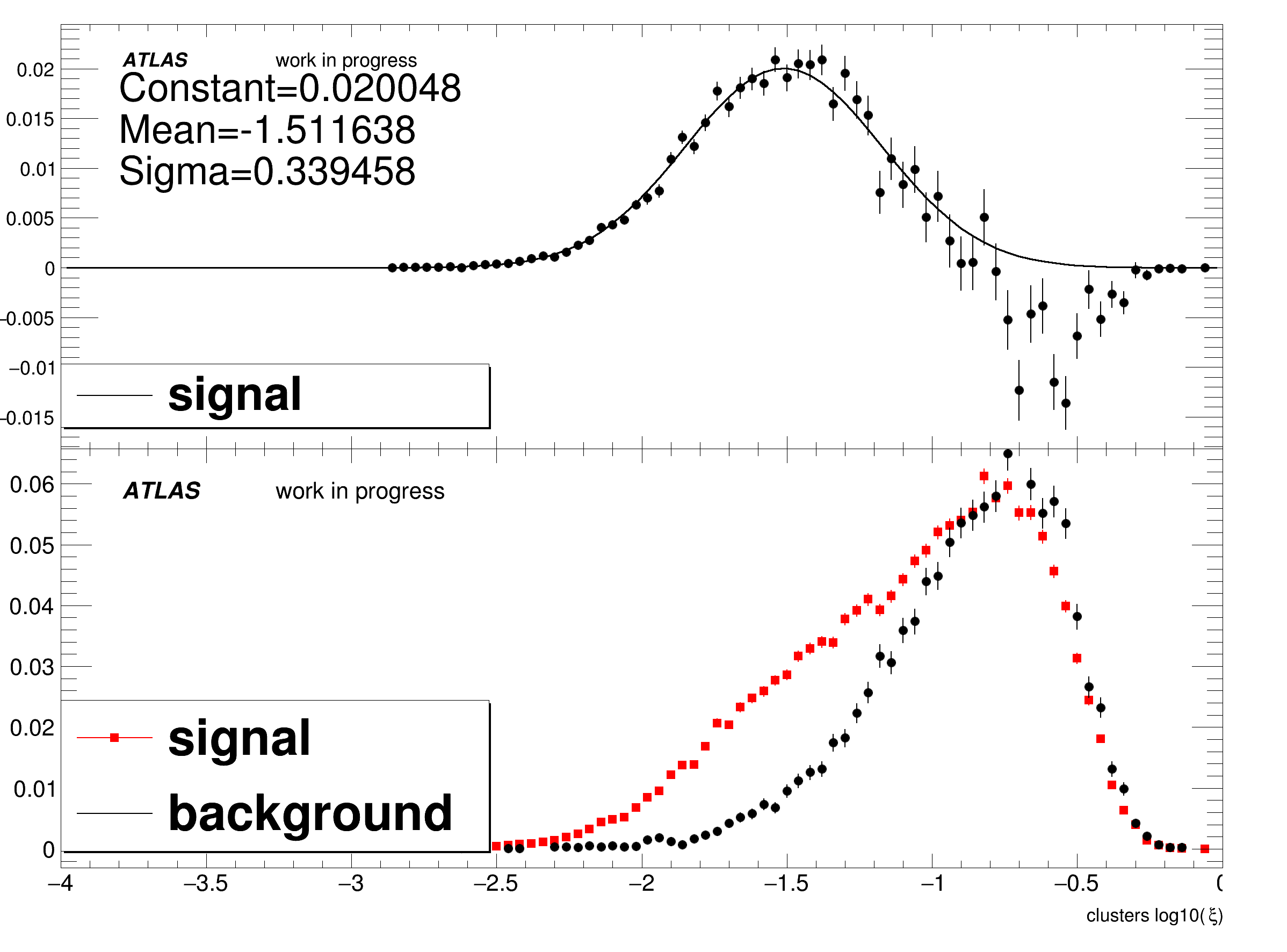}

\caption{\textbf{Left:} correlation between $\log_{10}(\xi_{cl})$ and proton $x$-position. \textbf{Right:} distributions of $\log_{10}(\xi_{cl})$ for all selected "signal" events (red dots, bottom plot) and "background" events (black dots, bottom plot). Their difference (top plot) is plotted with fitted Gaussian function.  }
\label{fig:bck_sub}

\end{figure}

The distributions of $\log_{10}(\xi_{cl})$ for the "signal" events (red dots, bottom plot) and "background" (black dots, bottom plot) samples are shown on the right bottom plot on Figure~\ref{fig:bck_sub}. The distributions were normalized to the integral of a 20 of last bins. It is visible that the both plots match very well for the larger $\xi_{cl}$ values and AFP events (red) have more events in the smaller $\xi_{cl}$ range. These additional events are expected to be the signal. 

In the next step, the two plots were subtracted and the Gaussian function was fitted to it (Figure~\ref{fig:bck_sub}, right top). In addition, these events were divided according to the range of protons $x$ position and for each considered range the Gaussian was fitted to subtracted plots. Then the mean values were drawn on the distribution~\ref{fig:bck_sub} (red dots, left plot) and the linear function was fitted to them. It is visible that the mean values of the fits are moving from smaller values of $\xi$ towards larger ones. This proves that there is a correlation between the central system and the forward proton, previously not visible due to background dominating the sample.

\section{Summary}

The analysis was done using 2017 low-$\mu$ data. The SD JJ events were selected using AFP+jet triggers with additional requirements on jets and protons and ensuring the good quality of studied objects. After such selection (described in Section~\ref{sec:sig_selection}), the background was still dominating the sample and no correlation between the central system and the forward proton was visible at first glance.

In the next step the ``background'' was identified by using random events without a requirement of forward proton but still requiring the presents of two jets. The ``signal'' events were then compared to the background distribution showing significant differences in the lower $\xi_{cl}$ regions. The Gaussian fit to the subtracted distributions of signal and background events with different proton $X_p$ position was calculated. This proves that there is a correlation between $\log_{10}(\xi_{cl})$ and proton $x$-position, however not visible due to large background.

\section*{Acknowledgements}
The work of Paula Erland has been partially supported by the `Diamentowy Grant' programme reg. no DI2016 013846 (0138/DIA/2017/46).

\ResCnt

\maketitle

\begin{abstract}
We review results of transition form factors for the $\gamma^* \gamma^* \to \eta_c$ couplings.
The electromagnetic form factors are calculated in the light-cone potential approach for several models of the $c\bar c$ interaction potential. The main ingredient of the formalism -- the light-cone wave function -- is obtained in two steps. Firstly, because the $\eta_c$ meson is regarded as a heavy quarkonium state composed of $c\bar c$, the radial part of the wave function can be found by solving Schr\"odinger equation. Then using the Terentev prescription, it is translated to light-cone ``radial'' wave functions. The space-like transition form factor is found for both virtual photons. The results for one virtual photon are compared to the BaBar experimental data. 
\end{abstract}

\section{Introduction}

Quarkonium can be regarded as a non-relativistic bound system with a heavy quark composition
with the same flavour that is $c\bar c$ or $b\bar b$.
The name is in the same spirit as positronium,
which consists of an electron and its antiparticle positron, thus classified as 'onium'.
The first discovery of the charmonium state ($J/\psi$) was announced in 1974 \cite{J_Observation, Psi_Observation}. The observation of a peak in the mass spectra in the $e^+ e^-$ and $  
p + Be \to e^+ + e^- + x$ reaction was evidence of the existence of the charm quark and initiated a new area of research. Charmonium states below the $D\bar D$ open-flavour threshold have narrow widths, which can be explained by the OZI-rule (Okubo-Zweig-Iizuka) that leads to the suppression of the strong decay channel ~\cite{OZI_rule,Donoghue:1992dd}.
Moreover, in the past, quarkonia states were used as tools to investigate the phenomenon of asymptotic freedom \cite{asymptotic_freedom}. 

In the literature, one can find many approaches to quarkonium production in high-energy collisions, such as the colour evaporation model or the colour-singlet model, and many others~\cite{Chapon:2020heu}. The colour singlet model is recognized as the first term of the Non-Relativistic Quantum Chromodynamics (NRQCD) and assumes that only the $Q\bar Q$ pair in the colour-singlet configuration could form a meson. Similarly, in our Light-Cone potential approach, we take into account only the first term in the light-cone Fock state expansion of the quarkonium state ${\cal Q}$
\begin{multline}
\big{|} {\cal Q} ; P_+, \vec{P}_{\perp}\big{>}= 
\sum_{i,j,\lambda, \bar \lambda}
\frac{\delta^i_j}{\sqrt{N_c}} \, 
\int {\frac{dz d^2\vec{k}_{\perp}}{ z(1-z) 16 \pi^3}} \Psi_{\lambda \bar \lambda}(z,\vec{k}_{\perp}) \\
\times \big{|} Q_{i \lambda}(z P_+ ,\vec{p}_{\perp Q})
	\bar Q^j_{\bar \lambda}((1-z)P_+,\vec{p}_{\perp \bar Q}) \big{>} + \dots\, ,
\end{multline}
here $P_{+}$, $P_{\perp}$ are momenta of the meson and transverse momenta of the quark/anti\-quark $\vec{p}_{\perp Q} = \vec{k}_{\perp} + z\vec{P}_{\perp}$, $\vec{p}_{\perp \bar Q} = -\vec{k}_{\perp} + (1 -z)\vec{P}_{\perp}$, and 
$z$ denotes the fraction of the longitudinal momentum carried by the quark, and $k_{\perp}$ is the relative momentum between the quark and the antiquark. We assume that the considered charmonium state is a pure $Q\bar Q$ state.
The radial part of the light-cone helicity wave function $\Psi_{\lambda \bar \lambda}$ can be found through the Schr\"odinger equation for $c \bar c$ interaction potential models. $J/\psi$ ($J^{PC} = 1^{--}$) as well as $\eta_c$ ($0^{-+}$) is described by the principal quantum number $n = 0$ and $l = 0$, thus they are $1S$ states, but due to the spin dependent forces $J/\psi$ is heavier.  
As for hydrogen atom, also for $c\bar c$ bound state, we can find the specific mass spectrum. 
We have analysed the results for five distinct potential models: harmonic-oscillator, logarithmic, power-like, Cornell, and Buchm\"uller-Tye \cite{Babiarz:2019sfa, Babiarz:2019mag}.
These potential models are fitted to the mass spectrum, and one of the parameter is the quark mass $m_c$. 

\section{Helicity amplitude for \texorpdfstring{$\gamma^* \gamma^* \to 0^{-+}$}{} and results}
The transition form factor provides information on how two photons can couple to the $c \bar c$ state. The transition form factor for two on-shell photons, $Q_1^2 = Q_2^2 = 0$ can be related to the decay width $\Gamma_{\gamma \gamma \to {\cal Q}}$. Usually, the dependence on the photon virtuality for the transition form factor is presented in the form $|F(Q^2,0)/F(0,0)|$. 
Due to charge parity conservation, only even charge parity meson can be considered via photon--photon fusion.
We start the derivation of the formula for our form factor
by writing the definition via helicity amplitude on the $\gamma^* \gamma^* \to 0^{-+}$ process \cite{Poppe:1986dq}:
\begin{equation}
M_{\mu \nu}(\gamma^* (q_1) \gamma^*(q_2) \to {\cal Q} )= -i 4 \pi \alpha_{em}\varepsilon_{\mu \nu \alpha \beta}q_1^{\alpha}q_2^{\beta} F(Q^2_1, Q^2_2)\,.
\label{eq:definition}
\end{equation}
Above $q_1$, $q_2$ are the four momenta of the incoming photons and $Q_1$, $Q_2$ stand for photon virtualities. In particular, we consider space-like photons, that is $Q_i^2 = -q_i^2 = \vec{q}\,^2_{i,\perp}$. 
We can unroll the process amplitude in terms of helicity wave function $\Psi_{\lambda \bar \lambda}$ and amplitude ${\cal A}^{\lambda \bar \lambda}$
\begin{eqnarray}
n^{+\mu} n^{-\nu} {\cal M}_{\mu\nu} = \frac{4\pi \alpha_{em} e^2_{Q}\, \mathrm{Tr} \mathbf{1}  _{\rm{color}}}{\sqrt{N_c}} \int \frac{dz d^2 \vec{k}_{\perp}}{z(1-z)16 \pi^3}
\sum_{\lambda \bar \lambda} \Psi^*_{\lambda \bar \lambda} n^{+\mu} n^{-\nu} {\cal A}_{\mu \nu}^{\lambda \bar \lambda} \, ,
\label{eq:matrix_element}
\end{eqnarray}
where the light-cone basis vectors are $n^{\pm}_{\mu} = 1/\sqrt{2}(1,0,0,\pm 1)$.
We can find the helicity amplitude $A^{\lambda \bar \lambda}_{\mu \nu}$ by evaluating quark/antiquark helicty spinors ${\bar u}_{\lambda}\, ,v_{\bar \lambda}$ \cite{helicity_spinors} with four momenta $\hat{p}_A,\, \hat{p}_B$ according to the Feynman rules, respectively for diagram A and B in Fig.~\ref{fig:helicity_ampl}, here $\hat{p}_{A} = p^{\mu}_A \gamma_{\mu}$
\begin{figure}[!htbp]
    \centering
\includegraphics[width = 0.9\textwidth]{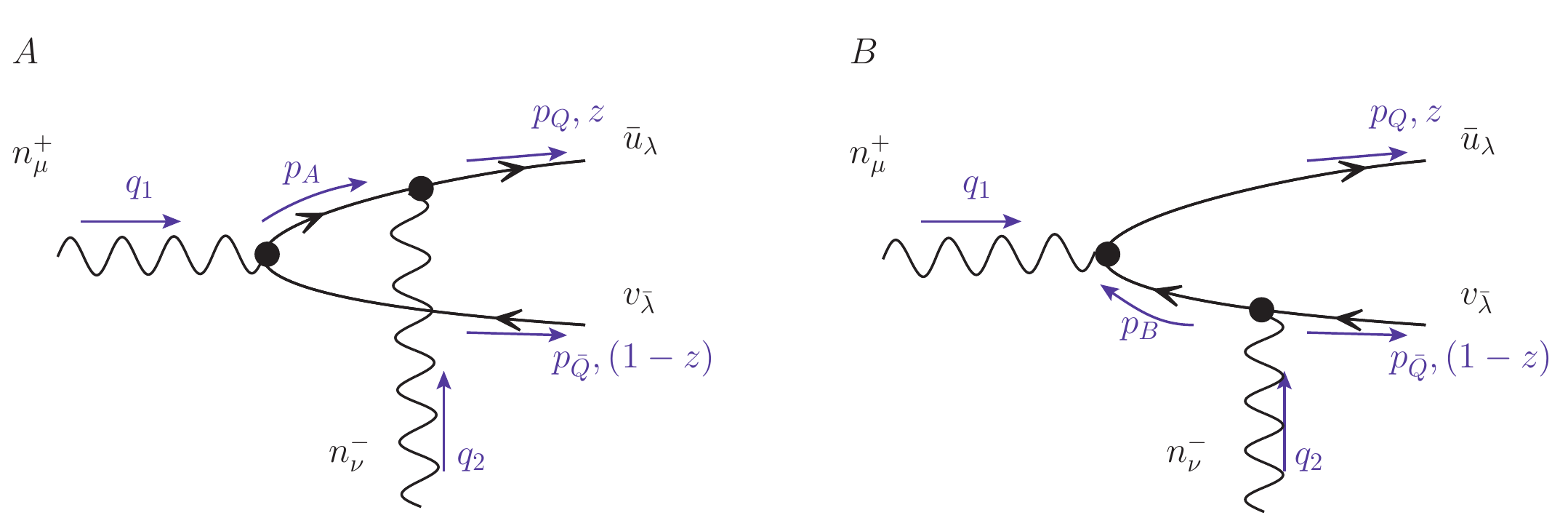}
    \caption{Generic diagrams of photon-photon reactions considered in calculations. }
    \label{fig:helicity_ampl}
\end{figure}
\begin{multline}
n^+_\mu n^-_\nu {\cal A}^{\lambda \bar \lambda} _{\mu \nu}\Big( \gamma^*(q_1) \gamma^*(q_2)\to Q_{\lambda}(z, \vec{p}_{\perp Q})\,  {\bar Q} _{\bar \lambda}(1-z, \vec{p}_{\perp \bar Q}) \Big) \\
= {\bar u}_{\lambda}(p_Q)\, \hat{n}^{+} \frac{\hat{p}_A+m_Q}{p^2_A -m_Q^2}\,  \hat{n}^{-}\, v_{\bar \lambda}(p_{\bar Q})\: + \:{\bar u}_{\lambda}(p_Q) \hat{n}^{-} \frac{\hat{p}_B+m_Q}{p^2_B - m_Q^2}\, \hat{n}^{+}\, v_{\bar \lambda}\,(p_{\bar Q}).
    \label{eq:helicity_amplitude}
\end{multline}
In Fig.~\ref{fig:helicity_ampl}~A and B, we present the u-channel and the t-channel exchange; the s-channel for photon-photon coupling is not allowed. 
In the case of the gluon in our colour-singlet approximation the s-channel is not involved, one could consider the colour-octet configuration. 
Now, we can insert Eq.~(\ref{eq:helicity_amplitude}) into Eq.~(\ref{eq:matrix_element}) and compare to the definition (Eq.~(\ref{eq:definition})), thus we obtain:
\begin{multline}
	F(Q_{1}^{2}, Q_{2}^{2}) = e_{c}^{2} \sqrt{N_{c}} 
	\cdot \int{ \frac{dz d^2 {\vec{k}_{\perp}}}{ z(1-z) 16 \pi^3}}  \psi(z,{\vec{k}_{\perp}})\\
	\times {\Big\{ } \frac{ 1-z}{({\vec{k}_{\perp}}- (1-z) \vec{q}_{2 \perp })^{2} + z (1-z) \vec{q}_{1 \perp}\,^{2} + m_{c}^{2}}\\
	+ \frac{z}{(\vec{k}_{\perp} + z \vec{q}_{2 \perp} )^2 + z (1-z) \vec{q}_{1\perp}\,^2 + m_c^2}
	{\Big \}}. \,
	\end{multline}
The main advantage of the formula above is that it incorporates not only the kinematics described in Eq.~\ref{eq:helicity_amplitude} but also some non-perturbative nature inscribed in the wave function of the charmonium state. The radial part of the wave function is included in the light-cone coordinates. In contrast, in the standard non-relativistic limit (NRQCD), only the radial part at origin $R(0)$ takes part in the form factor.
\begin{figure}[!htbp]
    \centering
    \includegraphics[width = 0.6 \textwidth]{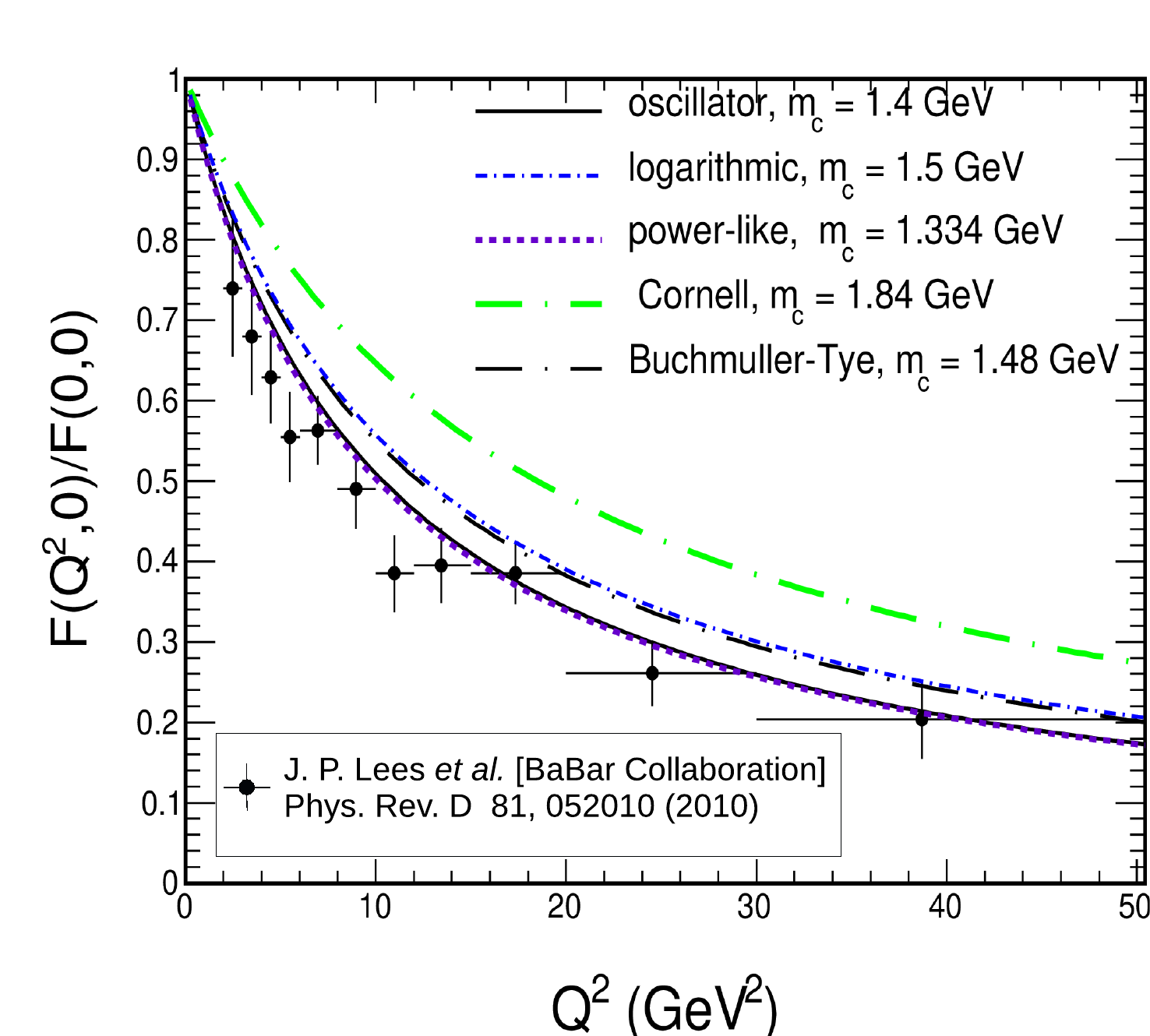}
    \caption{Normalized transition form factor for one real photon for five distinct potential models of $c\bar c$ interaction with the specified charm quark mass.}
    \label{fig:form_factor_one_real}
\end{figure}	
In Fig.~\ref{fig:form_factor_one_real} we present the normalised transition form factor at the on-shell point $F(Q^2,0)/F(0,0)$, with the wave function for several potential models. Different $m_c$ values are used in each of the models. For comparison, we plot the experimental data points from BaBar Collaboration \cite{BaBar:2010siw}.
A reasonable description of the data is obtained for the harmonic oscillator and power-like potential model. One would say that the physically justified/understood potential model is the Cornell model, which consists of the part acting on 'short distances' known as the Coulombic term and the linear part responsible for confinement. However, in our approach, a key role plays the quark mass. Thus, the best description gives the models with parameter $m_c$ around $1.3\,{\rm GeV}$.

\subsubsection*{Acknowledgement}
	This work was partially supported by the Polish National Science Center grant UMO-2018/31/B/ST2/03537.

\ResCnt

\maketitle

\begin{abstract}
This work discusses the methods of data analysis commonly used in research of $\Lambda_c^+$ decays in the LHCb Collaboration. The complexity of the following description was adjusted for undergraduate students with a particle physics background.
\end{abstract}

\section{Introduction}

The LHCb detector covers mostly research areas related to “b” quark physics at the LHC accelerator. It is a cone-shaped spectrometer, with particularly high vertex position resolution. During the first two data taking periods (Run 1 and Run 2), the LHCb collected 9$fb^{-1}$ of data \cite{lumi}, which contains information about the position, momentum, energy, and velocity of particles created in the collisions. Those data were processed, stored and analyzed. This work describes an example of such analysis done by the research team at IFJ. 

Standard Model (SM) is a main theory of high energy particle physics. It is an effective quantum field theory that describes particle interactions at the energy scale of GeV and TeV. Semileptonic $\Lambda_c$ decays can be used to test its predictions and look for new phenomena that are not described by it \cite{uurun1}. Those decays can occur as weak neutral current processes, as shown in Figure \ref{fig:diagrams}, which are strongly suppressed in SM. Measuring higher than expected rates of such decays can suggest contribution from outside the SM. It is also possible to look for a lepton family number violation, which directly breaks SM's rules. 
\begin{figure}[h!]
	\centering
	\includegraphics[width=0.45\textwidth]{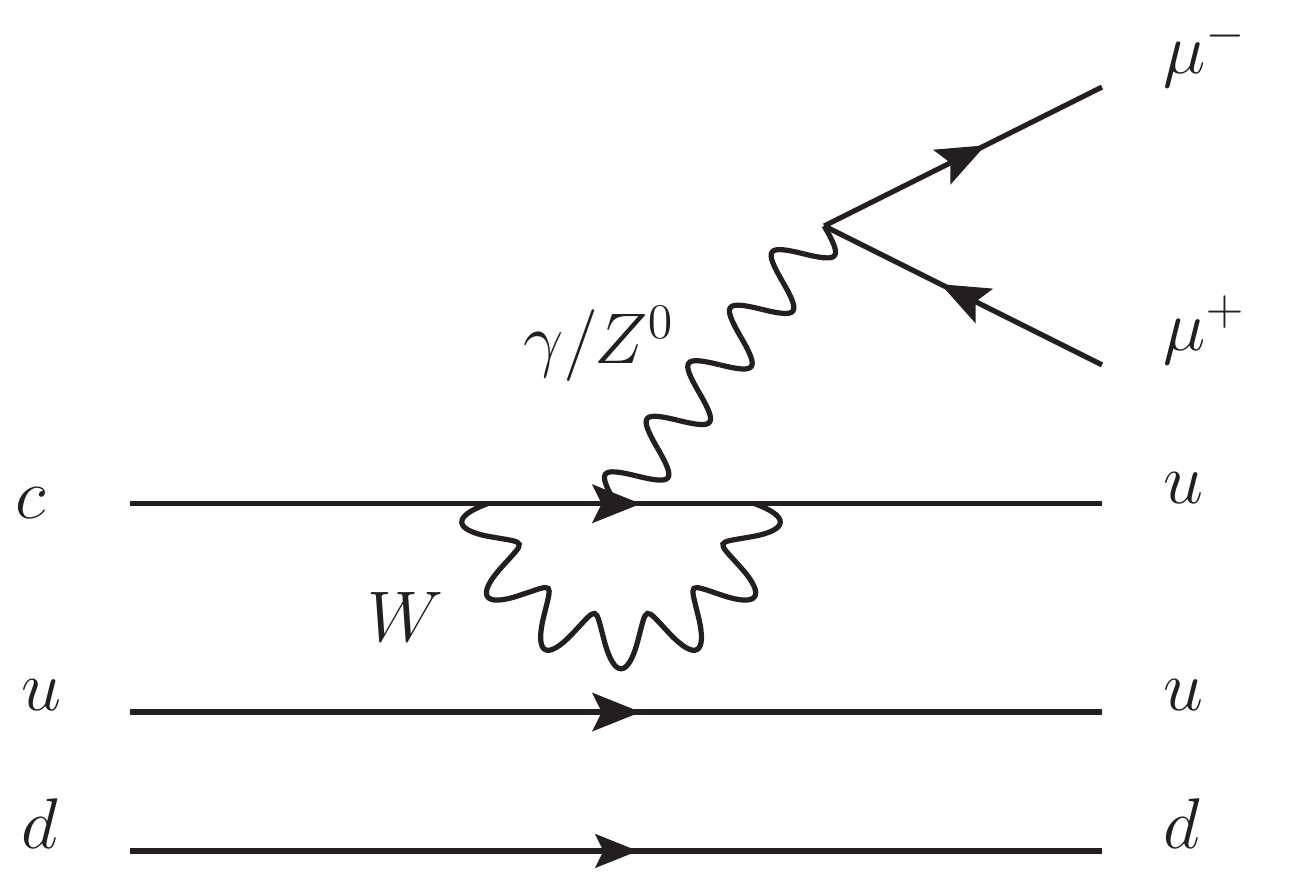}
\includegraphics[width=0.45\textwidth]{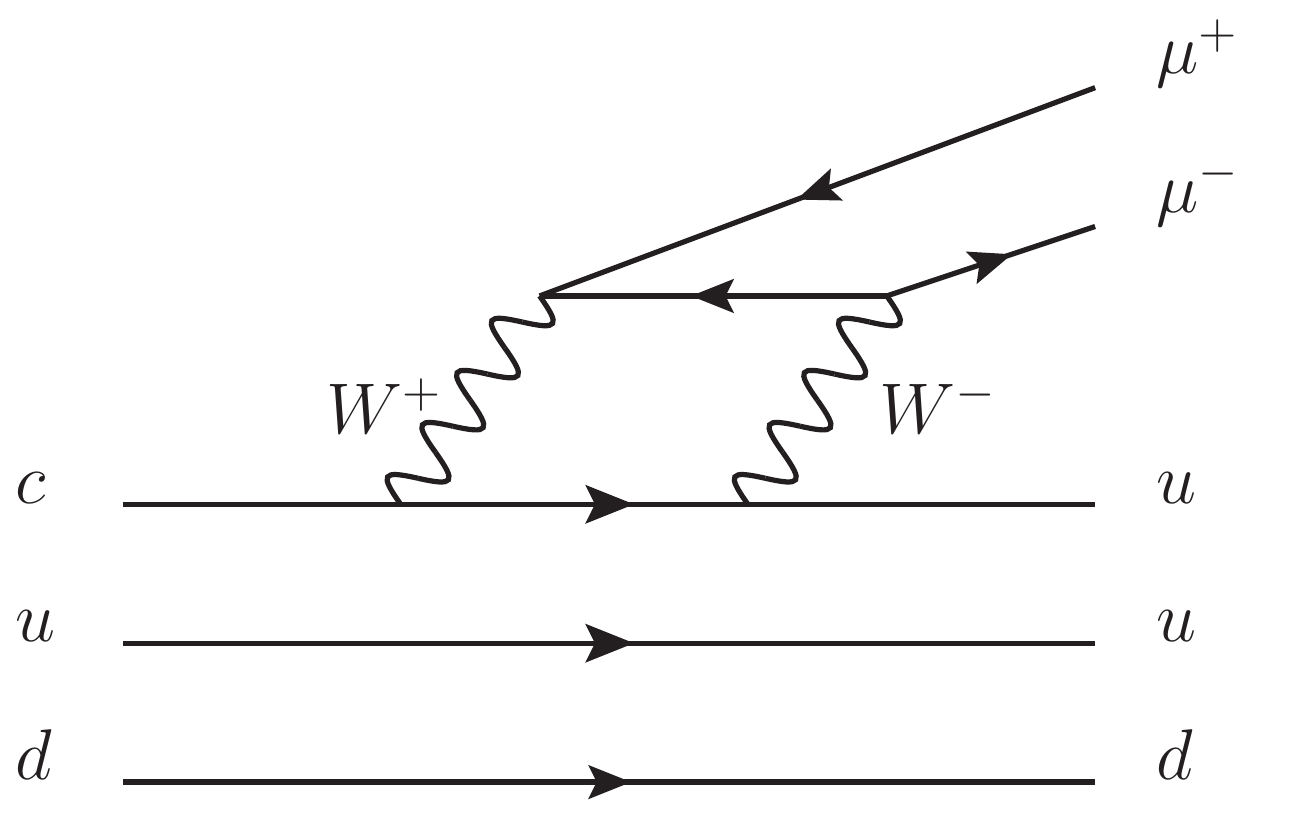}
    \caption{Example Feynman diagrams of $\Lambda_c$ decay.}
    \label{fig:diagrams}
\end{figure}
\section{Methods}
While colliding protons 40 million times per second \cite{rates}, only a small fraction of those collisions contain decay of interest, like in our case decay of $\Lambda_c$ to two leptons. To distinguish between the decays that we want to analyze (signal) and all the other decays (background), the data are filtered on many levels \cite{kit}. First, we want to check if a collision contains any interesting process. The simplest way to do it is to look at an amount of energy deposited in calorimeters and at the number of particles that were identified. It is also possible to put conditions based on the momentum of registered particles and use some multivariable analysis such as machine learning. 

After that, the reconstruction process takes place. Some particles decay before they can reach a detector, which means that only the decay products of those particles are registered. $\Lambda_c^+$ is such an example. It is decaying after less than $10^{-12}$s, which corresponds to around $0.1$mm of displacement inside LHCb. The decays that we studied are: $\Lambda_c \to p\mu\mu$, $\Lambda_c \to p\pi^0\mu\mu$, $\Lambda_c \to pe\mu$. All of them have to be reconstructed, to calculate the properties of the mother particle - $\Lambda_c$. 

Reconstructed data usually has a size suitable to analyze on a single machine, so at this stage, we can move from a distributed analysis to local processing. Before data can be thrown into an ML model, a distribution calibration is needed. ML classifiers are designed to distinguish between background and signal. To do that, we provided a simulated data that contains only the targeted signal. It takes a role of a reference for the supervised ML algorithms, which makes them sensitive to imperfections in the simulation. To combat that, we calibrate the simulation, by comparing it to the high statistic control data samples from more generic decays.

In our case, the analysis limitations mostly come from the experiment itself and simulation quality, rather than the properties of the used ML model. Nevertheless, we tested different algorithms and determined that the most suitable ML model for our case is boosted decision trees. Packages such as TMVA and XGBoost provided great results. Neural networks had similar values of area under ROC metric, but the training time was several times higher. To maximize the use of the available data, we train our models in folds and test the results with a cross validation approach.   

\begin{figure}[h!]
	\centering
	\includegraphics[width=0.65\textwidth]{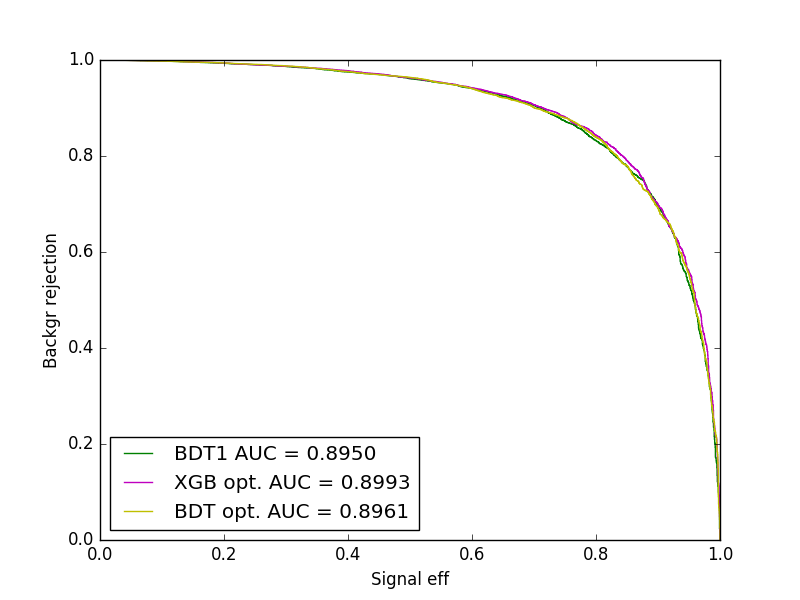}
    \caption{ROC curve for scikit-learn BDT a.}
    \label{fig:roc}
\end{figure}

This selection process can be tuned in many ways. We modified cuts, change processing and choose different variables for the ML model. On top of that, we use some automatic optimization for the final selection that minimized the expected upper limit on branching ratio. It could be possible to optimize our analysis in a way that would lead to a biased result. To avoid that, we used the “blind analysis” approach. This concept solves the bias issue, by removing the signal from the data during the analysis optimization phase. When all the preparations are finished, the analysis is presented to the LHCb collaboration. After a successful review process, the signal data can be processed without any modification to the selection procedure. This practice is important for the scientific quality of the  results.

\section{Conclusions}
The described methodology was presented during ``IFJ PAN PPSS Alumni Conference 2022''. The audience contained mostly undergraduate students. After the main part of the presentation, social aspects of proceeding a PhD in the particle physics was discussed.

\newpage
\mbox{\ }
\newpage

\ResCnt

\maketitle

\begin{abstract}
One of the possible applications of quantum computers in 
the near future are quantum simulations of physical  
systems. In this article simulations of loop quantum gravity,
which is a leading approach to quantum gravity, are briefly 
discussed. Here, the quantum geometry of space is represented 
by superposition of the so-called spin networks. A construction 
of quantum circuits that generate states of the Ising-type spin 
networks is described. The results of the implementation of the 
approach on the IBM superconducting quantum computers are presented.
\end{abstract}

\section{Introduction}

In recent years \cite{li2019quantum, cohen2021efficient, mielczarek2021prelude} 
the idea of performing quantum simulations of loop quantum gravity (LQG)
\cite{rovelli2008loop} has been developed. While at present, due to 
technical limitations, such simulations are possible to perform for simple 
systems only, the approach may provide a way to investigate Planck-scale 
degrees of freedom in the future. Because of the exponential growth of the 
dimension of the Hilbert space with the increase of the number of degrees 
of freedom, simulation of complex quantum gravitational systems is computationally a very demanding task.

The current fast progress in quantum computing technologies 
\cite{arute2019quantum} may open the possibility of simulating 
quantum gravitational systems unattainable to the most powerful 
classical supercomputers yet in this decade. Therefore, we should 
already prepare, test, and optimize quantum algorithms for future 
quantum simulations of quantum gravity. A side benefit of such 
an attempt is exploration of the quantum information structure 
spacetime.

Let us first introduce the very basics of quantum computations 
and spin networks in LQG. Quantum processors consist of qubits 
that are two-level quantum systems. The operations applied to 
qubits, called \textit{gates}, are $U(n)$ matrices, and the 
sequences of gates executed on quantum processors are represented 
diagrammatically by \textit{quantum circuits}, for example, Fig. 
\ref{fig:circ_node}, where each horizontal line represents a qubit.
At the end of the circuit the measurement is made, 
usually in $Z$-basis, i.e. in the basis of eigenvectors of 
the $\sigma_z$ Pauli $Z$ operator.

In LQG, geometry of space is represented by the so-called 
\textit{spin networks}. These are graphs with links labeled 
with fundamental representations of group $SU(2)$, i.e. spins, Fig. \ref{fig:spin_net}.
\begin{figure}
\begin{subfigure}{0.45\textwidth}
    \centering
    \includegraphics[scale=0.2]{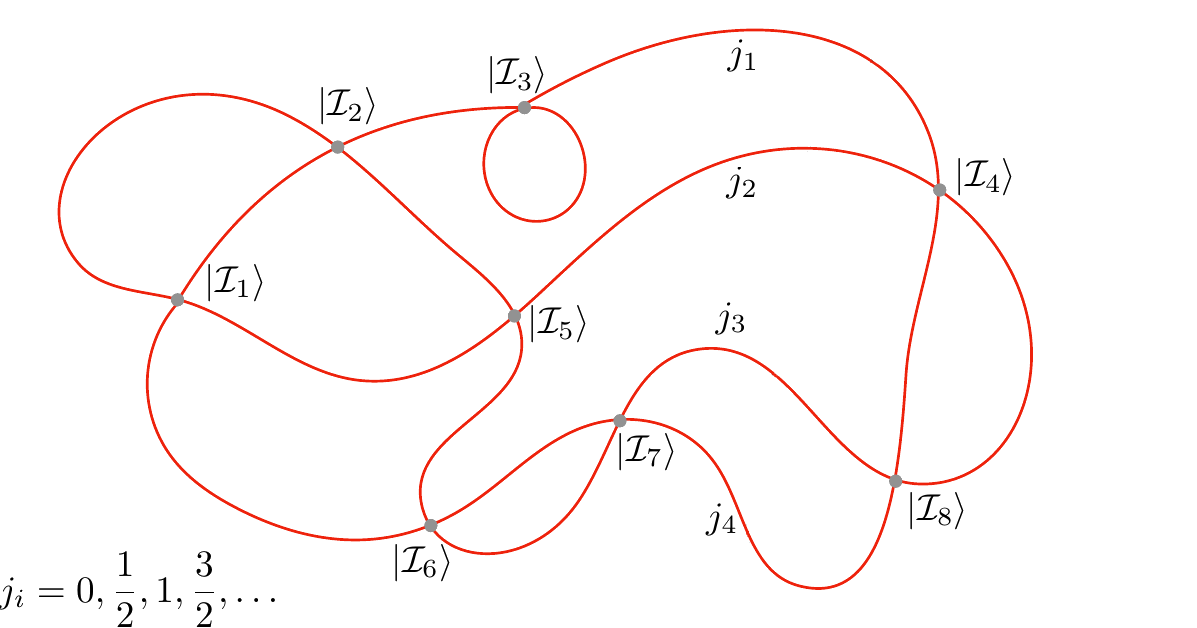}
    \caption{Example of four-valent spin network with eight nodes.}
    \label{fig:spin_net}        
\end{subfigure}
\hfill
\begin{subfigure}{0.45\textwidth}
            \centering
    \includegraphics[scale=0.25]{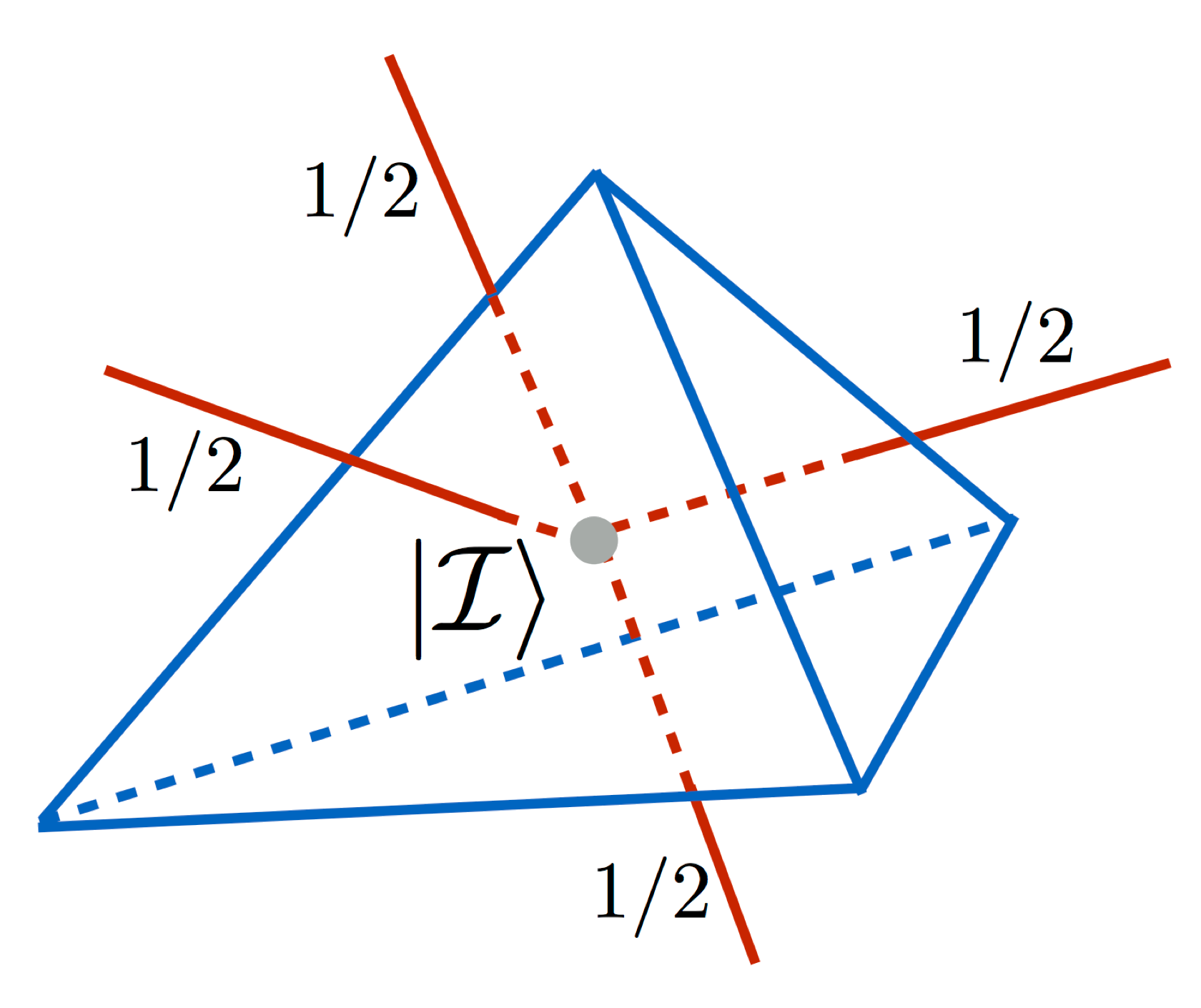}
    \caption{Single node dual to tetrahedron, here with spins equal to $\frac{1}{2}$.}
    \label{fig:node}
\end{subfigure}
\caption{Spin networks}
\end{figure}

With each node of the spin network, we associate the Hilbert 
space which is $SU(2)$-invariant subspace of the tensor product 
of the Hilbert spaces associated with spins that meet in a given node.
For the four-valent nodes:
\begin{equation}
\left|\mathcal{I}_n\right\rangle\in Inv_{SU\left(2\right)}\left(\mathcal{H}_{j_a}\otimes\mathcal{H}_{j_b}\otimes\mathcal{H}_{j_c}\otimes\mathcal{H}_{j_d}\right),
\label{eq:inv}
\end{equation}

These nodes are dual to polyhedrons and can be interpreted 
as the ``atoms'' of space. In the case of the four-valent 
spin network, the nodes are dual to tetrahedra, see Fig. \ref{fig:node}.
Condition (\ref{eq:inv}) is a consequence of the $SU(2)$ gauge 
invariance imposed by the Gauss constraint. The basis state of 
the whole spin network is given by the product over nodes:
\begin{equation}
    \left|\Gamma,j_l,\mathcal{I}_n\right\rangle = \bigotimes_n|\mathcal{I}_n\rangle.
    \label{eq:base}
\end{equation}

\section{Quantum computations}

In our attempt to simulations of spin networks we focus on the 
simplest case of four-valent networks with all spins equal 
$\frac{1}{2}$ - the Ising-type spin networks. In this case, Fig. \ref{fig:node}, 
Hilbert space associated with this node is two-dimensional.
In Ref. \cite{czelusta2021quantum} we proposed a quantum circuit, Fig. \ref{fig:circ_node}, generating state of a single node:
\begin{equation}
    |\mathcal{I} \rangle  = \frac{c_1}{\sqrt{2}} ( |0011\rangle +|1100\rangle )+ \frac{c_2}{\sqrt{2}} ( |0101\rangle +|1010\rangle )+ \frac{c_3}{\sqrt{2}} ( |0110\rangle +|1001\rangle )
    \label{eq:state_node}
\end{equation}
where $c_1$, $c_2$ and $c_3$ are complex parameters satisfying $c_2+c_2+c_3=0$.

\begin{figure}[ht!]
	\leavevmode
	\centering
	\Qcircuit @C=1em @R=0.7em {
			\lstick{\ket{0}} & \gate{H}  & \qw & \qw & \qw & \qw  & \ctrl{1} & \ctrl{2} & \ctrl{3} & \qw\\
			\lstick{\ket{0}} & \gate{U} & \ctrl{1} & \ctrlo{1} & \qw & \ctrlo{2} &  \targ & \qw & \qw & \qw\\
			\lstick{\ket{0}} & \qw & \gate{V} & \targ & \ctrlo{1} & \qw &  \qw & \targ & \qw & \qw\\
			\lstick{\ket{0}} & \qw & \qw & \qw & \targ & \targ &  \qw & \qw  & \targ & \qw\\
	}
	\caption{Circuit generating state (\ref{eq:state_node}) of single four valent node for spins $\frac{1}{2}$, $U$ and $V$ are matrices parameterized by $c_1$, $c_2$, $c_3$}
	\label{fig:circ_node}
\end{figure}
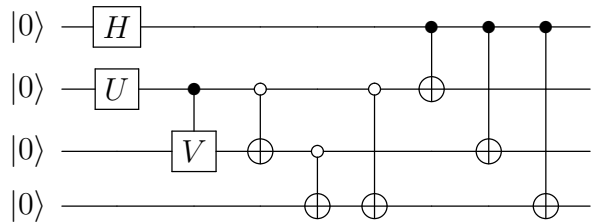

We executed this circuit on the IBM quantum processors \cite{webIBM} 
for six representative states of node. The fidelities computed between 
the obtained states and the exact theoretical ones are shown in Fig. \ref{fig:fidelities}.
The differences between processors are due to different architectures and error rates.
For an ideal processor, the fidelity of the obtained states would be 100\%.
\begin{figure}
    \centering
    \includegraphics[scale=0.35]{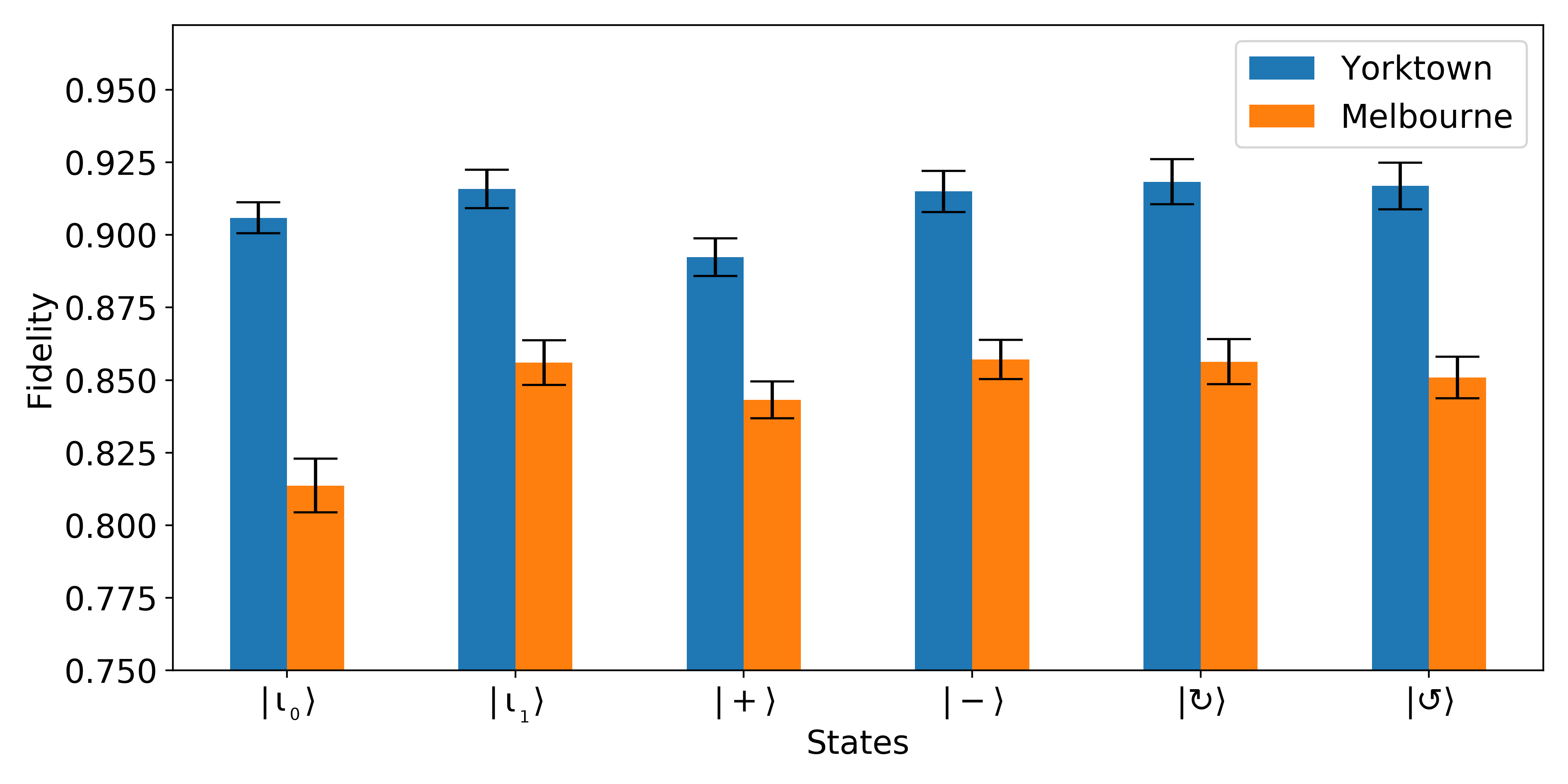}
    \caption{Fidelities of experimentally obtained states on two IBM processors.}
    \label{fig:fidelities}
\end{figure}

The classical phase-space structure that arises from spin networks 
is called a twisted geometry. A typical configuration from these 
geometries corresponds to a collection of uncorrelated tetrahedra. 
This is a consequence of the structure of the basis states of a spin 
network (\ref{eq:base}) that are unentangled. In our considerations, 
we will focus on geometries with a more rigid structure, i.e. vector 
geometries. In these geometries, the normals to the adjacent faces 
in neighbouring tetrahedra are back-to-back.

The vector geometry can be obtained \cite{baytacs2018gluing} using singlet 
states of spin $\frac{1}{2}$, that is, states $\left|\mathcal{B},\frac{1}{2}\right\rangle=\frac{1}{\sqrt{2}}\left(\left|01\right\rangle-\left|10\right\rangle\right)$, which are maximally entangled, on links
and then making a projection on a spin-network basis states:
\begin{equation}
	P_\Gamma=\sum_{j_l,\mathcal{I}_n}\left|\Gamma,\frac{1}{2},\mathcal{I}_n\right\rangle
  \left \langle \Gamma,\frac{1}{2},\mathcal{I}_n \right|  ,
	\qquad
	\left|\Gamma,\mathcal{B},\frac{1}{2}\right\rangle = P_\Gamma\bigotimes_l\left|\mathcal{B},\frac{1}{2}\right\rangle.
\end{equation}

We introduced a new circuit generating a single node, Fig. \ref{fig:Wacts},
\begin{figure}
\begin{subfigure}{0.55\textwidth}
    \centering
    \includegraphics[scale=0.35]{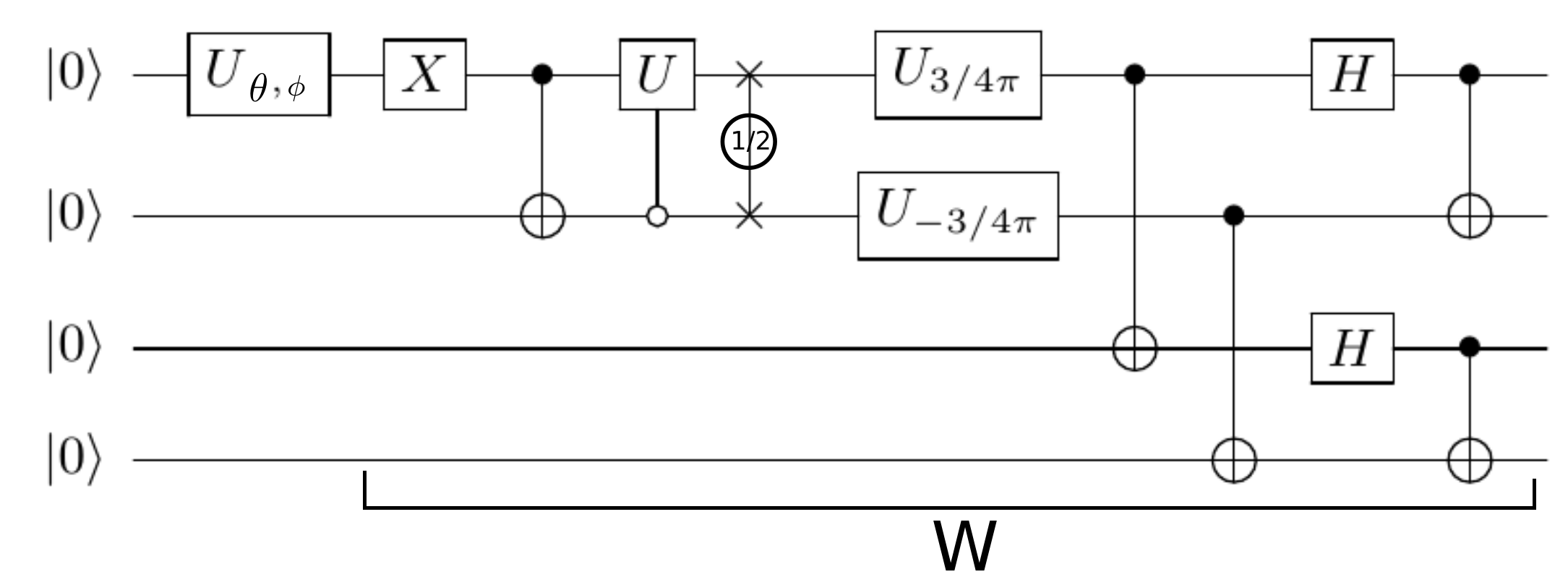}
    \caption{New circuit generating state of single node, 
    parametrization of state is only on first qubit, operator 
    $\hat{W}$ is fixed.}
    \label{fig:Wacts}
\end{subfigure}
\hfill
\begin{subfigure}{0.35\textwidth}
        \leavevmode
        \centering
        \Qcircuit @C=1em @R=0.65em {
            & \multigate{3}{W^\dagger} & \qw & \qw & \qw\\
            & \ghost{W^\dagger} & \qw & \ket{0} & \bra{0}\\
            & \ghost{W^\dagger}& \qw & \ket{0} & \bra{0}\\
            & \ghost{W^\dagger}& \qw & \ket{0} & \bra{0}\\
        }
        \caption{Projection operator on spin network basis states, expressed in one-qubit representation.}
        \label{fig:projection_circ}
\end{subfigure}
\caption{Projection method}
\label{fig:projection}
\end{figure}
which, in contrast to the previous one shown in Fig. \ref{fig:circ_node}, 
allows preparing an operator which projects state onto spin network basis 
states and expresses in a one-qubit representation, Fig. \ref{fig:projection_circ}.
One can show how it works on an example of a simple spin network with two 
nodes, i.e. dipole, Fig. \ref{fig:dipole}. Here, we need to use 8 qubits.
First, we prepare the links states, four pairs $|\mathcal{B},\frac{1}{2}\rangle$, and then we apply two operators \ref{fig:projection_circ}, as shown in 
Fig. \ref{fig:dipole_proj}. At the end, at two qubits, first and fifth, 
the physical state of the dipole is obtained.
\begin{figure}
\begin{subfigure}{0.4\textwidth}
    \centering
    \includegraphics[scale=0.2]{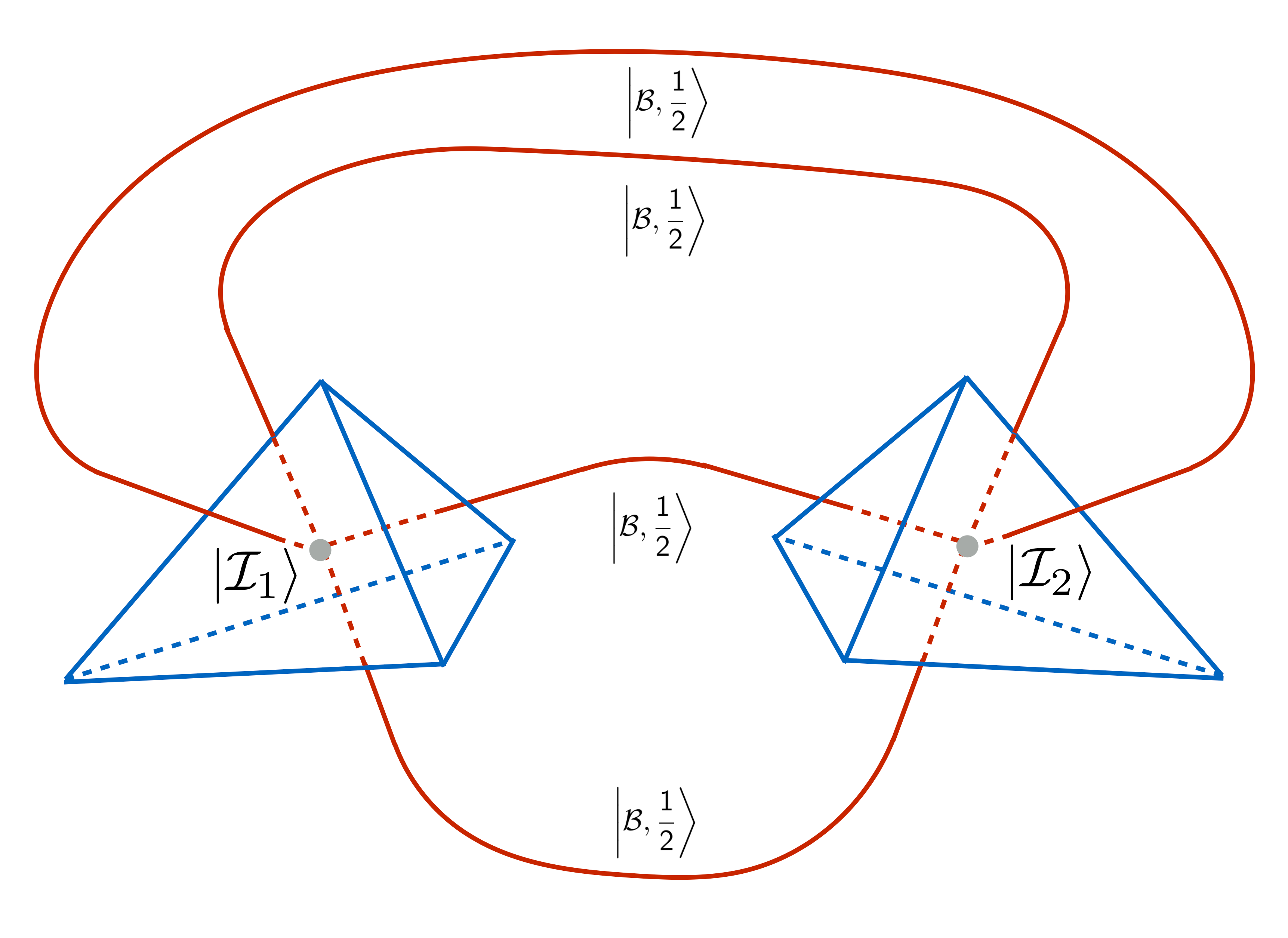}
    \caption{Dipole spin network.}
    \label{fig:dipole}
\end{subfigure}
\hfill
\begin{subfigure}{0.5\textwidth}
            \leavevmode
        \centering
        \Qcircuit @C=0.8em @R=0.2em {
            \lstick{\ket{0}} & \gate{X} & \gate{H} & \qw & \ctrl{4} & \qw  & \qw  & \qw & \multigate{3}{W^\dagger}&\qw&\qw\\
            \lstick{\ket{0}} & \gate{X} & \gate{H} & \qw  & \qw  & \ctrl{4} & \qw  & \qw & \ghost{W^\dagger}&\qw&\ket{0}\\
            \lstick{\ket{0}} & \gate{X} & \gate{H} & \qw  & \qw  & \qw  & \ctrl{4} & \qw & \ghost{W^\dagger}&\qw&\ket{0}\\
            \lstick{\ket{0}} & \gate{X} & \gate{H} & \qw  & \qw  & \qw  & \qw  & \ctrl{4} & \ghost{W^\dagger}&\qw&\ket{0}\\
            \lstick{\ket{0}} & \gate{X} & \qw & \qw  & \targ & \qw  & \qw  & \qw & \multigate{3}{W^\dagger}&\qw&\qw\\
            \lstick{\ket{0}} & \gate{X} & \qw & \qw  & \qw   & \targ & \qw & \qw & \ghost{W^\dagger} &\qw&\ket{0}\\
            \lstick{\ket{0}} & \gate{X} & \qw & \qw  & \qw  & \qw & \targ & \qw & \ghost{W^\dagger} &\qw&\ket{0}\\
            \lstick{\ket{0}} & \gate{X} & \qw & \qw  & \qw  & \qw  & \qw  & \targ & \ghost{W^\dagger} &\qw&\ket{0}\\
        }
        \caption{Projection of state $\bigotimes_{l\in\Gamma_2}\left|\mathcal{B},\frac{1}{2}\right\rangle$.}
        \label{fig:dipole_proj}
\end{subfigure}
\caption{Dipole spin network}
\end{figure}

Similarly, we can obtain the states of larger spin networks.
The states of open networks can also be constructed with this method.
Using pre-computed open networks, we can glue them together to 
easily obtain larger ones. For example, two open pentagrams 
can be glued into the decagram, as shown in Fig. \ref{fig:pentapenta}.
\begin{figure}
    \centering
    \includegraphics[scale=0.6]{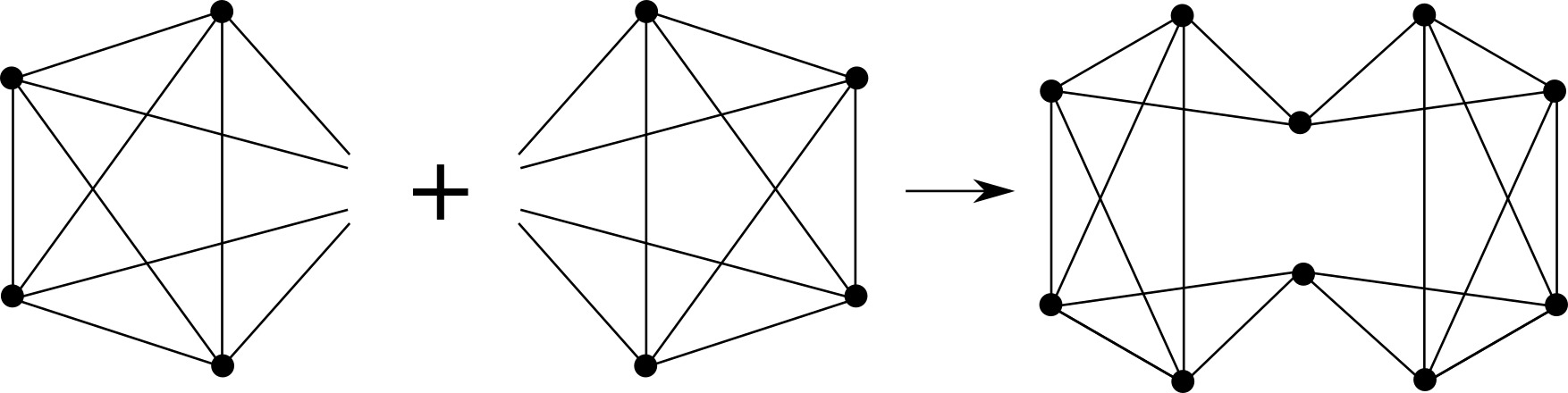}
    \caption{Scheme of obtaining decagram spin network from two open pentagrams.}
    \label{fig:pentapenta}
\end{figure}

\section{Summary}

In this work, we introduced quantum circuits that generate states 
of spin networks with fixed spins $\frac{1}{2}$ on links but with 
arbitrary topology. Prepared methods can allow us, for example, to 
compute spin foam amplitudes, which are hard to compute numerically 
using classical computers. The presented methods can be extended 
to cases with higher spins by combining a few physical qubits in 
one logical qudit, representing higher spin. Our method gives 
explicit recipes for circuits that can be executed on currently 
available quantum devices. Moreover, it gives us some new insight 
into the quantum information structure of quantum space-time.
\\

\textbf{Acknowledgements}: The research has been supported by the 
Sonata Bis Grant No. DEC-2017/26/E/ST2/00763 of the National Science
Centre Poland and by the Priority Research Area Digiworld under 
the program Excellence Initiative – Research University at the 
Jagiellonian University.

\ResCnt

\maketitle

\begin{abstract}
We examine the shear and bulk viscosities of the deconfined matter in pure Yang-Mills theory and QCD with light and strange quark
flavors using kinetic theory in the relaxation time approximation. The study is performed in the quasiparticle model which describes the QGP in terms of the dynamical quarks and
gluons dressed by the effective temperature-dependent masses. 
\end{abstract}

\section{Introduction}
A few decades of experimental and theoretical investigations have led to an agreement that a particular state of matter -- the quark-gluon plasma (QGP) -- is created at the heavy ion collisions~\cite{Andronic:2017pug}. It has been further confirmed experimentally that the QGP is a strongly-coupled fluid~\cite{Luzum:2008cw}, whose properties drastically change with temperature and time, as it expands and cools. Therefore it is of particular interest to study the dynamical and transport properties of the QGP in a self-consistent framework.

\section{Concept of quasiparticles}
The quasiparticle model (QPM)~\cite{Mykhaylova:2019wci,Mykhaylova:2020pfk,Bluhm:2006yh} assumes that as the particle of type $i$ propagates through the medium and interacts with the other constituents, it becomes dressed by the dynamically generated mass $m_i(T)$, which depends on the surroundings. Such particles are considered weakly-interacting quasiparticles with the interactions encoded in their effective masses. The approach then allows studying the QGP as the ideal system with massive constituents.

We assume the QGP to be in thermal and chemical equilibrium, therefore it is microscopically described by the standard Fermi-Dirac and Bose-Einstein statistics, $f_i^0= (\exp(E_i[G(T),T]/T)\pm 1)^{-1}$, for quarks and gluons, respectively. The subscript $i=g,l,s,(c)$ stands for gluons, light (degenerate up and down), strange, and eventually charm quarks in the system.  The energy of the on-shell quasiparticle reads $E_i[G(T),T]=\sqrt{p^2+m_i^2[G(T),T]}$, with a three-momentum $p$ and the effective medium-dependent mass defined as $m_i[G(T),T]=\sqrt{(m_i^0)^2+\Pi_i[G(T),T]}$.
$m_{i}^0$ is the bare mass of the quasiparticle, while $\Pi_i[G(T),T]$ is the dynamically generated self-energy that introduces the coupling and temperature dependence. We utilize the asymptotic forms of the gauge-independent hard thermal loop (HTL) self-energies~\cite{Bluhm:2006yh} at vanishing chemical potential ($\mu=0$). Further details, as well as numerical results for the quasiparticle masses, can be found in~\cite{Mykhaylova:2019wci}. The effective running coupling 
$G(T)$ is determined within the quasiparticle approach from the entropy density computed by lattice QCD (lQCD)~\cite{Mykhaylova:2019wci,Borsanyi:2016ksw}. In kinetic theory the total entropy density of the QGP reads~\cite{Mykhaylova:2019wci}
\begin{eqnarray}
s=\sum_{i=g,l,s,(c)} 2 d_i \int \frac{d^3p}{(2\pi)^3} \Big[  (1\pm f_i^0) \ln  (1\pm f_i^0) \mp f_i^0 \ln f_i^0  \Big], \label{s}
\end{eqnarray}
with the upper (lower) sign for bosons (fermions). The sum includes different terms, depending on the considered number of quark flavors. The degeneracy factor $d_i$ is related to spin and color of the quasiparticles, e.g. \mbox{$d_g=2(N_c^2-1)=16$} for gluons, while the factor of $2$ represents equal antiparticle contribution at $\mu=0$. Combining all the expressions given above, we equate Eq.~(\ref{s}) with the lQCD data for the entropy density illustrated in the left panel of Fig.~\ref{EntCoupl}. Since everything except the coupling is known, this allows us to deduce numerical values for the $G(T)$, shown in Fig.~\ref{EntCoupl} (right). 
\begin{figure}
	\centering
	\includegraphics[width=0.45\textwidth]{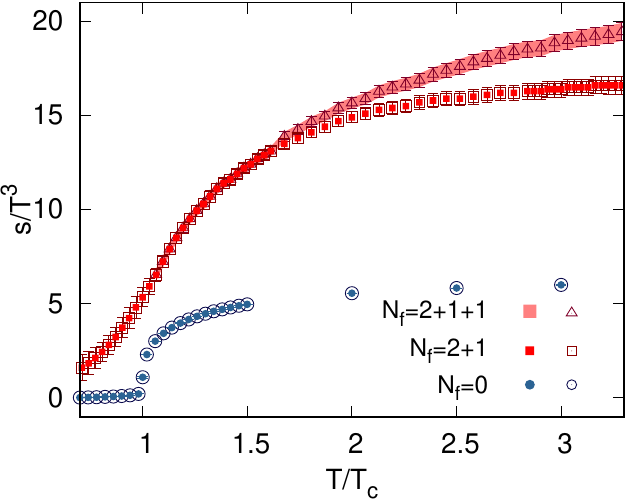}
	\includegraphics[width=0.45\textwidth]{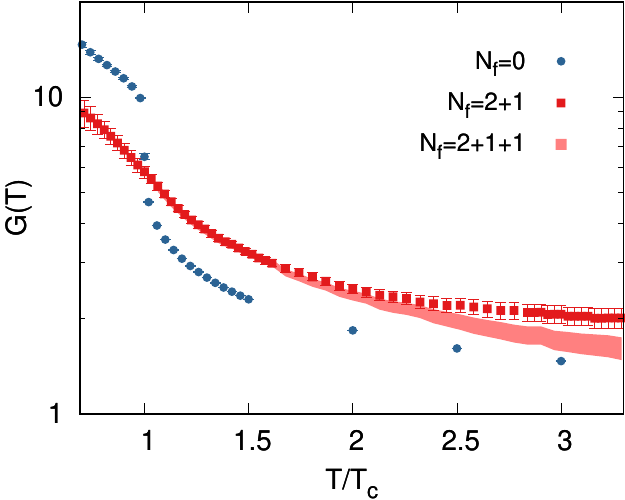}
	\caption{\label{EntCoupl} Scaled entropy density $s/T^3$ (left) as a function of the temperature, scaled by $T_c=260$ MeV for $N_f=0$ and $T_c=155$ MeV for $N_f=2+1(+1)$~\cite{Mykhaylova:2019wci}. Open symbols show the original lQCD data, while full symbols represent the results obtained in the quasiparticle model using the effective running coupling $G(T)$ (right) for different numbers of quark flavors.}
\end{figure}

Fig.~\ref{EntCoupl} (left) shows the matching of the QPM scaled entropy density to the lattice data when the effective coupling $G(T)$, see Fig.~\ref{EntCoupl}, is applied. The drastic change of the entropy density in pure Yang-Mills theory is related to the first-order phase transition and is also observed in the effective coupling. In QCD with $N_f=2+1(+1)$, eventually including thermalized charm quarks, the functions $s(T)$ and $G(T)$ are smooth due to the crossover.

\section{Transport parameters}
To demonstrate the application of the QPM, in this section we will discuss our numerical results for the specific shear and bulk viscosity, $\eta/s$ and $\zeta/s$, respectively. The shear viscosity coefficient reflects the reaction of the fluid to the longitudinal momentum modifications, as well as to the arising friction between the fluid's layers, while the bulk viscosity indicates the response to the expansion of the system's volume.
The transport parameters are usually presented as dimensionless ratios to ensure an adequate comparison of the results coming from different approaches for different systems. Moreover, each viscosity enters the hydrodynamic simulations exactly in a form of the specific ratio to the entropy density~\cite{Auvinen:2020mpc}.

Assuming that the QGP deviates from thermal equilibrium only slightly, we can determine the transport parameters via the Boltzmann kinetic equation. Our calculations are performed in the relaxation time approximation~\cite{Mykhaylova:2019wci}, which is valid in the QPM because the relaxation times of the quasiparticles are larger than the interparticle distance $d$, which scales with the number density of the quasiparticles as $d\sim n^{-1/3}$. 

The shear viscosity of the QGP for~$N_f=2+1$ is expressed as~\cite{Mykhaylova:2019wci}
\begin{equation}
\label{eta}
\eta =  \sum_{i=g,l,s} \frac{2d_i}{15 T} \int \frac{d^3p}{(2\pi)^3} \frac{p^4}{E_i^2}
\tau_i f_i^0 (1 \pm f_i^0),
\end{equation}
where the upper (lower) sign corresponds to fermions (bosons), while $\tau_i$ is the momentum-averaged relaxation time of each quasiparticle species~\cite{Mykhaylova:2019wci}. In pure Yang-Mills theory, the sum in Eq.~(\ref{eta}) reduces to a single term for gluons, with the corresponding coupling applied.

The bulk viscosity, including the speed of sound squared $c_s^2$~\cite{Mykhaylova:2020pfk}, reads
\begin{eqnarray}
\zeta=\sum_{i=g,l,s} \frac{2d_i }{T} \int \frac{d^3p}{(2 \pi)^3}f_i^0 (1\pm f_i^0) \frac{\tau_i}{E_i^2} \Big\{\Big(E_i^2-T^2 \frac{\partial \Pi_i(T)}{\partial T^2}\Big)c_s^2-\frac{p^2}{3}\Big\}^2. \label{zetaeq}
\end{eqnarray}

\begin{figure}
	\centering
	\includegraphics[width=0.46\textwidth]{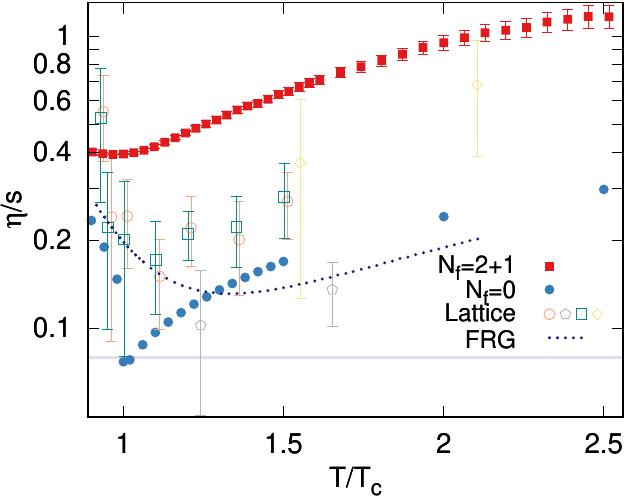}
	\includegraphics[width=0.53\textwidth]{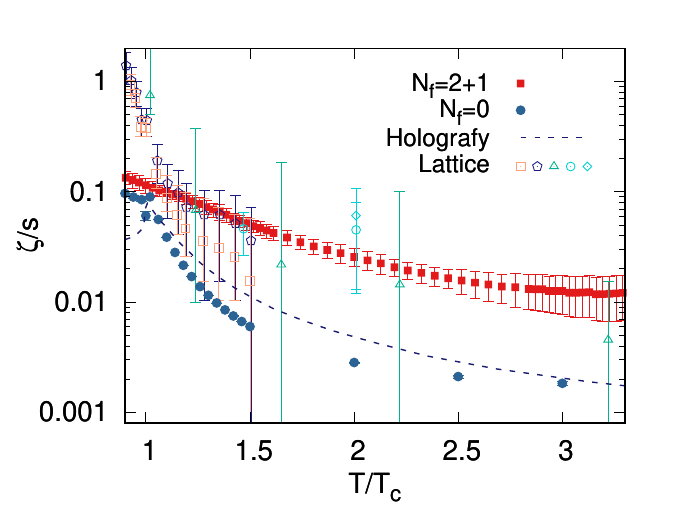}
	\caption{\label{Eta_zeta_fig} Specific shear viscosity $\eta/s$ (left) and the specific bulk viscosity $\zeta/s$ (right) as functions of $T/T_c$. The QPM results for the pure gluon plasma are shown by full circles, while the QGP with $N_f=2+1$ is represented by full squares. For comparison, the corresponding lQCD data are shown by open symbols, the horizontal line indicates the KSS-bound of $1/4\pi$, while the dotted one stands for the FRG result~\cite{Mykhaylova:2019wci}. On the right panel, alongside $\zeta/s$ computed in the QPM, the corresponding lattice data, as well as the holographic QCD outcome (dashed line) are shown~\cite{Mykhaylova:2020pfk}.}
\end{figure}

In Fig.~\ref{Eta_zeta_fig} (left), we show that the specific shear viscosity obtained by the QPM in hot QCD with $N_f=2+1$ is around 4 times larger than the result for the pure gluon plasma. This illustrates the impact of the quark quasiparticles on the transport properties of the QGP~\cite{Mykhaylova:2019wci}.
For the pure gluon plasma, $\eta/s$ exhibits a minimum around the first-order phase transition, reaching at $T_c$ the Kovtun-Son-Starinets (KSS) lower bound of $1/4\pi$, conjectured for all fluids in nature.  We additionally juxtapose our results with the lQCD and the functional renormalization group (FRG) outcomes~\cite{Mykhaylova:2019wci}.

The right panel of Fig.~\ref{Eta_zeta_fig} illustrates the specific bulk viscosity ratio, which in QCD with $N_f=2+1$ is larger than in pure Yang-Mills theory due to the presence of quarks in the system. The result for the gluon plasma is fairly consistent with the data sets from the lQCD, as well as with the anti-de-Sitter/conformal field theory correspondence (holographic QCD)~\cite{Mykhaylova:2020pfk}. The discrepancy between the QPM and holographic QCD in comparison to lattice data below $T_c$ is discussed in~\cite{Mykhaylova:2020pfk}. 

\section{Conclusions}
We have summarized the main features of the quasiparticle model (QPM), such as the dynamically generated masses and the effective coupling deduced from the lattice entropy density. Further, we have demonstrated the application of the QPM in the study of the specific shear and bulk viscosity of the hot deconfined matter with different numbers of quark flavors. Further possible extensions of the model may include the calculations at finite chemical potential $\mu$, the momentum anisotropy, the presence of the magnetic field, and many others.

\section{Acknowledgements}
This work was supported by the Polish National Science Center under the OPUS 16 Grant No.~2018/31/B/ST2/01663 and under the PRELUDIUM 20 Grant No.~UMO-2021/41/N/ST2/02615.

\ResCnt

\maketitle

\begin{abstract}
Monolithic CMOS sensors enable the development of detectors with a low material budget and a low fabrication cost. Moreover, using a small collection electrode results in a small sensor capacitance, a low analogue power consumption, and a large signal-to-noise ratio. These characteristics are very attractive in the development of new silicon sensors for charged particle tracking at future experiments. One of the goals of the Tangerine Project is to develop a test beam telescope setup consisting of detector prototypes designed in a novel \SI{65}{\nano\meter} CMOS imaging process. This contribution describes the first steps and results in the design of such a telescope using the Allpix Squared and Corryvreckan frameworks for simulation and analysis.
\end{abstract}

\section{Introduction}
Complementary Metal-Oxide Semiconductor (CMOS) monolithic sensors are very attractive for particle physics experiments.
In contrast to hybrid pixel detectors, monolithic pixel sensors contain the particle sensing and the signal processing elements integrated into one silicon block, reducing the material budget, the production cost, and the complexity. Moreover, CMOS pixel detector designs with a small collection electrode result in small sensor capacitance, low analogue power consumption, and a large signal-to-noise ratio.

The Tangerine Project (Towards Next Generation Silicon Detectors) exploits these capabilities to develop new silicon detectors that can be used at future lepton or electron-ion colliders. 
Currently, the project is investigating monolithic active pixel sensor (MAPS) in a novel \SI{65}{\nano\meter} CMOS imaging process, being the first application of this technology in particle physics and thus demonstrating its capabilities. 


The primary initial goal of the Tangerine project is to develop a test beam telescope that can be used at the DESY II Test Beam facility \cite{testbeamdesy}. 
This work summarizes the first simulations of such a setup, with a focus on optimizing the telescope geometry with respect to the tracking resolution.
The sensors developed to be used in the telescope planes have a target position resolution $\sim$ \SI{3}{\micro\meter}, a time resolution below \SI{10}{\nano\second}, and a material budget of less than \SI{0.05}{\percent} of a radiation length. 
Three different sensor layouts are available, with different doping profiles affecting charge collection, the so-called \textit{standard} \cite{standard}, \textit{n-blanket} \cite{snoeys} and \textit{n-gap} \cite{munker} layouts. 
Sensor simulations, as well as lab characterization and test beam measurements of the first prototype test chip, are ongoing \cite{hakan}.


\section{Simulation flow}
The small collection electrode design results in a strongly non-linear and complex electric field that needs to be simulated precisely. 
To do so, generic doping profiles are used to perform three-dimensional Technology Computer-Aided Design (TCAD) simulations of pixel cells. By numerically solving Poisson's equation for electrostatic potential, TCAD provides highly accurate electric fields inside the pixel cell.

The generic doping profiles and TCAD electric fields are imported into Allpix$^2$ \cite{allpix}. The Allpix$^2$ framework allows for complete and fast Monte Carlo simulations of a test beam telescope, including charge deposition, carrier propagation and simplified front-end electronics response.
Six parallel telescope planes are created each with a matrix of 1024$\, \times \,$1024 pixels.
In the work presented here, the sensors simulated have a pixel size of 20$\, \times \,$\SI{20}{\micro\meter^2}. 
To study the telescope resolution at the device under test (DUT) position, a silicon box with a thickness of \SI{50}{\micro\meter} is placed in the center of the setup, from which the Monte Carlo truth position can be compared with the reconstructed track intercept.
A beam of \SI{5}{\giga\eV} electrons with a Gaussian profile is simulated, crossing the setup in the z-direction, perpendicular to the sensor planes.
For each telescope geometry presented in Section \ref{sec:results}, \SI{250}{000} single-electron events are simulated.
The Corryvreckan framework is used to analyze the data \cite{corry}.

\section{Track reconstruction and residuals}

To reconstruct the trajectory of the incoming relativistic particle beam through the telescope planes, the General Broken Lines track model (GBL) is used \cite{gbl}. 
For each event, the following information is available: the Monte Carlo truth data, the reconstructed center-of-gravity position of the clusters in the telescope planes, and the reconstructed track intercepts at all the planes.  

By computing the difference between the track intercept and the associated cluster center for a set of tracks, the biased residual distributions in x and y with respect to the telescope reference plane are obtained. 
Figure \ref{fig:residuals} shows the residual distributions for the first and third telescope planes in the x (similar in the y) direction. The standard deviation of those distributions defines the biased residual width. 

\begin{figure}[htbp]
\centering
\begin{subfigure}{.5\textwidth}
  \centering
  \includegraphics[width=\linewidth]{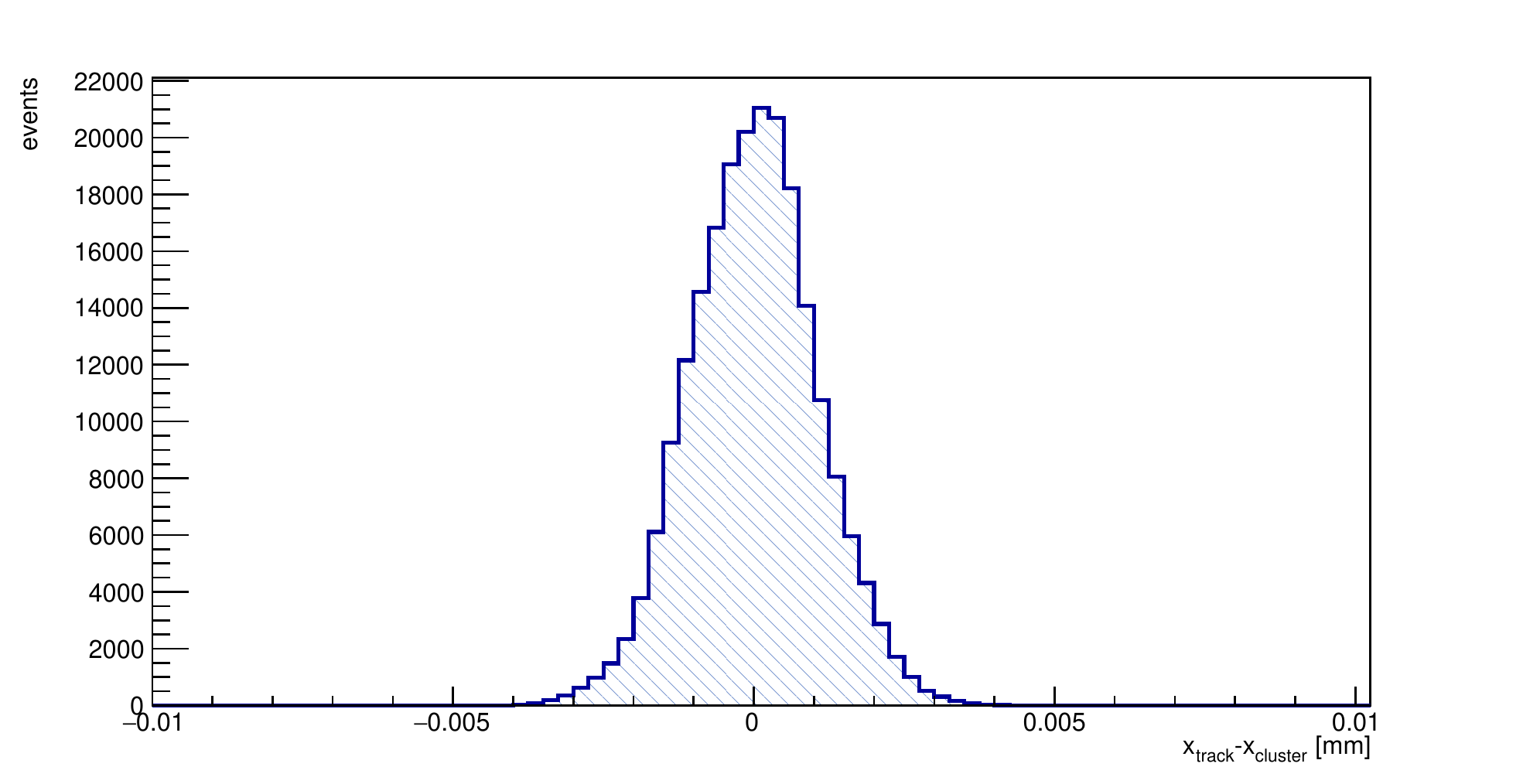}
  \put(-93,75){\tiny $\sigma = 1.271 \pm 0.002$ \SI{}{\micro\meter}}
  \caption{}
  \label{fig:telescope0}
\end{subfigure}%
\begin{subfigure}{.5\textwidth}
  \centering
  \includegraphics[width=\linewidth]{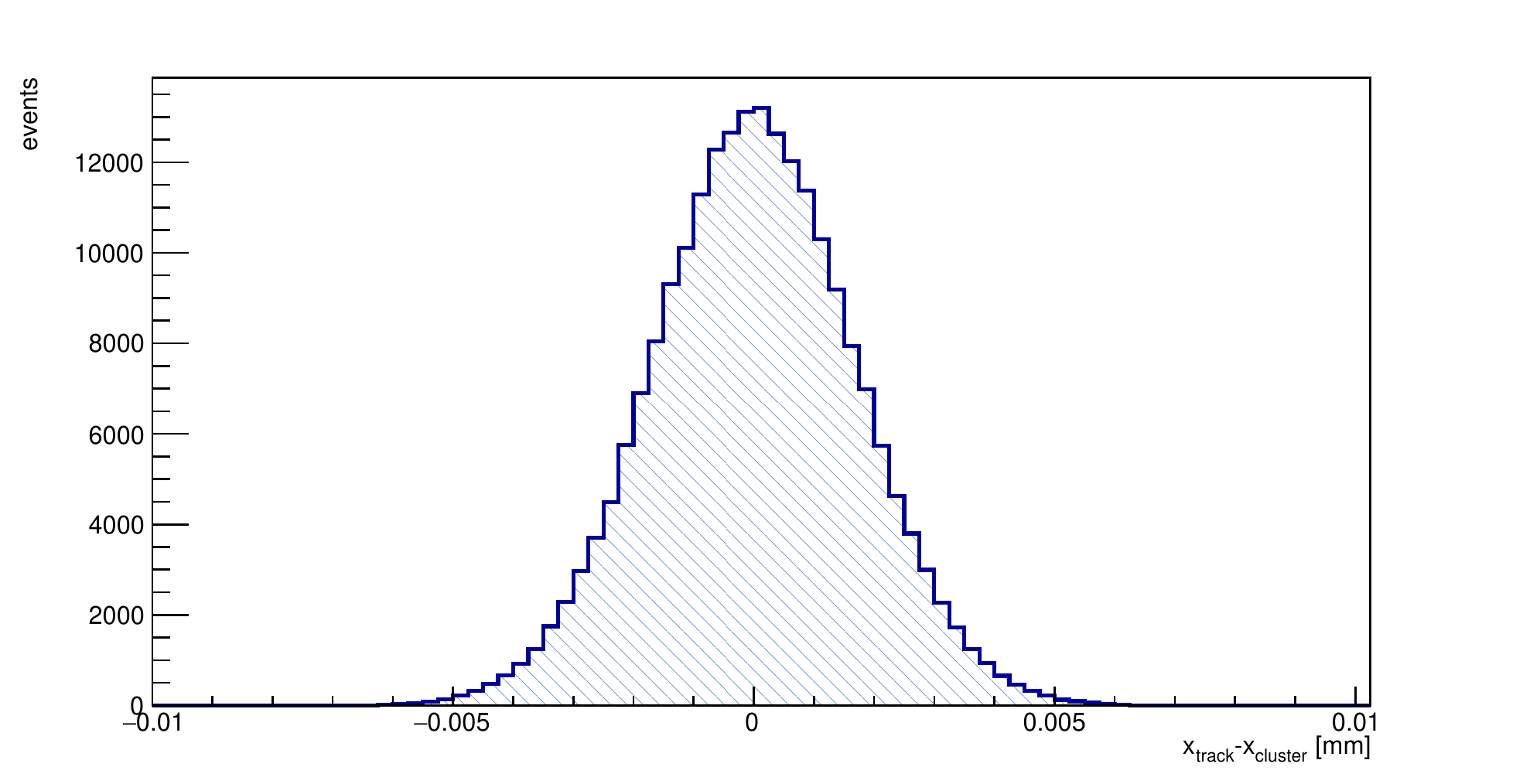}
  \put(-93,75){\tiny $\sigma = 2.025 \pm 0.003$ \SI{}{\micro\meter}}
  \caption{}
  \label{fig:telescope2}
\end{subfigure}
\caption{\small Residual distribution in x between the track intercept and the center-of-gravity position of the cluster for the first (a) and third (b) telescope planes.}
\label{fig:residuals}
\end{figure}

The telescope resolution at the DUT position is obtained from the residual distribution width between the Monte Carlo truth position and the track intercept with the DUT (excluded in track reconstruction).
In this way, the telescope resolution at the DUT position is independent of the intrinsic resolution of the DUT.

\section{Telescope resolution}\label{sec:results}

Figure \ref{fig:biased} shows the telescope resolution at the different planes separated by a distance $dz$ of \SI{150}{\milli\meter}. 
The red curve represents the biased residual width between the cluster center and the Monte Carlo truth positions, i.e, the intrinsic resolution of the sensors. 
In blue, the biased residual width between the cluster center and the track intercept on the planes is shown. It can be seen that the telescope resolution slightly deteriorates from the DUT towards the second and fifth telescope planes (the residuals get larger). 
The small biased residuals in the first and last reference planes are a consequence of the GBL track reconstruction algorithm: for the outermost planes, the kink angle is not available, so the algorithm only minimizes the distance between the track and cluster positions, and the residuals get smaller.
For the same reason, the biased residual widths between the track intercept and the Monte Carlo position (in green) converge in the outermost planes to the intrinsic resolution of the sensors. 
Since the cluster center is used for tracking, and not the Monte Carlo truth position, the distance between the track intersection and the cluster center is in general smaller than the separation between the Monte Carlo position and the track intercept, resulting in smaller residuals.
\vspace{-0.6cm}
\begin{figure}[htbp]
\begin{minipage}[t]{0.5\linewidth}
    \centering
    \includegraphics[width=1\textwidth]{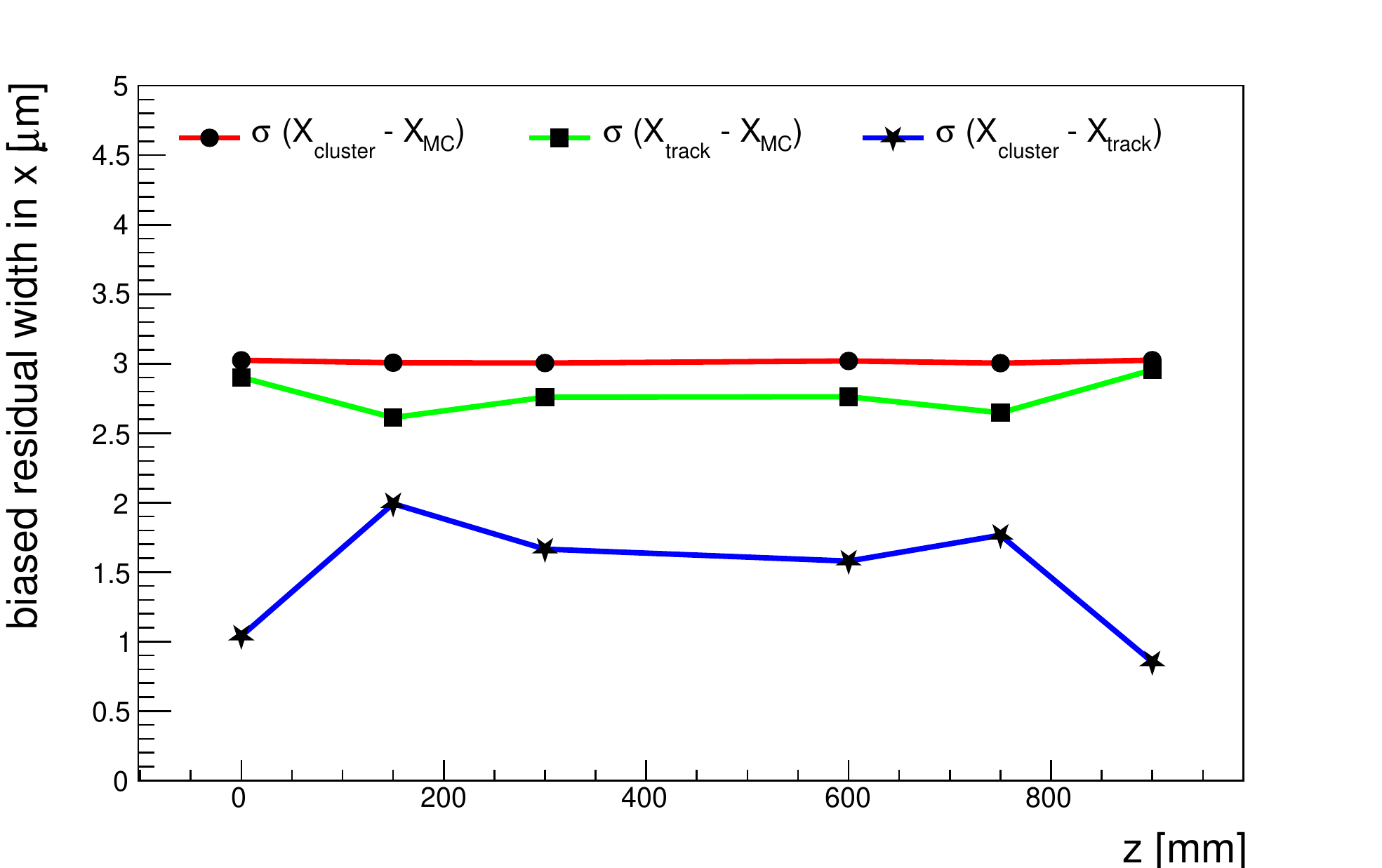}
    \caption{\small Biased residual widths between the cluster center and the Monte Carlo position, track intercept and Monte Carlo position, and cluster center and track intercept for each of the telescope planes.}
    \label{fig:biased}
\end{minipage}
\hspace{0.1cm}
\begin{minipage}[t]{0.5\linewidth} 
    \centering
    \includegraphics[width=1\textwidth, height=4.8cm]{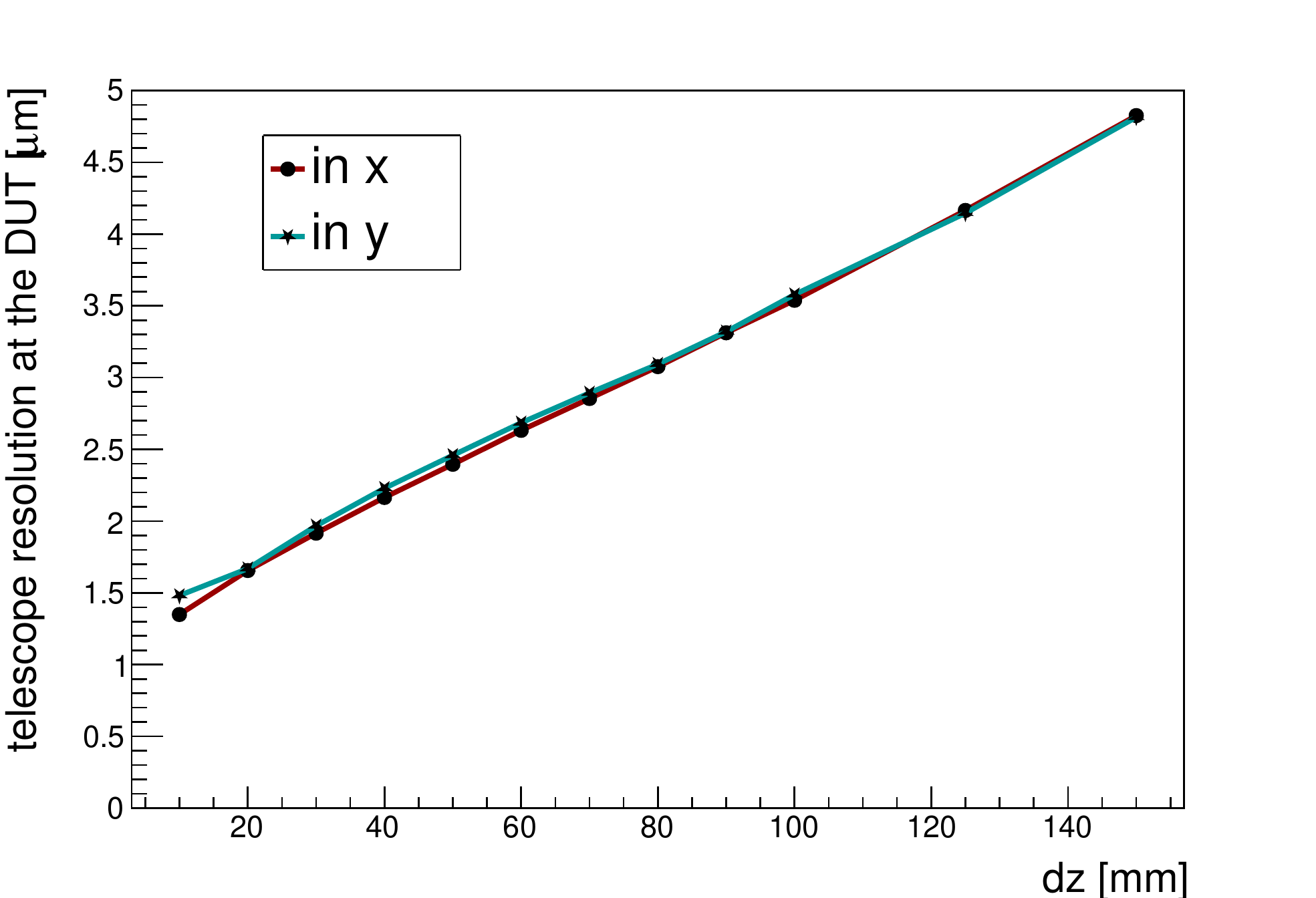}
    \caption{\small Telescope resolution at the DUT position for different distances between telescope planes.}
    \label{fig:distance}
\end{minipage}        
\end{figure} 
\vspace{-0.4cm}

The telescope resolution in x and y at the DUT position is shown in Figure \ref{fig:distance} for different distances between telescope planes $dz$. The telescope resolution improves linearly when this separation is reduced. The fact that the projections in x and y are not perfectly overlapping is an effect of the random misalignment of the planes introduced in the simulations, as in a real scenario.

Figure \ref{fig:layouts} shows the biased residual widths between the cluster center and the track intercept for the three different sensor layouts, and for all the telescope planes at $dz = $ \SI{125}{\milli\meter}. 
As expected from an improvement of the charge-weighted cluster center reconstruction for cluster sizes larger than one, the standard layout has the best tracking resolution. 
However, this sensor layout is expected to have the lowest efficiency compared to the other two layouts. 
Less charge sharing is achieved in the n-blanket and n-gap layouts, resulting in worse position resolution, but they are expected to have a higher detection efficiency  \cite{hakan}. 

\begin{figure}[htbp]
\begin{minipage}[t]{0.5\linewidth}
    \centering
    \includegraphics[width=1\textwidth]{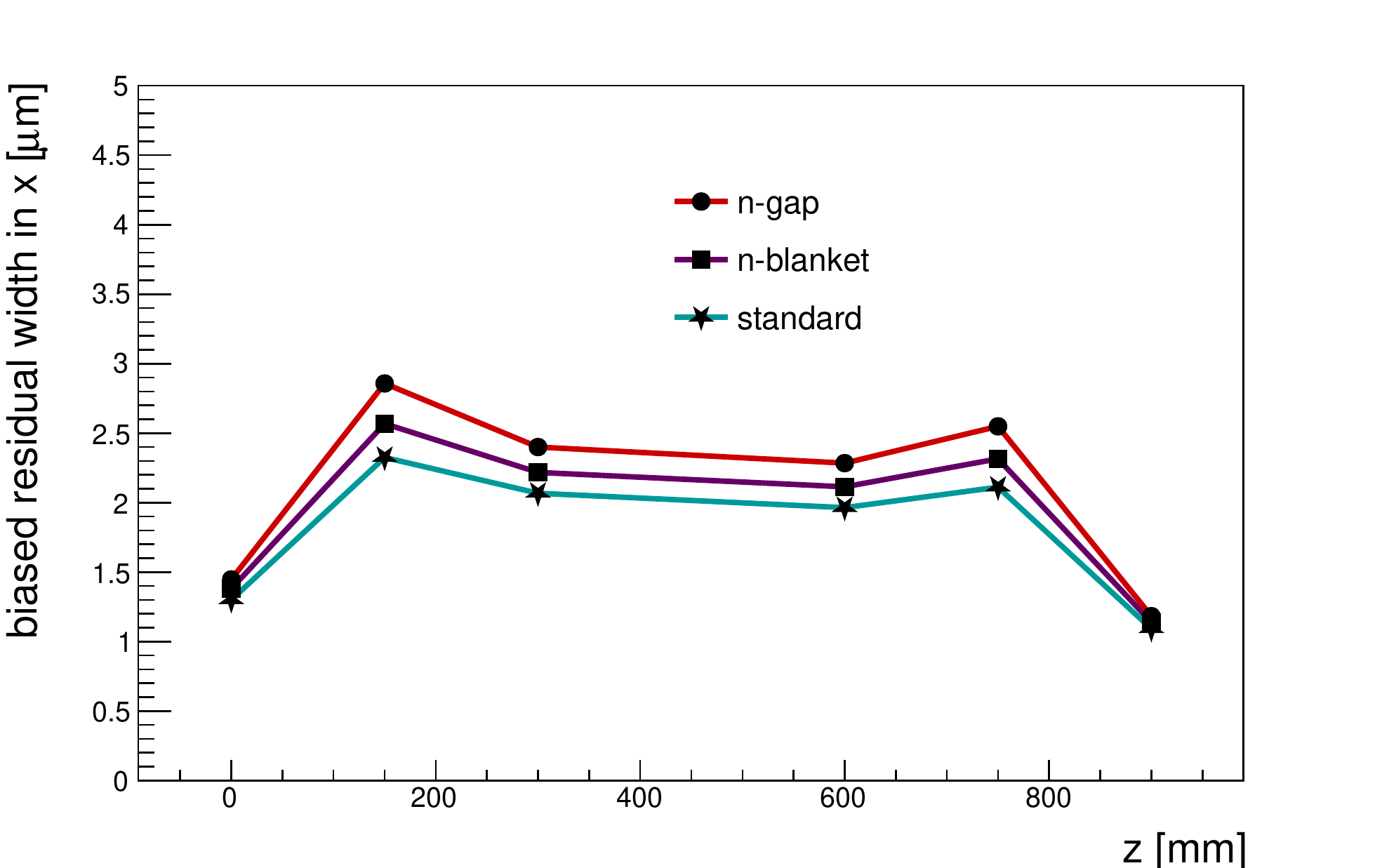}
    \caption{\small Biased residual widths for the three different pixel layouts.}
    \label{fig:layouts}
\end{minipage}
\hspace{0.1cm}
\begin{minipage}[t]{0.5\linewidth} 
    \centering
    \includegraphics[width=1\textwidth]{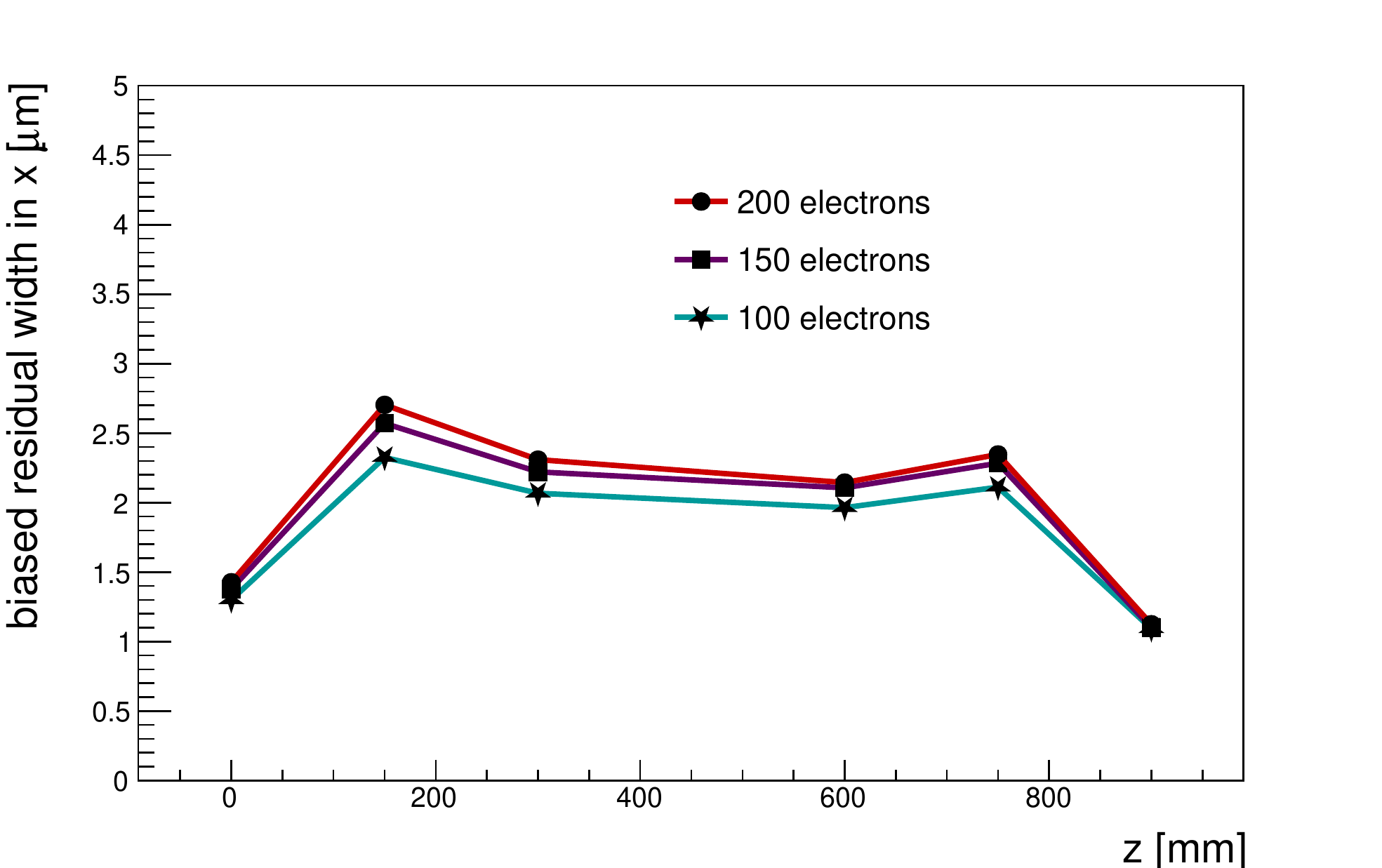}
    \caption{\small Biased residual widths for three different detection thresholds with the standard layout.}
    \label{fig:threshold}
\end{minipage}        
\end{figure} 

The same effect is observed in Figure \ref{fig:threshold}, where three different detection thresholds are compared for the standard layout. By increasing the detection threshold, the cluster size is reduced, and the charge-weighted cluster reconstruction is applied to a small fraction of events. Therefore, at higher thresholds, the telescope resolution deteriorates.

\section{Conclusions and outlook}
Starting from three-dimensional TCAD simulations of the pixel cells, a full test beam telescope based on a novel \SI{65}{\nano\meter} CMOS imaging technology has been successfully simulated with Allpix$^2$. Different sensor layouts and telescope geometries have been tested.
Continuous development is ongoing to include the full digitization stage in the simulations. 
Tracking performance at the DUT position for different geometries, sensor layouts, detection thresholds, material budget of the DUT and beam energy are some of the planned studies.

\ResCnt

\maketitle

\begin{abstract}
Electrons are commonly produced in collisions at the Large Hadron Collider~(LHC) both directly at the interaction point and as a result of particle decay. They are a key component in high-energy physics due to their presence in various signal and background processes. Electron reconstruction and identification prove challenging in heavy-ion collisions due to high detector occupancy. Therefore, electron performance evaluation is crucial for precision measurements of the Standard Model and searches for new phenomena beyond it. The studies aim at electron reconstruction, identification, isolation, and trigger performance in p+Pb collisions at $\sqrt{s_{\mathrm{NN}}}=8.16$~TeV collected by the ATLAS experiment in 2016. A total integrated luminosity in data amounts to 165~nb$^{-1}$. The Tag-and-Probe method is used, allowing for the measurement of electron efficiency independently in data and Monte Carlo~(MC) simulation.
\end{abstract}

\section{Tag-and-Probe method}
\label{sec:TaP}

Tag-and-Probe~\cite{bib:TaP} is a well-established method used for electron efficiency measurements in the ATLAS experiment~\cite{bib:atlas}. The total electron efficiency $\varepsilon_\mathrm{total}$ is a product of four efficiencies, related to electron reconstruction $\varepsilon_\mathrm{reco}$, identification  $\varepsilon_\mathrm{id}$, isolation  $\varepsilon_\mathrm{iso}$ and trigger $\varepsilon_\mathrm{trig}$:
\begin{equation} \label{eq:total}
\varepsilon_\mathrm{total} = \varepsilon_\mathrm{reco} \cdot \varepsilon_\mathrm{id} \cdot \varepsilon_\mathrm{iso} \cdot \varepsilon_\mathrm{trig}.
\end{equation}
The method uses electron pairs from a well-known $Z\rightarrow~e^+e^-$ resonance decay. One electron~(tag) is required to meet strict selection criteria, while another electron~(probe) serves as an unbiased object. Exemplary distributions of invariant mass $m_{ee}$ from the $Z\rightarrow~e^+e^-$ process are presented in Figure~\ref{fig:zmass}.

\begin{figure}[htb]
	\centering
	\includegraphics[width=0.42\textwidth]{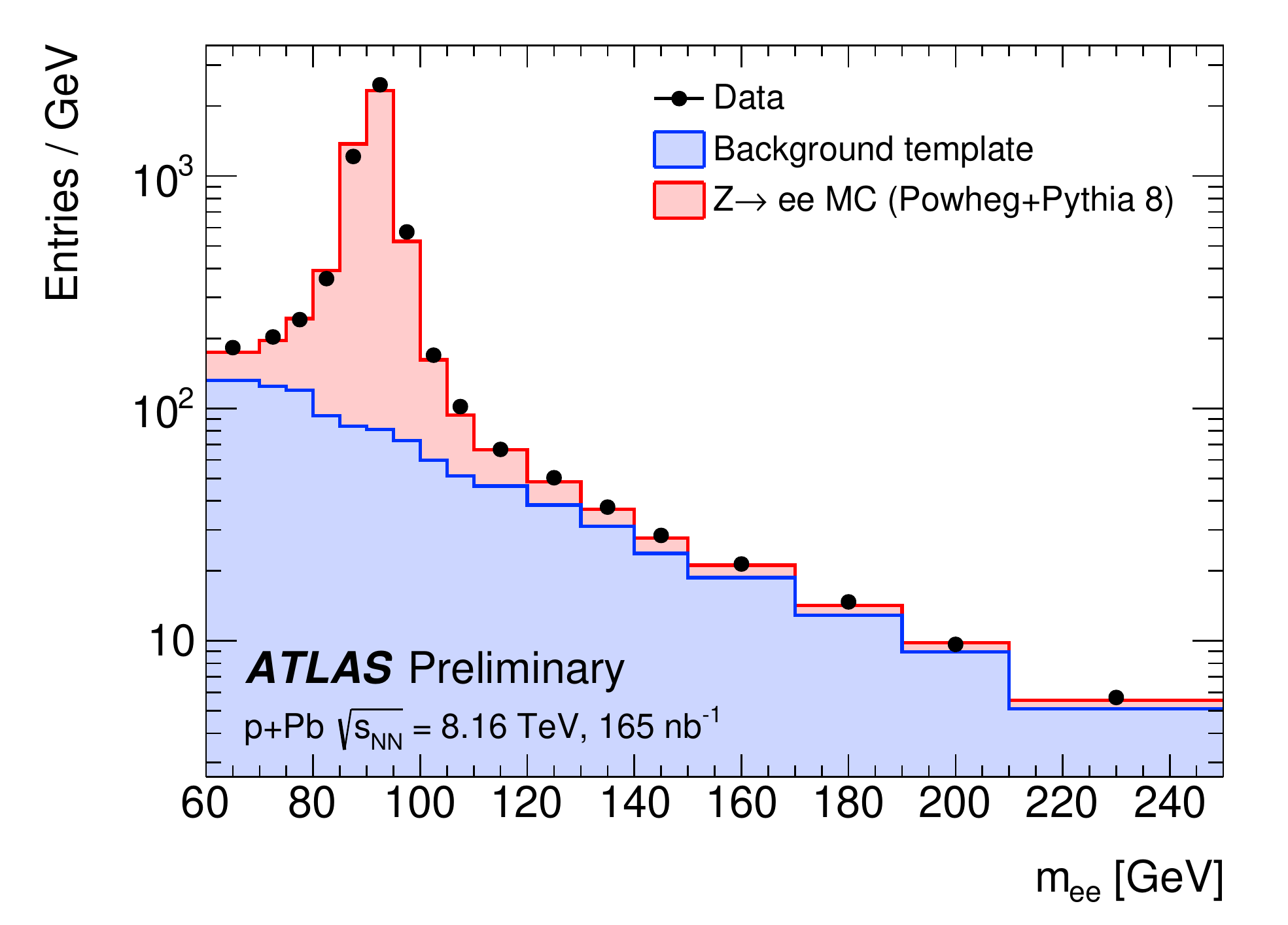} \hspace{0.08\textwidth}
	\includegraphics[width=0.42\textwidth]{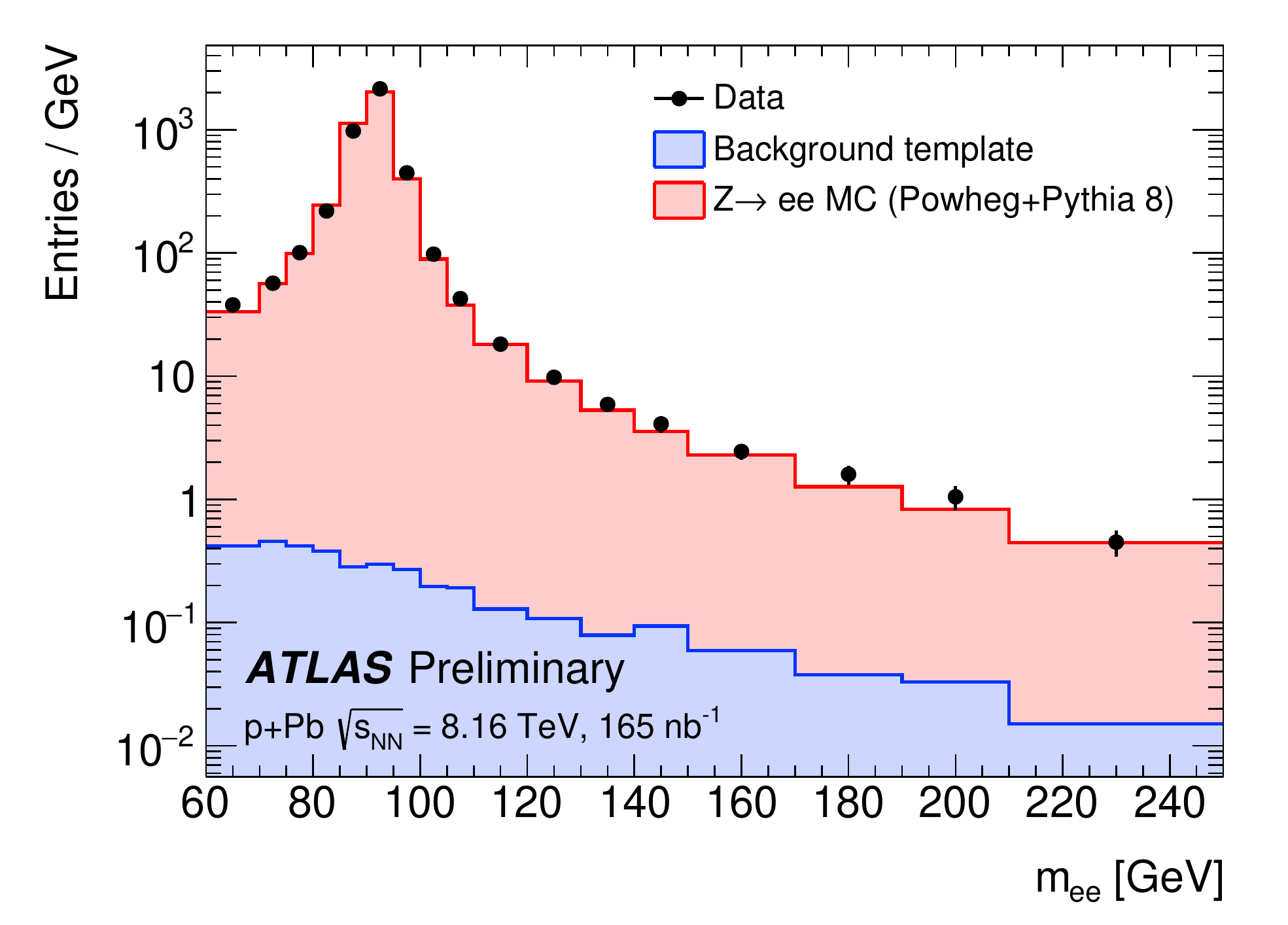}
	\caption{Invariant mass distributions of opposite-sign electron pairs with reconstructed~(left) and identified probes~(right) in 2016 p+Pb data (points) and in MC simulation for the signal (red) and backgrounds (blue) processes~\cite{bib:plots}.}
	\label{fig:zmass}
\end{figure}

\section{Electron reconstruction}
\label{sec:reco}

A passing electron leaves hits in the Inner Detector, which can be reconstructed as a track, and deposits the remaining energy in the electromagnetic~(EM) calorimeter, creating a cluster of energy. The purpose of electron reconstruction is to match the track with the cluster in the EM calorimeter. The reconstruction efficiency is estimated as the ratio of the number of reconstructed electron candidates to the number of EM clusters~\cite{bib:reco}.

Figure~\ref{fig:reco} shows electron reconstruction efficiency as a function of electron transverse energy $E_\mathrm{T}$ and pseudorapidity $\eta$. The efficiency rises with $E_\mathrm{T}$ from 93\% at $E_\mathrm{T}=15$~GeV and reaches the plateau with 98\% at $E_\mathrm{T}=50$~GeV. Higher efficiency is observed for central electron pseudorapidities and decreases to 93\% for $|\eta|>1.37$. No significant deviation from unity is found for data-to-MC ratios.

\begin{figure}[htb]
	\centering
	\includegraphics[width=0.37\textwidth]{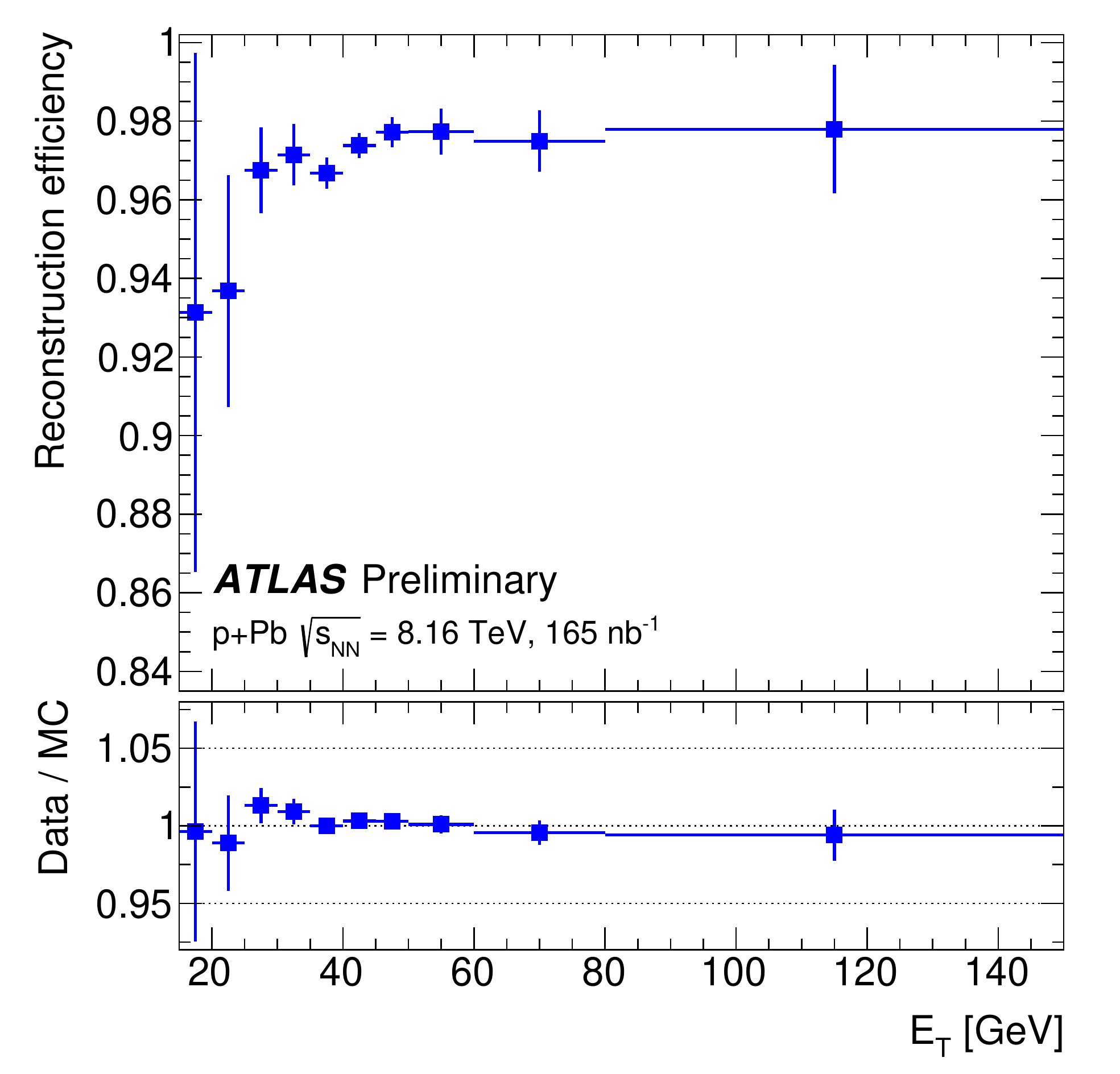} \hspace{0.1\textwidth}
	\includegraphics[width=0.37\textwidth]{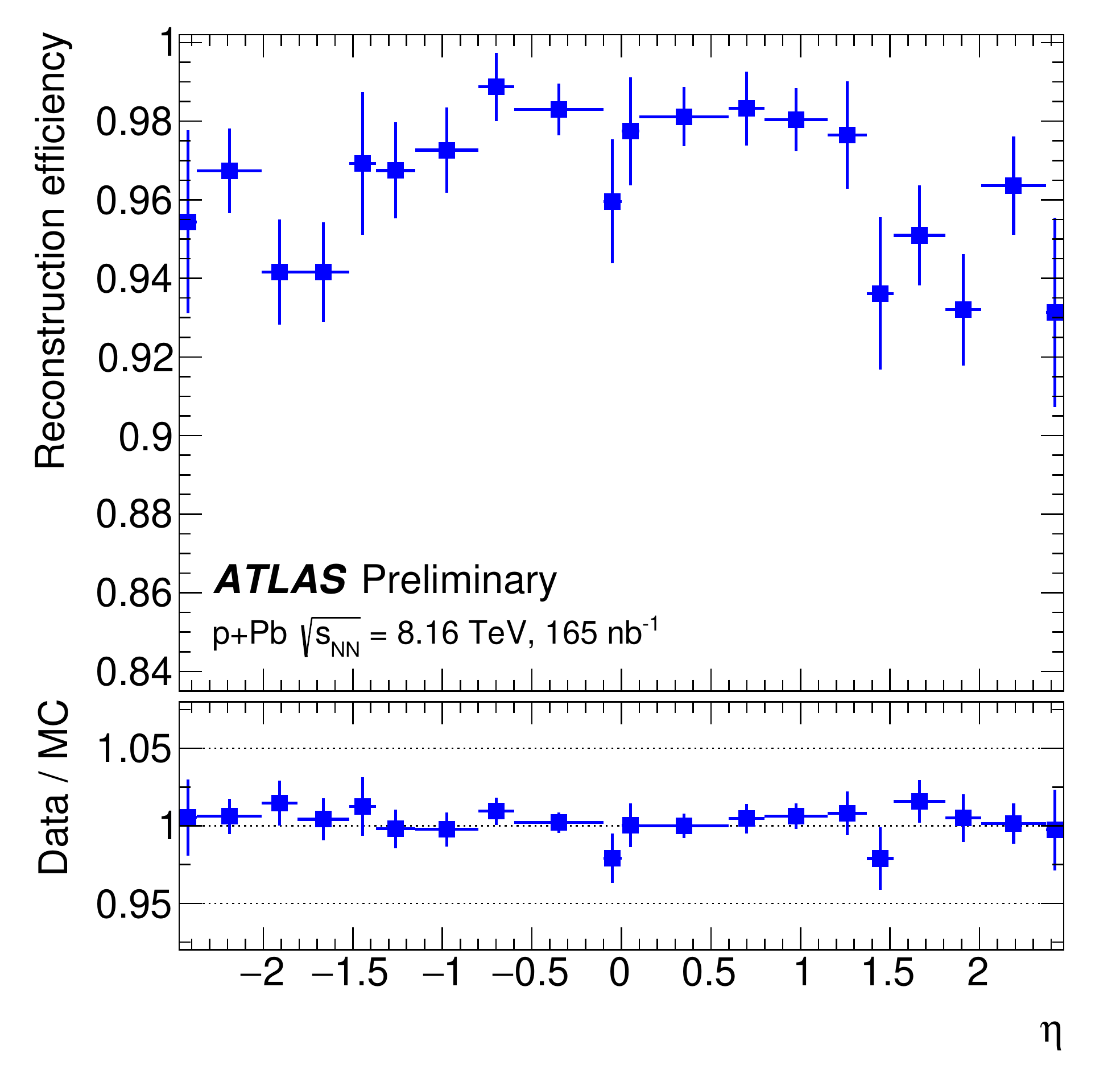}
	\caption{The electron reconstruction efficiency as a function of electron $E_\mathrm{T}$~(left) and $\eta$~(right) evaluated in 2016 p+Pb data. The bottom panels show the data-to-MC ratio. Error bars represent the total uncertainties composed of statistical and systematic components added in quadrature~\cite{bib:plots}.}
	\label{fig:reco}
\end{figure}

\section{Electron identification}
\label{sec:id}

The application of an electron identification algorithm allows to determine whether electrons originate from signal or background processes. Four identification selections are provided by the algorithm, referred to as Loose, LooseAndBLayer, Medium and Tight~\cite{bib:TaP}. The identification efficiency is defined as the ratio of the number of identified electrons to the number of reconstructed electrons in the $Z\rightarrow~e^+e^-$ sample.

Electron identification efficiency as a function of $E_\mathrm{T}$ and $\eta$ for the four working points is presented in Figure~\ref{fig:id}. The efficiency increases with electron energy from 82\%~(68)\% for Medium~(Tight) at $E_\mathrm{T}=15$~GeV and reaches the plateau with 92\%~(87\%) for $E_\mathrm{T}$ at around 60~GeV for Medium~(Tight). Data-to-MC ratios drop below unity at $|\eta|>1$.

\begin{figure}[htb]
	\centering
	\includegraphics[width=0.37\textwidth]{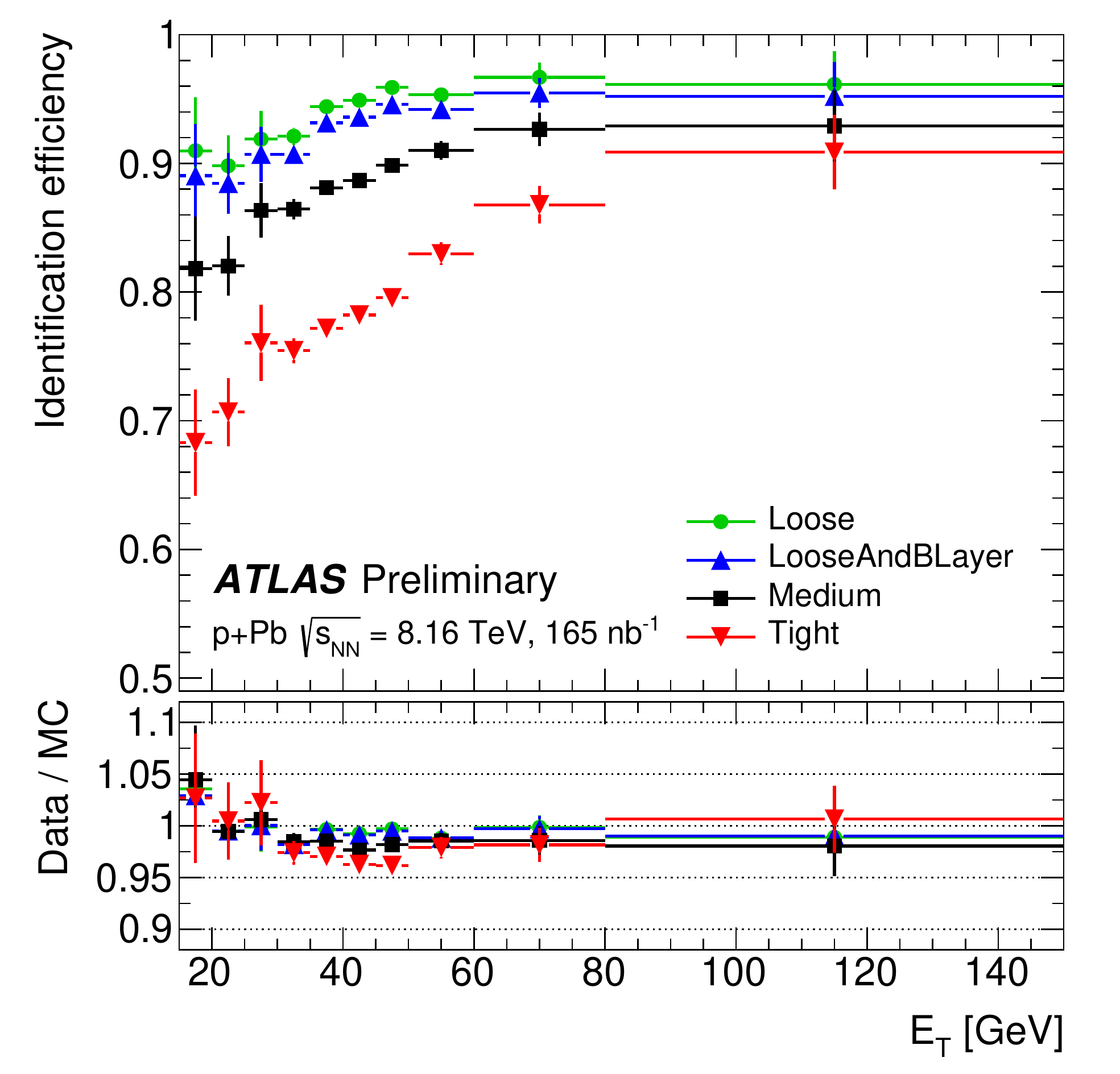} \hspace{0.1\textwidth}
	\includegraphics[width=0.37\textwidth]{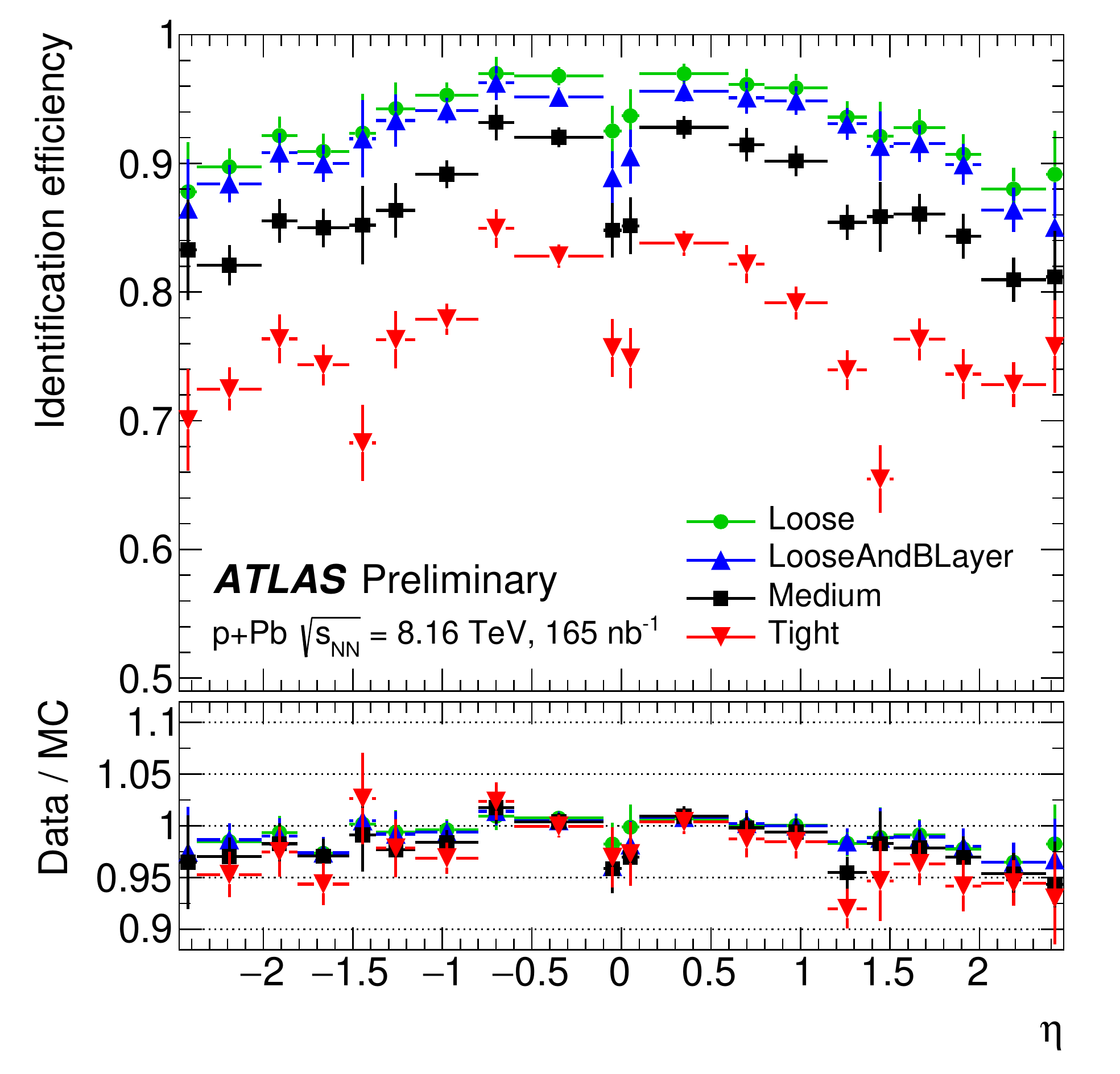}
	\caption{The electron identification efficiency as a function of electron $E_\mathrm{T}$~(left) and $\eta$~(right) evaluated in 2016 p+Pb data for four working points. The bottom panels show the data-to-MC ratios. Error bars represent the total uncertainties composed of statistical and systematic components added in quadrature~\cite{bib:plots}.}
	\label{fig:id}
\end{figure}

\section{Electron isolation}
\label{sec:iso}

Isolation criteria are used to further distinguish signal electrons from background processes. Various isolation working points are defined using track isolation $p^\mathrm{varcone}_\mathrm{T}$ and calorimeter isolation $E^\mathrm{cone}_\mathrm{T}$~\cite{bib:TaP}. The isolation efficiency is evaluated as the ratio of the number of isolated electrons to the number of electrons passing Medium identification requirements in the $Z\rightarrow~e^+e^-$ sample.

Electron isolation efficiency as a function of $E_\mathrm{T}$ and $\eta$ for four selections is shown in Figure~\ref{fig:iso}. The efficiency varies between 65--96\% at electron $E_\mathrm{T}=15$~GeV for the studied working points. Data-to-MC ratios deviate from unity up to 8\% at the lowest $E_\mathrm{T}$ and highest $|\eta|$.

\begin{figure}[htb]
	\centering
	\includegraphics[width=0.37\textwidth]{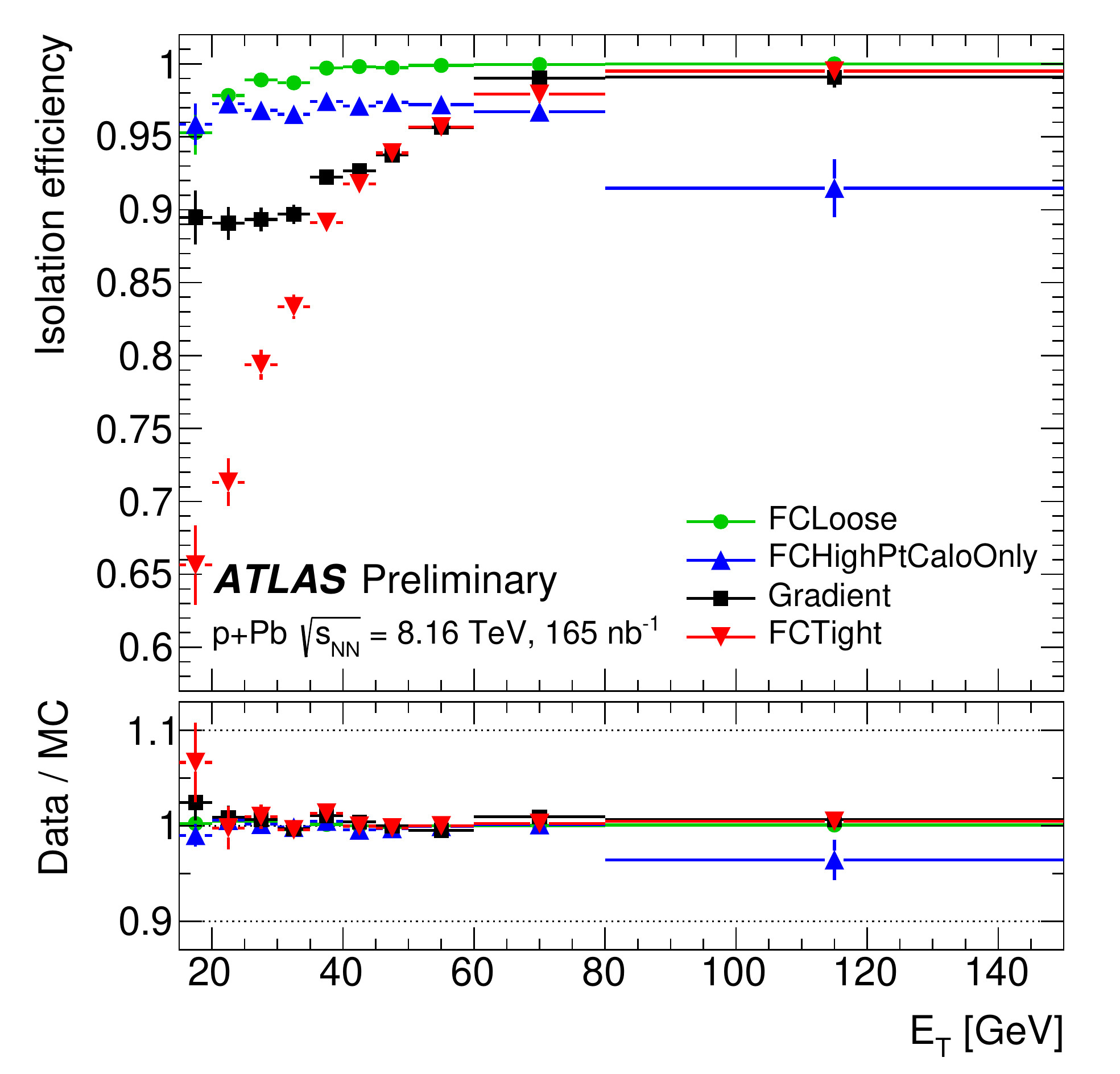} \hspace{0.1\textwidth}
	\includegraphics[width=0.37\textwidth]{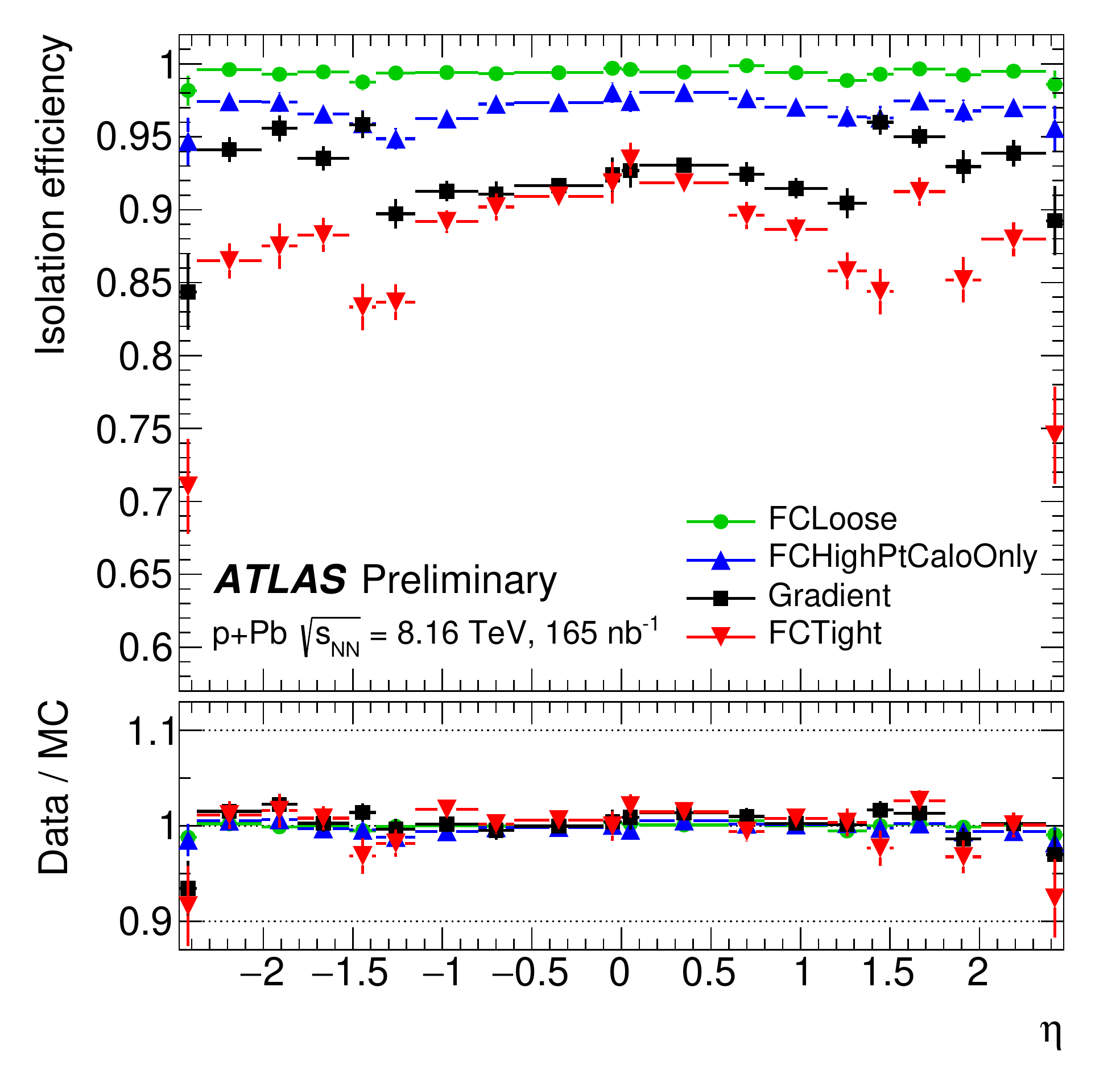}
	\caption{The electron isolation efficiency as a function of electron $E_\mathrm{T}$~(left) and $\eta$~(right) evaluated in 2016 p+Pb data for four working points. The bottom panels show the data-to-MC ratios. Error bars represent the total uncertainties composed of statistical and systematic components added in quadrature~\cite{bib:plots}.}
	\label{fig:iso}
\end{figure}
\vspace{-0.4cm}
\section{Electron trigger}
\label{sec:trig}

The ATLAS trigger system is used to select the most interesting events with a maximum rate of $\sim$2~kHz out of all events delivered by the LHC with the rate of $\sim$40~kHz. The trigger efficiency is determined as the ratio of the number of triggered electron candidates to the number of electrons with Medium identification and Gradient isolation requirements in the $Z\rightarrow~e^+e^-$ sample. In this study, the efficiency is measured for a single electron trigger with the $E_\mathrm{T}=15$~GeV threshold and Loose identification criteria~\cite{bib:trig}.

Figure~\ref{fig:trig} shows electron trigger efficiency as a function of $E_\mathrm{T}$ and $\eta$. The efficiency increases with electron energy from 82\% for $E_\mathrm{T}$ at 15~GeV and reaches the plateau at $E_\mathrm{T}=40$~GeV with 98\%. A deviation from unity up to 5\% is observed for data-to-MC ratios at $E_\mathrm{T}<20$~GeV, $\eta \approx 0$ and in the calorimeter transition region ($\eta \approx -1.5$).

\begin{figure}[htb]
	\centering
	\includegraphics[width=0.37\textwidth]{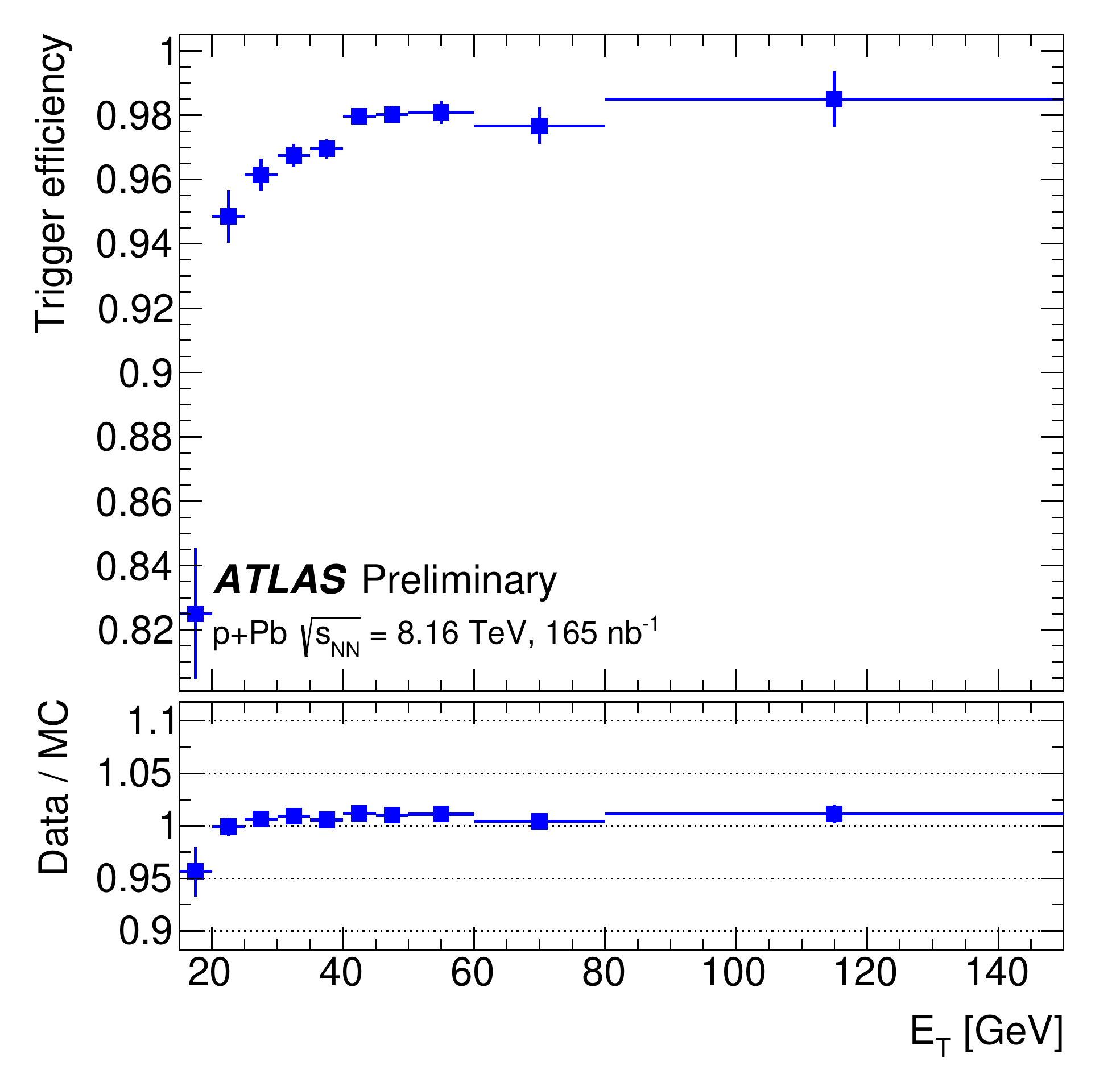} \hspace{0.1\textwidth}
	\includegraphics[width=0.37\textwidth]{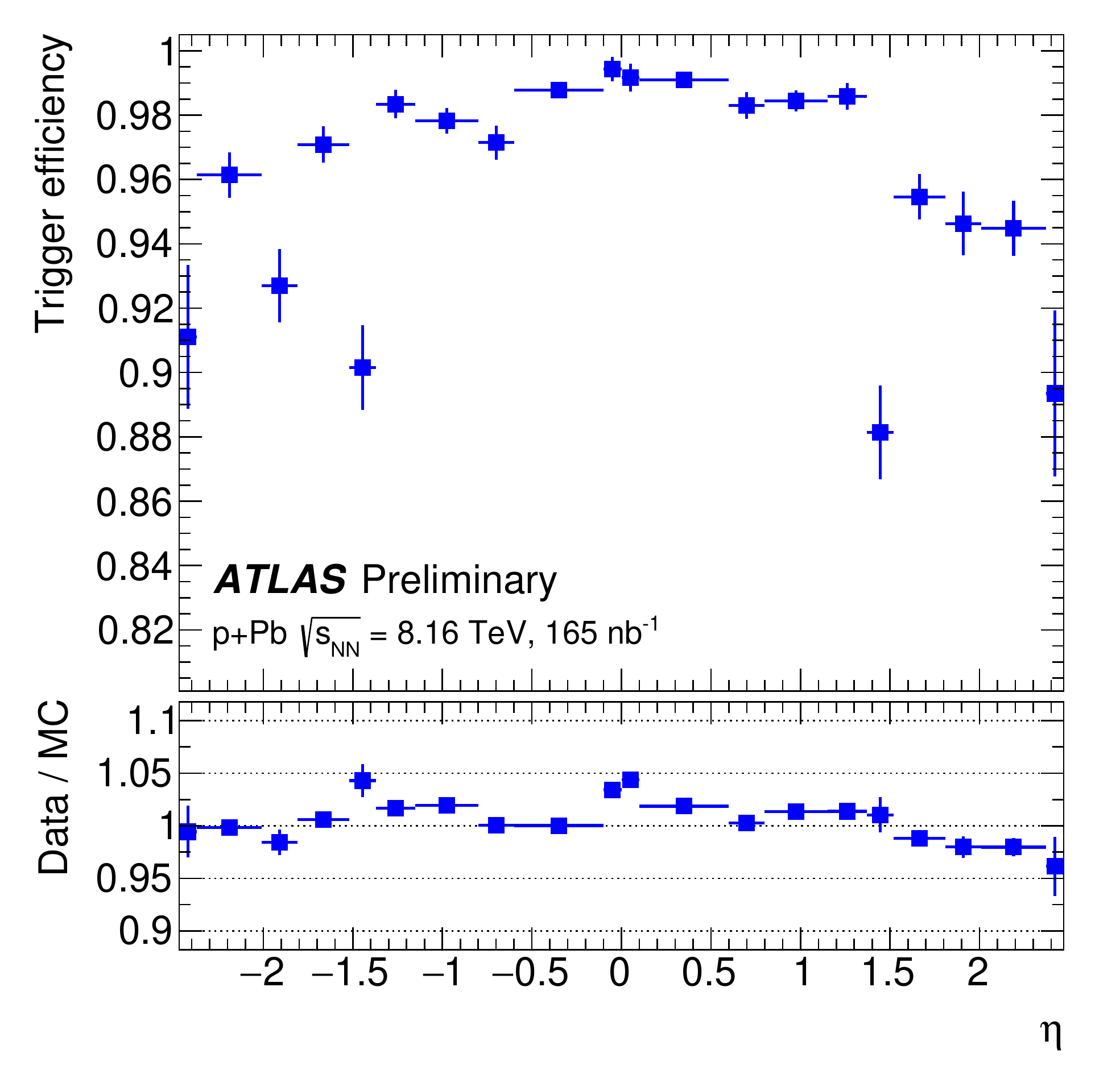}
	\caption{The electron trigger efficiency as a function of electron $E_\mathrm{T}$~(left) and $\eta$~(right) evaluated in 2016 p+Pb data. The bottom panels show the data-to-MC ratio. Error bars represent the total uncertainties composed of statistical and systematic components added in quadrature~\cite{bib:plots}.}
	\label{fig:trig}
\end{figure}
\vspace{-0.4cm}
\section{Conclusion}
\label{sec:conclusion}

The electron efficiencies have been measured in p+Pb collisions at $\sqrt{s_{\mathrm{NN}}}=8.16$~TeV collected by ATLAS in 2016, corresponding to a total integrated luminosity of 165~nb$^{-1}$. Four efficiencies, related to electron reconstruction, identification, isolation and trigger, have been determined in data and MC simulation. Data-to-MC ratios have been derived and are currently used as input in physics measurements with electrons in the form of a multiplicative correction to account for mismodelling of the detector in MC simulation.
\vspace{-0.4cm}
\subsection*{Acknowledgements}
\label{sec:acknowledgements}

This work was partly supported by the program „Excellence initiative - research university” for the AGH University of Science and Technology, the National Science Centre, Poland, grant 2020/37/B/ST2/01043 and PL-Grid Infrastructure.

\ResCnt

\maketitle

\begin{abstract}
The beauty-to-charm physics concerns e.g.\ matter-antimatter asymmetries properties determination, thus its elaborative studies are conducted at the LHCb experiment. One of them is ongoing as the ``First observation and branching fraction measurement of the $\Lambda_{b}^{0} \rightarrow D_{s}^{-} p$ decay" which I have been working on with the B2OpenCharm group. A data-driven approach is applied to the decay studies, in order to establish its branching fraction. This includes toy generation process, important for systematic uncertainty determination. The process of Monte Carlo sample generation is discussed with an example of pull value calculation, based on the Central Limit Theorem properties. The study is to be published in Autumn 2023.
\end{abstract}


The LHCb experiment was designed for studies of beauty physics, which is the physics of $b$-particles (containing a $b$ quark). Their decays provide an insight into the phenomenon of matter-antimatter asymmetries and so-called $CP$-symmetry breaking. Examples are the beauty to charm hadron decays where the $b$ to $c$ quark transition takes place. In a given talk, I have described the study of those decays that I had started contributing to with my Master's thesis at University of Warsaw and have been working on as a part my PhD programme at IFJ PAN, Cracow. Its results will be published in ``First observation and branching fraction measurement of the $\Lambda_{b}^{0} \rightarrow D_{s}^{-} p$ decay" study of the LHCb collaboration in Autumn 2023.

\section{The decay of $\Lambda_{b}^{0} \rightarrow D_{s}^{-} p$}

Why does $\Lambda_{b}^{0} \rightarrow D_{s}^{-} p$ decay study deserve to be considered as a separate analysis? There are a few observations to support this. The first one can be is shown at the Feynman diagram of the decay in Fig.~\ref{fig:lb2dsp}.

\begin{figure}
\centering
\includegraphics[width=0.45\textwidth]{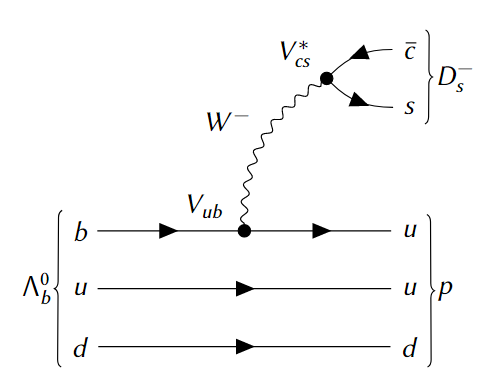}
\caption{\label{fig:lb2dsp}The Feynman diagram of the $\Lambda_{b}^{0} \rightarrow D_{s}^{-} p$ decay.}
\end{figure}

This is a beauty-to-charm decay, as the $\Lambda_{b}^{0}$ baryon with a $b$ quark decays to a $D_{s}^{-}$ meson with a $c$ quark, thus it is a decay that can be measured by the LHCb as explained earlier. $\Lambda_{b}^{0} \rightarrow D_{s}^{-} p$  constitutes a background
to other analyses such as the study of $CP$ violation with $B_{s}^{0} \rightarrow D_{s}^{\mp} K^{\pm}$ decays \cite{CP2018}, where $\Lambda_{b}^{0}$ is one of the contributing backgrounds in the signal region. Estimating the yield (number of candidates) of the decay in question in the samples considered by many LHCb analyses is crucial for those analyses. It reduces the systematic uncertainties associated with the background estimation for other decays, improving the results they deliver.

Second reason, why the $\Lambda_{b}^{0} \rightarrow D_{s}^{-} p$ study is valuable can be found at Fig.~\ref{fig:lb2dsp}, at the interaction vertices, to be exact. When one calculates the branching fraction $\mathcal{B}$ of this decay, it will be proportional to the absolute value squared of the $V_{ub}$ element of the Cabibbo–Kobayashi–Maskawa (CKM) matrix, that describes the strength of the flavour-changing weak interaction. Knowing as precise values as possible of the matrix elements is desired for the $CP$-violation studies. Thus obtaining the $\mathcal{B}(\Lambda_{b}^{0} \rightarrow D_{s}^{-} p)$ opens up a new window for $V_{ub}$ parameter estimation and improving its observed value.

Moreover, another argument why to consider the mentioned beauty-to-charm decay is for the comparison of non-factorizable effects in baryonic and mesonic decays. $\Lambda_{b}^{0} \rightarrow D_{s}^{-} p$ is a decay with a meson-hadron interaction where, due to the presence of quarks in both final-state particles, the gluon exchanges are present. A similar methodology was already described for exclusive, nonleptonic $B$ meson decays~\cite{Beneke:2000ry} and applying it for the $\Lambda_{b}^{0} \rightarrow D_{s}^{-} p$ may improve models that have been used so far.

\section{The $BR(\Lambda_{b}^{0} \rightarrow D_{s}^{-} p)$ analysis}
The analysis is ongoing in a $BR(\Lambda_{b}^{0} \rightarrow D_{s}^{-} p)$ subworking group, which is a part of the B2OpenCharm working group of LHCb experiment. The study's mission is to establish first estimation of the $\Lambda_{b}^{0} \rightarrow D_{s}^{-} p$ branching fraction measurement. I am among the proponents with Jordy Butter, Niels Tuning, Sevda Esen, Agnieszka Dziurda and Tomasz Szumlak. This is a Polish-Dutch collaboration between the IFJ PAN and the NIKHEF group.

\begin{figure}
\centering
\includegraphics[width=0.9\textwidth]{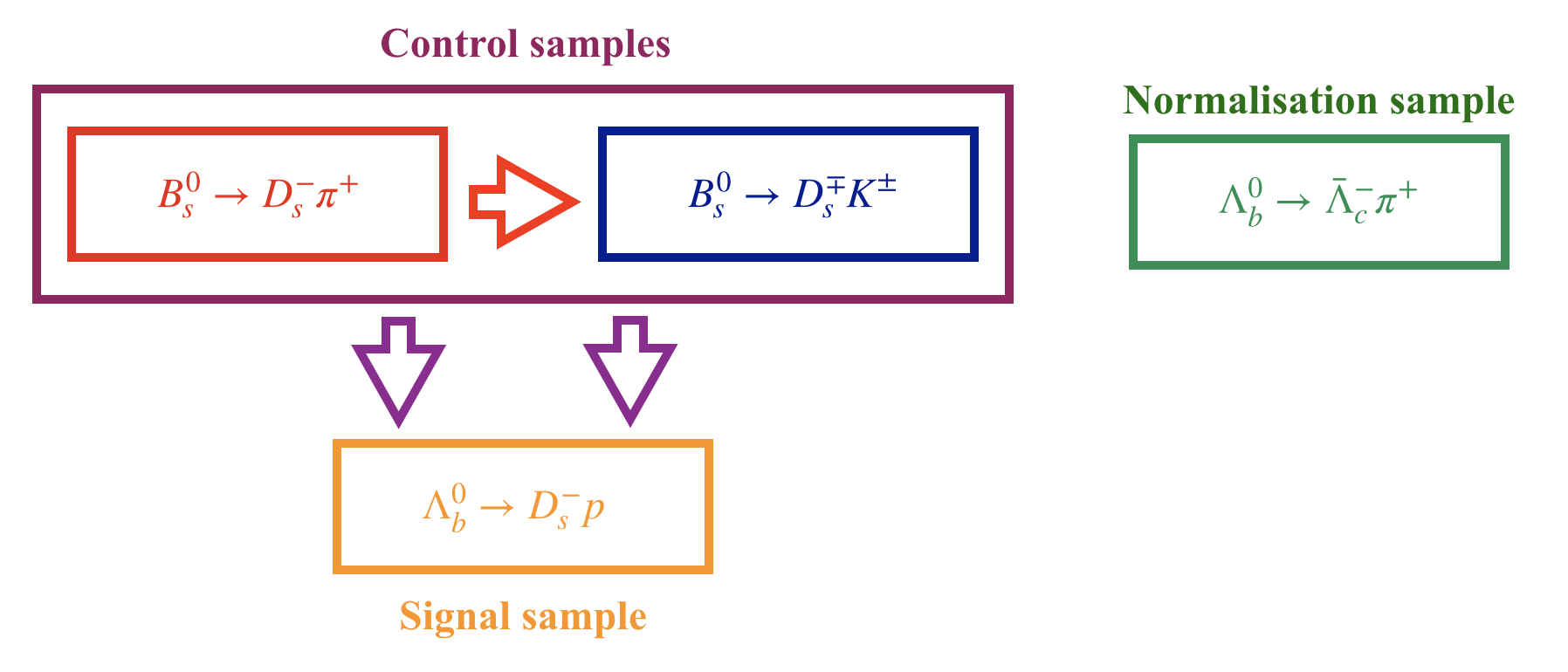}
\caption{\label{fig:analysis-strategy}Schematic overview of the analysis strategy. Invariant mass fits are performed to the additional control modes $B^{0}_{s} \rightarrow D^{-}_{s} \pi^{+}$ and $B^{0}_{s} \rightarrow D^{\mp}_{s} K^{\pm}$, which allow to calculate the contributions from $D^{-}_{s} \pi^{+}$-like and $D^{\mp}_{s} K^{\pm}$-like decays in the $\Lambda_{b}^{0} \rightarrow D_{s}^{-} p$ sample in a data-driven way.}
\end{figure}

The general idea that is applied is shown in Fig.~\ref{fig:analysis-strategy}. The group has turned to the data-driven approach to calculate the contributions from the $B^{0}_{s} \rightarrow D^{\mp}_{s} h^{\pm}$ decays into the signals invariant mass spectra, where $h$ is an additional hadron. The analysis considers data samples of Run2 data taking period of LHCb (2015-2018) due to larger dataset and cleaner selection than for Run1 (2012-2015). Two control samples are used  $B^{0}_{s} \rightarrow D^{-}_{s} \pi^{+}$ and $B^{0}_{s} \rightarrow D^{\mp}_{s} K^{\pm}$. From them, the constraints on the yields and shapes of all $D^{-}_{s} \pi^{+}$-like and $D^{\mp}_{s} K^{\pm}$-like decays can be made and passed onto the $\Lambda_{b}^{0} \rightarrow D_{s}^{-} p$ signal sample consideration. This way, the obtained result is based more on the data than the MC (Monte Carlo generated) samples and does not introduce the additional data/MC corrections. Additionally, the $\Lambda_{b}^{0} \rightarrow \Lambda_{c}^{-} \pi^{+}$ sample is used for normalization (it is a decay with a similar kinematics and a well-determined branching fraction). Calculating branching ratio of $\Lambda_{b}^{0} \rightarrow D_{s}^{-} p$ and $\Lambda_{b}^{0} \rightarrow \Lambda_{c}^{-} \pi^{+}$ gets rid of most systematic effects for the uncertainty calculations, e.g.\ due to a similar kinematics.

\section{Toy generation}

From my contributions to the analysis a few can be distinguished: modelling of the signal and background shapes for the decays, invariant mass fits, systematic studies and validation. For most of those issues, the tool of toy sample generation is invaluable. In order to apply it for the considered case, one should make an invariant mass fit to data with a considered theoretical model of the backgrounds and signal particles invariant mass distribution. This way, the estimates of yield values for all decays can be obtained. Decay examples are $B^{0}_{s} \rightarrow D^{-}_{s} \pi^{+}$, $B^{0}_{s} \rightarrow D^{\mp}_{s} K^{\pm}$, $\Lambda_{b}^{0} \rightarrow \Lambda_{c}^{-} \pi^{+}$ and $\Lambda_{b}^{0} \rightarrow D_{s}^{-} p$. Using obtained values and distributions shapes one can generate pseudorandom samples for a given seed. In order to do that, for a new MC sample, the set of yield values is drawn from Poisson distributions which have the center values given by the yields from the data sample. In such a way, one may obtain a ``fake" sample that will have similar physical properties as the ``original" data sample, but will be different in a manner that simulates statistical fluctuations. This process can be repeated for many different seed numbers, creating a set of MC samples. Such samples may later be used for the systematic studies.
An example of doing that is to define a pull $p$ of parameter $N$ as in Eq.~\eqref{eq:pull}:

\begin{equation}
p(N) = \dfrac{N_{gen} - N_{fit}}{\sigma_{N; fit}},
    \label{eq:pull}
\end{equation}

for statistical uncertainty of a considered parameter denoted as $\sigma_{N}$ and for subscripts $gen$ and $fit$ denoting values \textbf{gen}erated for the MC sample and obtained in a \textbf{fit} to this sample. Now the pull of e.g.\ the signal yield may be calculated for each MC sample. The process of generating MC samples and obtaining pulls may be automated as shown schematically in Fig.~\ref{fig:scheme}.

\begin{figure}
\centering
\includegraphics[width=\textwidth]{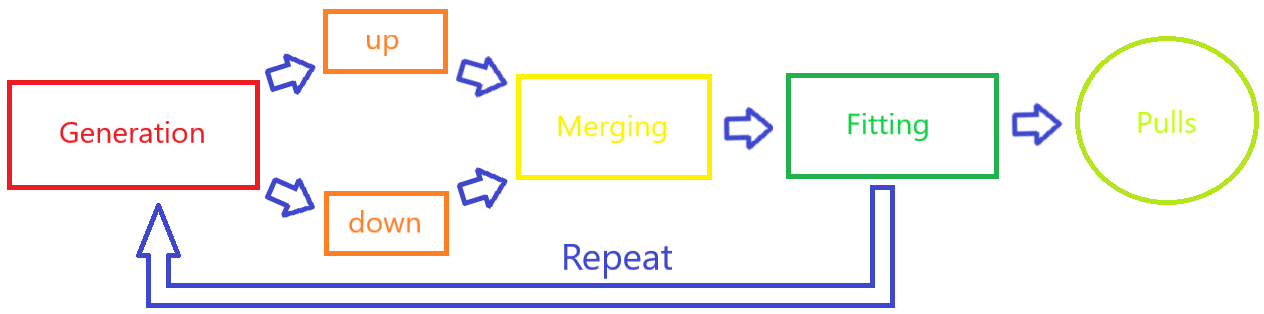}
\caption{\label{fig:scheme}Schematic overview of a step-by-step process of MC samples generation for both magnet polarities, merging, fitting and obtaining pull values for considered observable. This process can be automated and repeated for different seed numbers.}
\end{figure}

The main reason to apply this procedure for the analysis is its priceless contribution to the process of determining the systematical uncertainty of obtained signal yield value in a fit to data. This can be done via plotting the histogram of pull values obtained in consecutive fits to MC samples. According to the Central Limit Theorem, the distribution of pull values for an ideal theoretical model (no mismatches) should be described by a Gaussian function with a mean of zero and the standard deviation of one. An exemplary distribution of pull values, for an initially considered model of $\Lambda_{b}^{0} \rightarrow D_{s}^{-} p$ signal and its backgrounds is shown in Fig.~\ref{fig:Lb2Dsp-pull}. The shift between the obtained mean of fitted Gaussian function and the theoretically expected value of zero introduces a bias, that is proportional to the systematic uncertainty of the yield value obtained in a fit to the data.

\begin{figure}
\centering
\includegraphics[width=0.9\textwidth]{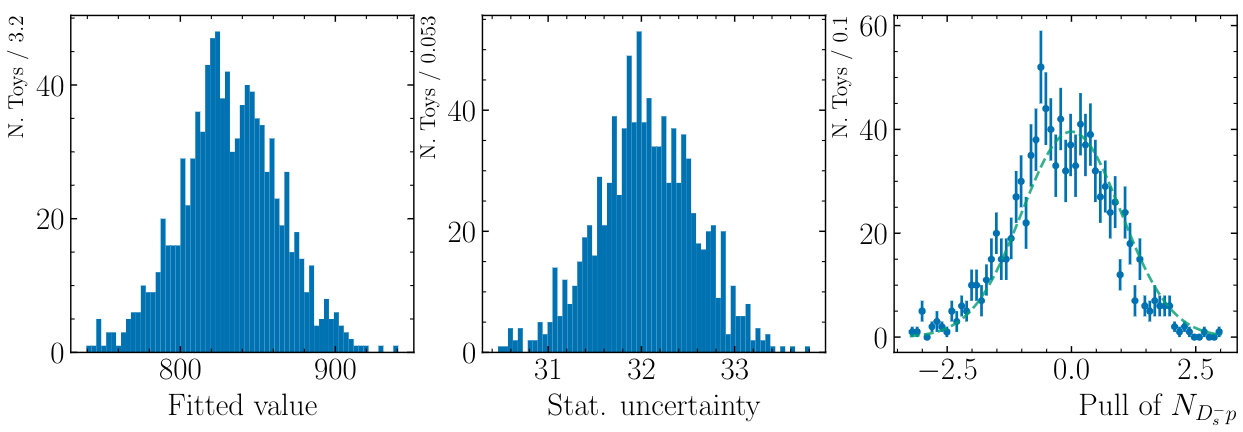}
\caption{\label{fig:Lb2Dsp-pull}Pull values of the $\Lambda_{b}^{0} \rightarrow D_{s}^{-} p$ signal yield $N_{D_{s}^{-} p}$ obtained from the fits to a thousand MC samples for an initial distributions model. Starting from the left: histograms of the fitted values, statistical uncertainties and the pulls. To compare, on the third panel the standard normal distribution is drawn using dashed line.}
\end{figure}

In general, systematic studies have been relying on this procedure to check different theoretical models, in order to determine which one is the most stable (obtained pull histogram can be described by a Gaussian and its properties closely follow the theoretically predicted standard normal distribution) and which parameters of the invariant mass model need to be tuned for the systematics to be improved. The example would be shifting certain parameter of the model, e.g.\ by its standard deviation, to check its influence on the yield bias.

All the mentioned tools have been found to be essential for the ``First observation and branching fraction measurement of the $\Lambda_{b}^{0} \rightarrow D_{s}^{-} p$ decay" study of the LHCb collaboration that should be published in Autumn next year.

\ResCnt

\maketitle

\begin{abstract}
Until recent times, it was believed that lepton-pair production due to photon clouds interactions in the ultrarelativistic heavy-ion collisions was dominant mainly in the ultraperipheral collisions, where the incoming nuclei do not touch. However, it was realised that when we lift this restriction in more central collisions, this $\gamma\gamma$ fusion mechanism is also predominant at low pair transverse momenta $P_t$. We use the Wigner distribution formalism to perform calculations and distributions of differential cross-section as a function of invariant mass $W$, and total cross-section as a function of centrality for the $PbPb \rightarrow PbPb e^{+}e^{-}$ process. The same cuts and collision energy ($\sqrt{s_{NN}}=5.02$ TeV) as in the ALICE data have been used for the calculations. Excellent agreement of theory and experiment has been achieved. We also show that the ratio of $e^{+}e^{-}$ production is roughly the same both inside and outside the nuclei. 
\end{abstract}

\section{Introduction}
In the ultrarelativistic heavy ion collisions, the ions are surrounded by a large coherent photon clouds which interact with each other. One of the mechanisms of creation of dilepton pairs is photon-photon fusion coming from these clouds. It has been shown that lepton pairs produced from the $\gamma\gamma$ mechanism play a dominant role at a very low pair transverse momenta $P_T$ of the photons \cite{2021paper}. 

Until recently, it was believed \cite{2019paper} that this was only occurring in the ultraperipheral collisions, where the incoming nuclei do not touch. These are quantified with the impact parameter $b>2R$, where $R$ is the nuclear radius. However, it was been found that if we lift this restriction and move to more central collisions, where $b<2R$, the dilepton production from $\gamma\gamma$ fusion also occurs and is predominant \cite{2021paper,our_paper} (again, at very low $P_T$).

In this paper, we study the invariant mass distributions of lepton pairs (specifically $e^+e^-$) produced in the ultrarelativistic heavy-ion collisions at a very low pair transverse momenta, $P_T \leq 0.15$ GeV, and using the kinematical cuts from ALICE data. We focus on the $PbPb \rightarrow PbPb e^{+}e^{-}$ process. Comparison with the mentioned data is shown and discussed.

\section{Methodology}
\let\thefootnote\relax\footnote{Work done under supervision of Dr. Mariola Klusek-Gawenda from IFJ PAN, Krakow.}
\addtocounter{footnote}{-1}\let\thefootnote\svthefootnote
At ultrarelativistic speeds, quantities such as the mass of the nuclei become hard to measure at colliders directly. We use the invariant mass of the nuclei, \textit{i.e.} their energy in the rest frame, to perform our calculations. This is a very useful quantity because the mass of the products of the collisions is the same as the rest mass of the decaying nuclei \cite{invariantmass_manchester}.

The formalism used for the calculations, based on the Wigner distributions, is explained in \cite{2021paper}. It allows to calculate the centrality dependence of lepton pair production coming from the $\gamma\gamma$ mechanism. 

\section{Results}
When we move from the ultra-peripheral collisions to more central ones, we allow the incoming nuclei to collide, letting $b<2R$ when integrating over the impact parameter $b$ \cite{our_paper}. We were interested in investigating where the dilepton production from $\gamma\gamma$ fusion is dominant: inside our outside the nuclei? For that, we have calculated the distributions for differential cross-sections for the $PbPb \rightarrow PbPb e^{+}e^{-}$ process as a function of invariant mass $W$ for two fixed centrality ranges (see Fig \ref{firstplots}). 

\begin{figure}[!htbp]
  \begin{subfigure}{0.5\textwidth}
    \includegraphics[width=0.95\linewidth]{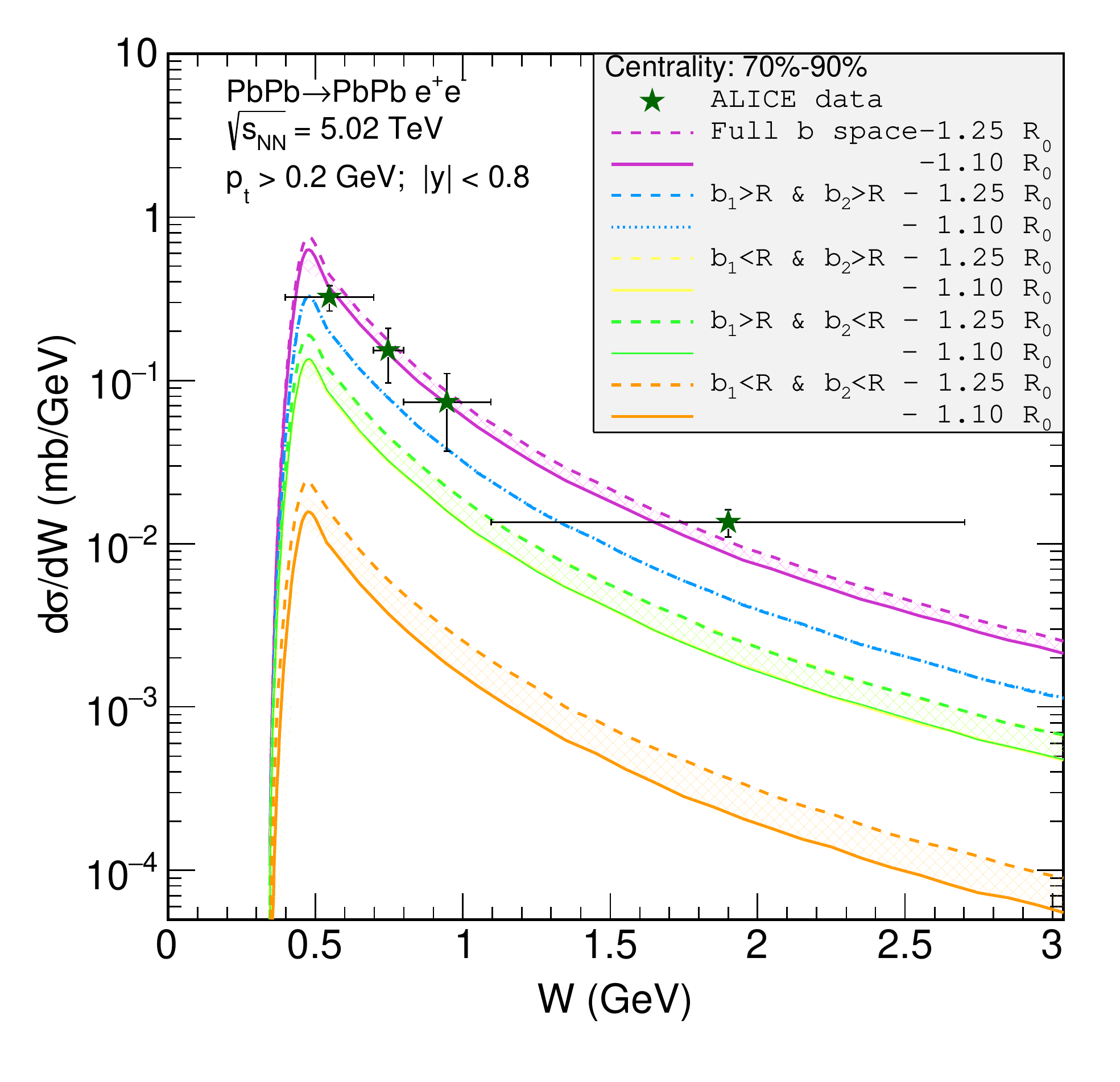}
    \caption{}
  \end{subfigure}
  \begin{subfigure}{0.5\textwidth}
    \includegraphics[width=0.95\linewidth]{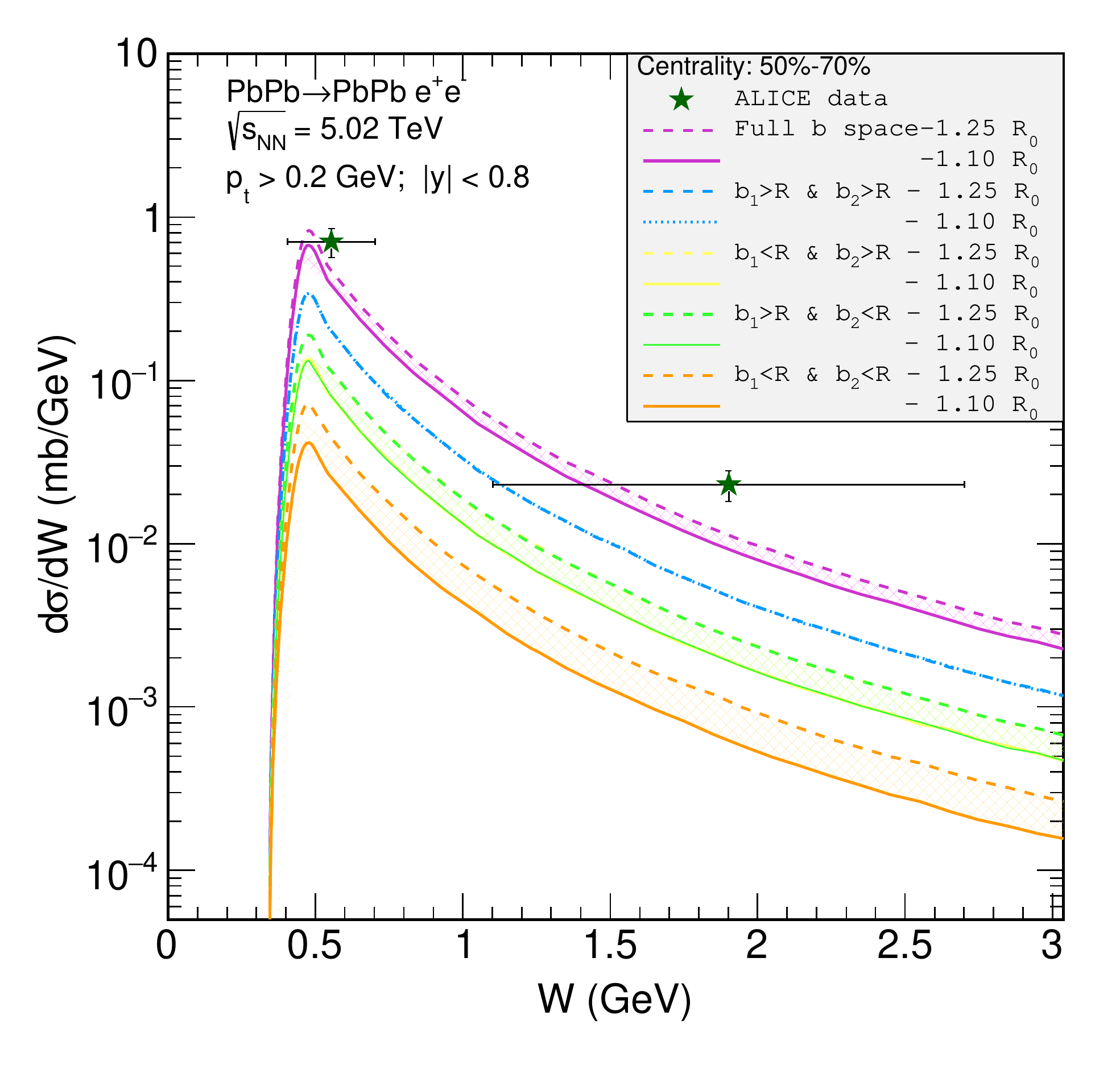}
    \caption{}
  \end{subfigure}
\caption{Differential cross sections as a function of invariant mass $W$. Centrality ranges are (a) 70-90\% and (b) 50-70\%. Both ranges are made with the same kinematical limitations and $R_0$ parameters. Dashed lines refer to $R_0=1.25$ fm and solid lines to $R_0=1.1$ fm. ALICE data from \cite{alice_data} is shown as green stars. Yellow lines are not visible because they overlap with the green ones, as they represent virtually the same physical situation.}
\label{firstplots}
\end{figure}

For studying where the lepton pairs are produced, we calculated, for the first time, the values of cross-sections for different regions in the impact parameter space. The four considered cases are:
\begin{enumerate}
    \item full impact parameter space,
    \item outside of both nuclei: $b_{1}>R_{1}, b_{2}>R_{2}$,
    \item inside each nucleus, excluding the area of overlap: $b_{1}<R_{1}, b_{2}>R_{2}$ or $b_{1}>R_{1}, b_{2}<R_{2}$, 
    \item overlapping area: $b_{1}<R_{1}$, $b_{2}<R_{2}$. 
\end{enumerate}

\noindent It seemed interesting to take into account the definition of the nuclear radius:
\[R=R_0A^{1/3},\]
because the value of the parameter $R_0$ varies \cite{nuclear_radius_def}. We have used the range $R_0 = (1.10 - 1.25)$ fm, which give $R \approx (6.52 - 7.41)$ fm for $^{208}$Pb \cite{our_paper}.

Figure \ref{firstplots} shows the differential cross-sections as a function of the invariant mass of low $P_t$ lepton pairs. The same kinematical cuts and collision energy ($\sqrt{s_{NN}}=5.02$ TeV) as in the ALICE data \cite{alice_data} have been used, to compare theory and experiment. As one can judge from the above figure, a good agreement has been achieved with the data, implying that the $\gamma\gamma$ fusion mechanism dominates for dilepton production both at peripheral (a) and semi-central (b) collisions. We notice how in the case (a) with bigger centrality, the cross-section is drastically reduced in condition (4), \textit{i.e.} the overlapping area between the nuclei.

In Table \ref{thetable}, we compare the cross-sections of the electron-positron pair production both inside (conditions 3 + 4) and outside (condition 2) the nuclei, considering as well the values of the nuclear radius mentioned above. We have found that, when taking averages, the ratio of the cross-sections is approximately equal to 1. This means that $e^+e^-$ pairs are produced roughly at the same ratio inside and outside the nuclei. This is a new, and also an important result.

\begin{wrapfigure}{r}{0.5\linewidth}
    \centering
    \includegraphics[width=1.0\linewidth]{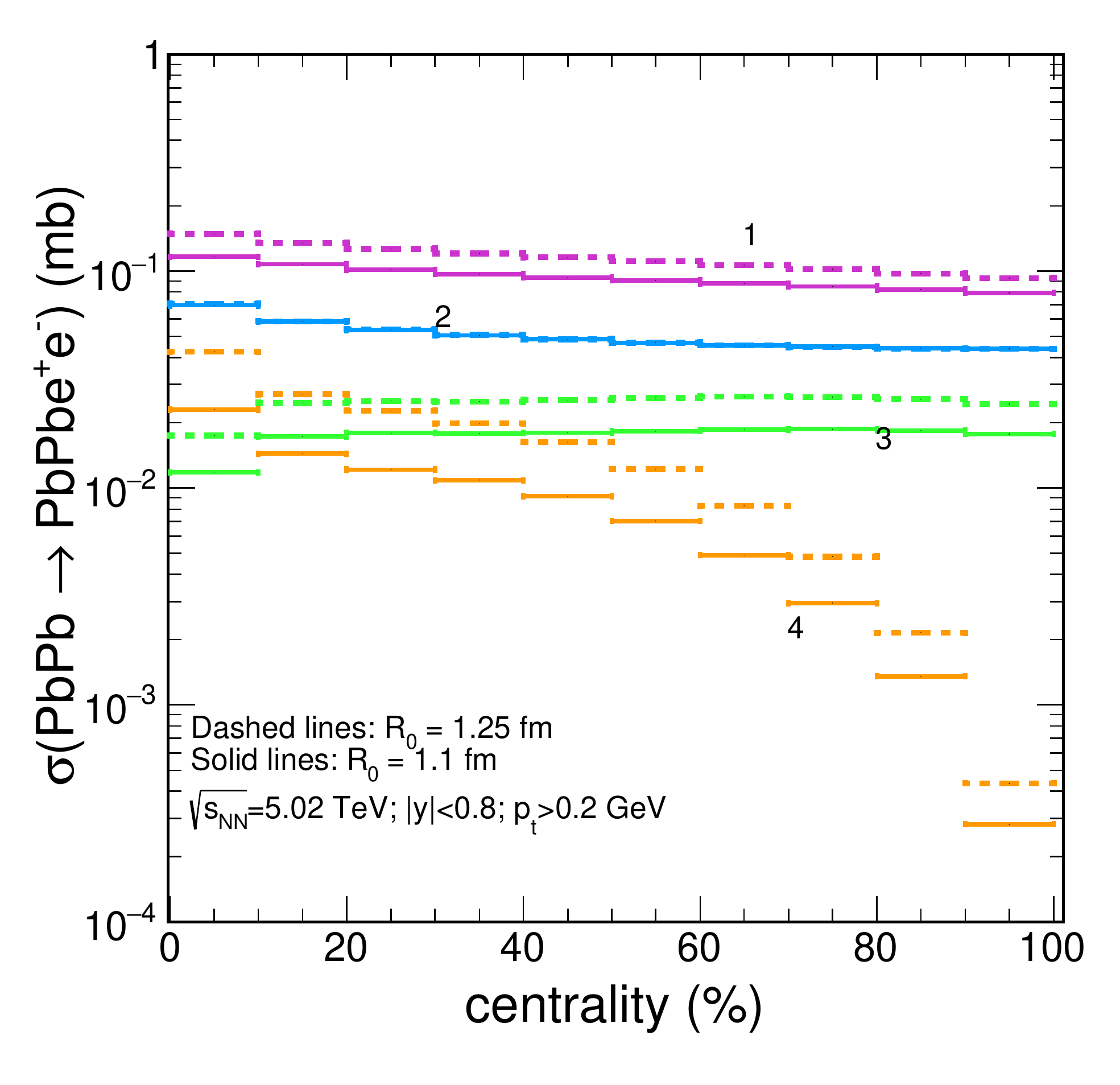}
    \caption{Total cross-section dependence on centrality for $PbPb \rightarrow PbPb e^{+}e^{-}$ process. Solid lines refer to $R_0=1.1$ fm, and dashed lines to $R_0=1.25$ fm. Kinematical limitations are the same as in Fig \ref{firstplots}. Yellow lines overlap with the green ones.}
    \label{centrality_one}
\end{wrapfigure}

It was also interesting to check the dependence of the cross-sections (both total and differential) on the $R_0$ parameter. We have found that when it is small, lepton pairs are rather produced outside the nuclei. However, when we look at larger values of radius, they are more likely to be produced inside. This result shows the sensitivity of the cross-section to the radius value used. Therefore, it seems reasonable to use a lead nucleus radius in the range of $R=(6.52-7.41)$~fm, rather than a fixed value, \textit{i.e.} $R=7$~fm.

To generalise results a bit, we have also performed calculations of the total cross-section for the $PbPb \rightarrow PbPb e^{+}e^{-}$ process relative to centrality (fig \ref{centrality_one}). We have considered ten centrality ranges. Again, we notice the drastic reduction in the cross-section for the case of the overlapping nuclei (condition 4) as centrality increases. 

\begin{table}[tbp]
\centering
	    \caption{Total cross-section values for the two centrality ranges studied in Fig \ref{firstplots} (left: 70-90\%, right: 50-70\%). Inside the nuclei refers to condition 3+4 and outside to condition 2 (see text). We have taken into account the different values of the nuclear radius, characterised by the $R_0$ parameter.}
	\label{thetable}
		{\footnotesize{\begin{tabular}{|c|c|c|c|c|} \hline
		\multicolumn{5}{|c|}{Total cross section (mb)} \\ \hline
		    \multirow{2}{*}{} & 
		   \multicolumn{2}{|c|}{\textbf{Centrality : 70-90\%}} & \multicolumn{2}{|c|}{\textbf{Centrality : 50-70\%}} \\
		   
			 & $R_0 = 1.10$ fm &$R_0 = 1.25$ fm  & $R_0 = 1.10$ fm & $R_0 = 1.25$ fm  \\
			\hline
			Inside nuclei & 0.07842 & 0.11090 & 0.08568 & 0.12528 \\
			\hline  
			Outside nuclei & 0.08876 & 0.08852 & 0.09245 & 0.09231 \\
			\hline
			Full b space & 0.16717 & 0.19942 & 0.17813 & 0.21760 \\
			\hline
		\end{tabular}}}
\end{table}

\section{Conclusions}
We have calculated the differential cross-sections distributions of the $e^+e^-$ pairs for the $PbPb \rightarrow PbPb e^{+}e^{-}$ process as a function of the invariant mass $W$. Calculations have been perfomed using the same kinematical cuts and collision energy ($\sqrt{s_{NN}}=5.02$ TeV) as in ALICE data \cite{alice_data}. We have also considered a range of values for the nuclear radius of Pb: $R=(6.52-7.41)$~fm. 

Comparison between theory and experiment has been done checking with the data, and good agreement has been achieved. This implies that the $\gamma\gamma$ fusion process dominates for the production of low pair transverse momenta dileptons both at peripheral and semi-central collisions (see Fig \ref{firstplots}). It was also concluded that the ratio of production of the $e^{+}e^{-}$ pairs was roughly the same both inside and outside the nuclei (see Table \ref{thetable}). In addition, we have calculated, for the same process, the total cross-sections dependent on ten centrality ranges (see fig \ref{centrality_one}). We notice in both types of distributions (Figures \ref{firstplots} and \ref{centrality_one}) a drastic reduction of both differential and total cross-section for larger values of centrality.

Finally, we checked the dependence of the cross-sections on the $R_0$ parameter. For smaller values of the nuclear radius, lepton pairs were rather produced outside the nuclei. Similar, but opposite case was for larger values of radius. We conclude by emphasising the importance of using a radius range rather than a fixed value.




\end{papers}

\end{document}